%% file: main.tex
\documentclass[
    aps,
    twocolumn,
    superscriptaddress,
    nofootinbib,
    floatfix,
    citeautoscript,
    10pt,
]{revtex4-2}
\DeclareUnicodeCharacter{200B}{}

\usepackage[dvipsnames]{xcolor}
\usepackage{amsthm}
\usepackage{thmtools}
\usepackage{mathtools}
\usepackage{bm} 
\usepackage{bbm} 
\usepackage{braket}
\usepackage{dsfont}
\usepackage{amsmath}
\usepackage{amssymb}
\usepackage{tikz}
\usetikzlibrary{quantikz}
\usepackage[most]{tcolorbox}
\usetikzlibrary{arrows, positioning}
\usepackage{graphicx}
\usepackage{todonotes}

\usepackage[ruled, linesnumbered, vlined]{algorithm2e}
\SetArgSty{textnormal}

\usepackage[colorlinks]{hyperref}
\hypersetup{
    colorlinks  = true,
    citecolor   = PineGreen,
    linkcolor   = MidnightBlue,
    urlcolor    = PineGreen
}
\usepackage{cleveref}
\crefalias{figure*}{figure}

\usepackage{enumitem}

\usepackage{etoolbox}
\apptocmd{\sloppy}{\hbadness 10000\relax}{}{}

\usepackage[table]{xcolor}
\usepackage{booktabs}
\usepackage{multirow}

\usepackage{svg}

\renewcommand\thesection{\arabic{section}}
\renewcommand\thesubsection{\thesection.\arabic{subsection}}
\renewcommand\thesubsubsection{\thesubsection.\arabic{subsubsection}}

\makeatletter
\def\p@section{}        
\def\p@subsection{}     
\def\p@subsubsection{}  
\makeatother

    \DeclarePairedDelimiter{\norm}{\lVert}{\rVert} 

        \newcommand*{\N}{\mathbb{N}}

        \newcommand*{\idty}{\mathds{1}}
        \newcommand*{\modulo}{\, \text{mod} \,}

        \newcommand{\x}{\text{X}}
        \newcommand{\y}{\text{Y}}
        \newcommand{\z}{\text{Z}}
        \newcommand{\rx}{\text{RX}}
        \newcommand{\ry}{\text{RY}}
        \newcommand{\rz}{\text{RZ}}
        \newcommand{\cz}{\text{CZ}}
        \newcommand{\cnot}{\text{CNOT}}
        \newcommand*{\controlled}{C} 

        \newcommand*{\gatecount}{\ensuremath{\mathbf{G}}} 
        \newcommand*{\depth}{\ensuremath{\mathbf{D}}} 

        \newcommand{\parameters}{\bm{\theta}} 
        \newcommand{\variance}{\mathbb{V}} 
        \newcommand{\expectation}{\mathbb{E}} 

        \newcommand{\modelparams}{(F, h, c, b)} 
        \newcommand{\parameterspacelorenz}{\Omega_{\text{L96}}} 
        \newcommand{\nodesignpointsmodel}{n_{\text{design}}^{\text{L96}}} 

        \newcommand{\observations}{\bm{Z}_{\text{obs}}} 
        \newcommand{\trueobservations}{\observations^{\text{true}}} 
        \newcommand{\implausibilitythreshold}{T_{\text{impl}}} 
        \newcommand{\maxnoimplausibilities}{n_{\text{impl}}^{\text{max}}} 
        \newcommand{\designpoints}{\Theta_{\text{design}}} 
        \newcommand{\gp}{\text{GP}} 
        \newcommand{\nroy}{\text{NROY}} 
        \newcommand{\candidates}{\Theta_{\text{cand}}} 
        \newcommand{\candidate}{\bm{\theta}_{\text{cand}}} 
        \newcommand{\nodesignpoints}{n_{\text{design}}} 
        \newcommand{\targets}{f_{\text{targ}}} 
        \newcommand{\observationalvariance}{V_e} 
        \newcommand{\emulatorvariance}{V_{\eta}} 
        \newcommand{\nosamplepoints}{n_{\text{smpls}}} 
        \newcommand{\nosamplepointsforwave}{\nosamplepoints^{(w)}} 

        \newcommand{\itkernel}{k^{\text{IT}}} 
        \newcommand{\itkernelmatrix}{K^{\text{IT}}} 
        \newcommand{\rmkernel}{k^{\text{RM}}} 
        \newcommand{\rmkernelmatrix}{K^{\text{RM}}} 

        \newcommand{\noclusters}{n_{\text{clusters}}^{\text{max}}} 
        \newcommand{\convergencethresholdmetrics}{t_{\text{conv}}} 
        \newcommand{\nohmrepetitionsmin}{n_{\text{repeat}}^{\text{min}}} 
        \newcommand{\nohmrepetitionsmax}{n_{\text{repeat}}^{\text{max}}} 
        \newcommand{\implausibilitythresholdmax}{\implausibilitythreshold^{\text{max}}} 
        \newcommand{\implausibilitythresholdmin}{\implausibilitythreshold^{\text{min}}} 
        \newcommand{\maxnowaves}{n_{\text{waves}}^{\text{max}}} 
        \newcommand{\implausibilitythresholdmindecayfactor}{\lambda_{\text{impl}}^{\text{min}}} 
        \newcommand{\maxruntime}{\tau^{\text{max}}_{\text{HM}}} 
        \newcommand{\trainkernelonlyonce}{\chi_{\text{single-train}}} 
        \newcommand{\randomnessseed}{s_{\text{rand}}} 

        \newcommand{\nohmrepetitions}{n_{\text{repeat}}} 
        \newcommand{\distancerescaled}{d_{\text{resc}}} 
        \newcommand{\meandistancerescaled}{\overline{\distancerescaled}} 
        \newcommand{\optimaldistancerescaled}{\distancerescaled^\ast} 
        \newcommand{\nowaves}{n_{\text{waves}}} 
        \newcommand{\meannowaves}{\overline{\nowaves}} 
        \newcommand{\optimalnowaves}{\nowaves^\ast} 

        \newcommand{\npqcreferenceparamy}{\theta_r^{\text{y}}}

        \newcommand{\dfine}{d-fine GmbH, Frankfurt, Germany}
        \newcommand{\LUH}{Leibniz Universität Hannover, Institut für Theoretische Physik, Hanover, Germany}
        \newcommand{\planqc}{PlanQC GmbH, Garching near Munich, Germany}
        \newcommand{\UniHamburg}{University of Hamburg, Institut für Quantenphysik, Hamburg, Germany}
        \newcommand{\DLR}{Deutsches Zentrum für Luft- und Raumfahrt, Institut für Physik der Atmosphäre, Oberpfaffenhofen, Germany}
        \newcommand{\UniBremen}{University of Bremen, Institute of Environmental Physics (IUP), Bremen, Germany}
        \newcommand{\EqualContributionAndEmail}{These authors contributed equally to this work. \\ Corresponding author is \href{mailto: paul.christiansen@d-fine.com}{PJC}.}

\makeatletter
\@ifclasswith{revtex4-2}{linenumbers}{
    \NewCommandCopy\oldequation\equation
    \NewCommandCopy\endoldequation\endequation
    \renewenvironment{equation}{\linenomath\oldequation}
    {\endoldequation\endlinenomath}

    \NewCommandCopy\oldalign\align
    \NewCommandCopy\endoldalign\endalign
    \renewenvironment{align}{\linenomath\oldalign}
    {\endoldalign\endlinenomath}
}{}
\makeatother

\begin{document}

\title{Quantum Bayesian Optimization for the Automatic Tuning of Lorenz-96 as a Surrogate Climate Model}

\author{Paul J. Christiansen}
\thanks{\EqualContributionAndEmail}
\affiliation{\dfine}
\affiliation{\LUH}

\author{Daniel Ohl de Mello}
\thanks{\EqualContributionAndEmail}
\affiliation{\dfine}

\author{Cedric Brügmann}
\affiliation{\dfine}

\author{Steffen Hien}
\affiliation{\dfine}

\author{Felix Herbort}
\affiliation{\planqc}
\affiliation{\UniHamburg}

\author{Martin Kiffner}
\affiliation{\planqc}

\author{Lorenzo Pastori}
\affiliation{\DLR}

\author{Veronika Eyring}
\affiliation{\DLR}
\affiliation{\UniBremen}

\author{Mierk Schwabe}
\affiliation{\DLR}

\date{\today}

\begin{abstract}
    In this work, we propose a hybrid quantum-inspired heuristic for automatically tuning the Lorenz-96 model -- a simple proxy to describe atmospheric dynamics, yet exhibiting chaotic behavior. Building on the history matching framework by \textcite{lguensat_semi-automatic_2023}, we fully automate the tuning process with a new convergence criterion and propose replacing classical Gaussian process emulators with quantum counterparts. We benchmark three quantum kernel architectures, distinguished by their quantum feature map circuits. A dimensionality argument implies, in principle, an increased expressivity of the quantum kernels over their classical competitors. For each kernel type, we perform an extensive hyperparameter optimization of our tuning algorithm. We confirm the validity of a quantum-inspired approach based on statevector simulation by numerically demonstrating the superiority of two studied quantum kernels over the canonical classical RBF kernel. Finally, we discuss the pathway towards real quantum hardware, mainly driven by a transition to shot-based simulations and evaluating quantum kernels via randomized measurements, which can mitigate the effect of gate errors. The very low qubit requirements and moderate circuit depths, together with a minimal number of trainable circuit parameters, make our method particularly NISQ-friendly.
\end{abstract}


\maketitle

\section{Introduction}\label{section:Introduction}
\input{SECTIONS/01_Introduction}

\section{Preliminaries}\label{section:Preliminaries}
\input{SECTIONS/02_Preliminaries/01_Lorenz-96-Model}
\input{SECTIONS/02_Preliminaries/02_GPs}
\input{SECTIONS/02_Preliminaries/03_History-Matching}

\section{Previous work}\label{section:PreviousWork}
\input{SECTIONS/03_Previous-Work/01_Semi-Automatic-Tuning-of-L96}
\input{SECTIONS/03_Previous-Work/02_QGP-Regression-for-Bayesian-Optimization}

\section{Methods}\label{section:Methods}
\input{SECTIONS/04_Methods/01_Automatic-Tuning-of-L96}
\input{SECTIONS/04_Methods/02_Quantum-Kernel-Evaluation-Methods}
\input{SECTIONS/04_Methods/03_Quantum-Kernel-Architectures}

\section{Results}\label{section:Results}
\input{SECTIONS/05_Results/01_HPO-via-Optuna}
\input{SECTIONS/05_Results/02_Performance-Comparison}
\input{SECTIONS/05_Results/03_Towards-Real-Quantum-Hardware}


\section{Conclusion}\label{section:Conclusion}
\input{SECTIONS/06_Conclusion}

\vspace*{-2.5pt}

\begin{acknowledgments}

\vspace*{-2.5pt}

This project was made possible by the DLR Quantum Computing Initiative and the Federal Ministry of Research, Technology and Space; \href{qci.dlr.de/projects/klim-qml}{qci.dlr.de/projects/klim-qml}. This work used resources of the Deutsches Klimarechenzentrum (DKRZ) granted by its Scientific Steering Committee (WLA) under project ID 1179. PJC thanks Tobias J. Osborne for helpful discussions. V.E. was additionally supported by the Deutsche Forschungsgemeinschaft (DFG, German Research Foundation) through the Gottfried Wilhelm Leibniz Prize awarded to Veronika Eyring (Reference No. EY22/2-1).

\end{acknowledgments}

\twocolumngrid

\bibliography{KLIM-QML}


\appendix
\renewcommand\thesection{\Alph{section}}                 
\renewcommand\thesubsection{\thesection.\arabic{subsection}}    
\renewcommand\thesubsubsection{\thesubsection.\arabic{subsubsection}} 

\section{NPQC Shift Factors}\label{appendix:NPQCShiftFactors}

\input{APPENDIX/01_NPQC}

\section{HPO Results} \label{appendix:HPO}
\input{APPENDIX/02_HPO}

\section{HM Results of Best NPQC Trial} \label{appendix:HM}
\input{APPENDIX/03_HM}

\end{document}

%% file: SECTIONS/01_Introduction.tex
Climate models are steadily improving, yet uncertainties and errors remain \cite{eyring_ai-empowered_2024}. This is largely due to the fact that a significant part of the underlying processes occurs on a spatial scale that is too small to be resolved by global models. Capturing the influence of these effects on the model’s resolved variables still mainly relies on schemes to represent them as simplified parametric functions or \textit{parameterizations} \cite{stensrud_parameterization_2007}. As the underlying parameters are not fully determined by observations, uncertainties are generally associated with them (and the structure of the equations) \cite{schneider_earth_2017}. The process of tuning these model parameters is still largely manual, relying on intuition and domain expertise of experienced modelers \cite{hourdin_art_2017,mauritsen_tuning_2012,schmidt_practice_2017,giorgetta_icon_2018,mignot_tuning_2021}. However, with the continuous growth of climate models in terms of sophistication and complexity, the urge for automated tuning schemes is stronger than ever. 

Considering that climate models usually involve dozens of parameters that interact in complex, nonlinear ways, methods based on machine learning (ML) lend themselves to approaching automation. Over the last years, different ML-assisted strategies have been proposed in this regard to foster automatic parameter tuning in the climate context \cite{hourdin_art_2017,hourdin_toward_2023,jebeile_machine_2023,bonnet_tuning_2024,elsayed_leveraging_2023}.

Generally, these frameworks can be divided into two classes. One of them consists of a rapid optimization of a cost function that calculates the discrepancy between a limited set of observations and the results produced by model simulations, potentially taking into account the sensitivity of the model with regard to the different parameters \cite{bellprat_objective_2012,zhang_automatic_2015}. The other class relies on some form of uncertainty quantification and Bayesian inference. In most cases, the central concept consists of using a surrogate model instead of the expensive-to-evaluate \textit{global circulation model (GCM)} in combination with a Bayesian optimization scheme to efficiently explore the parameter landscape and find the optimal parameter set, while respecting the various uncertainties associated with observations, the GCM, and the emulator.
In essence, this can be expressed as solving a problem of the form
\begin{equation}
    \bm{\theta}^\ast = \underset{\bm{\theta} \in \Omega}{\arg \min} \; g(\bm{\theta})
\end{equation}
with $g(\bm{\theta})$ being the black-box function to optimize, in this case, the deviation between the model to be tuned and some observed ground truth. Here, $\bm{\theta}$ is a point in the parameter space $\Omega$.\footnote{Throughout this work, we will denote vectors with more than one component by bold letters.}

With the recent development of increasingly powerful machine learning tools, especially the latter class of tuning schemes has gained momentum for being closely related to the principle of learning from data (either from observations or high-resolution simulations) \cite{schneider_earth_2017}. Since drawing samples from $g(\bm{\theta})$ typically requires running the global climate model for a sufficient time to allow comparing averages against observed values, any such sample is associated with significant computational resources. For this reason, one strives to find an appropriate surrogate function $f(\bm{\theta})$ that approximates $g(\bm{\theta})$ such that both functions are minimized by the same parameter set $\bm{\theta}^\ast$, while being less expensive to evaluate.
Since the optimal form of $f(\bm{\theta})$ is not known a priori, a common approach is to start with a general ensemble of possible functions and use successive evaluations of $g(\bm{\theta})$ to narrow down the functional form iteratively -- a process known as \textit{Bayesian optimization}. Popular choices for such emulators are \textit{Gaussian processes (GPs)} thanks to being efficient to evaluate while also providing information about the amount of uncertainty associated with the functional form \cite{rasmussen_gaussian_2006}.

Given the urgency of improving the climate models and the rapid progress in quantum computing, it is worth exploring the use of quantum devices in this field already now \cite{schwabe_opportunities_2025}. In this work, we explore the potential of quantum machine learning within a parameter tuning framework. Specifically, we investigate how quantum kernel methods \cite{schuld_supervised_2021}, and in particular, quantum-enhanced Gaussian processes (QGPs) \cite{rapp_quantum_2024}, can be used within a Bayesian optimization framework to find the ideal parameter settings for a given model. 
We argue that QGPs are well-suited for this task, as (i) their underlying quantum feature maps allow for an increased expressivity compared to classical transformations due to an exponentially larger feature (Hilbert) space dimension, and (ii) they do not require extensive training periods like quantum neural networks \cite{beer_training_2020,mcclean_barren_2018,cybulski_impact_2023}. Also, their very limited qubit requirement makes them, in principle, amenable to current and near-future NISQ hardware.

We verify our idea by applying it to a well-studied toy model in the context of parameter tuning and climate modeling: the Lorenz-96 (L96) model \cite{lorenz_predictability_1995}, which can be seen as a strongly simplified atmospheric model. As a tuning scheme, we use \textit{history matching (HM)} \cite{thomas_nonlinear_1972}, which is commonly employed for more or less advanced climate models \cite{lguensat_semi-automatic_2023,williamson_history_2013,williamson_identifying_2015}. For this, we can build on an already existing framework developed by \textcite{lguensat_semi-automatic_2023}. On the classical side, we will refine and extend it in several aspects. Most importantly, we propose a convergence criterion that turns the partly manual HM procedure into a fully automatic process. Then, to strengthen our approach, we benchmark three quantum kernel architectures, differing in how they encode points from the parameter space as states in a Hilbert space. To ensure a robust comparison to the canonical classical RBF kernel, we perform an extensive hyperparameter optimization (HPO) via \texttt{Optuna} \cite{akiba_optuna_2019} for each of the four kernels. Based on the best hyperparameter configurations, we investigate various HM properties and the obtained solutions. Finally, we discuss two strategies to make the transition from a quantum-inspired approach using statevector simulation to executing quantum circuits on real quantum hardware. More specifically, we numerically investigate a statistical ansatz as an alternative quantum kernel evaluation method, as well as the effect of shot noise due to a finite number of measurements. 

In \cref{section:Preliminaries}, we walk through the classical building blocks of our algorithm, including the L96 model in \cref{subsec:Lorenz96Model}, Gaussian processes in \cref{subsec:GPs}, and history matching in \cref{subsec:HistoryMatching}. \Cref{section:PreviousWork} provides a brief overview of the foundational works underpinning this study, specifically the contributions of \textcite{lguensat_semi-automatic_2023} (\cref{subsec:SemiAutomaticTuningofL96}) and the QGP approach by \textcite{rapp_quantum_2024}. \Cref{section:Methods} introduces the key concepts used in the tuning process. This includes classical extensions of \cite{lguensat_semi-automatic_2023} in \cref{subsec:AutomaticTuningOfL96}, as well as the quantum kernel architectures and evaluation methods detailed in \cref{subsec:QuantumKernelArchitectures,subsec:QuantumKernelEvaluationMethods}, respectively. \Cref{section:Results} presents the analysis of hyperparameter optimization with \texttt{Optuna} (\cref{subsec:HPOviaOptuna}), a comprehensive performance comparison (\cref{subsec:PerformanceComparison}) and an outline of our pathway towards a NISQ implementation (\cref{subsec:TowardsRealQuantumHardware}). Finally, \cref{section:Conclusion} summarizes our conclusions and offers an outlook on future research directions.

%% file: SECTIONS/02_Preliminaries/01_Lorenz-96-Model.tex
\subsection{Lorenz-96 Model} \label{subsec:Lorenz96Model}

The \textit{Lorenz-96 (L96) model}, introduced by \textcite{lorenz_predictability_1995} as part of his portfolio of forced dissipative systems with quadratic nonlinear terms \cite{lorenz_deterministic_1963}, is one of the simplest models to describe atmospheric dynamics. In particular, it can have a variable number of dimensions, exhibits chaos for suitably chosen parameter configurations, and - in its full version - covers two different timescales of evolution. Both are coupled linearly to each other:
\begin{subequations} \label{eq:Lorenz96}
\begin{align}
    \frac{\mathrm{d}X_k}{\mathrm{d}t} &= - X_{k-1}\left(X_{k-2} - X_{k+1}\right) - X_k + F - \frac{hc}{b} \sum_{j=1}^J Y_{j,k} \label{eq:L96SlowVariables} \\
    \frac{\mathrm{d}Y_{j,k}}{\mathrm{d}t} &= - c \, b \, Y_{j+1, k} \left(Y_{j+2,k} - Y_{j-1,k}\right) - c \, Y_{j,k} + \frac{hc}{b} X_k. \label{eq:L96FastVariables}
\end{align}
\end{subequations}
for $k \in \{1,...,K\}=:[K]$, periodic boundary conditions
\begin{equation}\label{eq:L96PeriodicBoundaries}
    X_{k+K}=X_k \quad \text{and} \quad Y_{j+J,k}=Y_{j,k} \; , \; Y_{j,k+K}=Y_{j,k}
\end{equation}
and model parameters $\modelparams$. The L96 model \eqref{eq:Lorenz96} amounts to $K$ slowly varying components \eqref{eq:L96SlowVariables} and $J\,K$ fast evolving ones \eqref{eq:L96FastVariables}, making $K(J+1)$ variables in total. The first summands on the right-hand sides (RHSs) of \cref{eq:L96SlowVariables,eq:L96FastVariables} are advection terms, and the linear self-dependence induces diffusion in the system. The slow variables are subject to a forcing $F$. The parameter $h$ solely controls the coupling strength, while $c$ and $b$ correspond to temporal-scale and spatial-scale ratios, respectively \cite{lorenz_predictability_1995,lguensat_semi-automatic_2023}. 

The periodic boundary conditions \eqref{eq:L96PeriodicBoundaries} promote the interpretation of the components being arranged in a latitude circle. A common choice is $K=36$ and $J=10$ \cite{lguensat_semi-automatic_2023,rasp_coupled_2020}, corresponding to a discretization of the latitude circle into 10-degree wide sections, each being decomposed into small 1-degree subpartitions. On the other hand, the configuration
\begin{equation}\label{eq:L96ParameterTruth}
    F = 10, \quad h = 1, \quad  c = 10, \quad  b = 10
\end{equation}
is ubiquitous in the literature \cite{schneider_earth_2017,lguensat_semi-automatic_2023,lorenz_predictability_1995,rasp_coupled_2020} for exhibiting behavior closest to the atmosphere. This setting corresponds to a factor of ten between the fluctuations of the fast and the slow timescale, as well as to the inverse in terms of amplitude. Nevertheless, the L96 model should not be thought of as describing the real atmosphere; it is rather (one of) the simplest formulations that still manages to resemble its chaotic dynamics to a small extent. L96 can hence serve as a simplistic proxy for climate models in cases where the qualitative behavior shall be investigated at low simulation cost, making it a suitable surrogate model for our work. 

%% file: SECTIONS/02_Preliminaries/02_GPs.tex
\subsection{Gaussian Processes} \label{subsec:GPs}

Gaussian processes are the generalization of Gaussian distributions to the infinite-dimensional function space. More formally, a \textit{Gaussian process (GP)} is a collection of random variables, for which any finite subset follows a joint (multivariate) Gaussian distribution \cite[Definition 2.1]{rasmussen_gaussian_2006}. The random variables are here given by the function values $h(\bm{x})$ at a point $\bm{x}$ in a continuous, potentially multi-dimensional domain $\Omega$. As a subcategory of stochastic processes, GPs are often defined over time, which is, however, not obligatory and will not be the case in our setting. Like Gaussian distributions, GPs are fully determined by a mean function $m(h(\bm{x})) \equiv m(\bm{x})$ and a covariance function or \textit{kernel} $k(h(\bm{x}), h(\bm{x}^\prime)) \equiv k(\bm{x}, \bm{x}^\prime)$, 
%
\begin{align*}
    m(\bm{x}) &= \expectation[h(\bm{x})] \, ,\\
    k(\bm{x}, \bm{x}^\prime) &= \expectation\left[\left(h(\bm{x}) - m(\bm{x})\right) \left(h(\bm{x}^\prime) - m(\bm{x}^\prime)\right)\right] \, ,
\end{align*}
%
where $\expectation[\cdot]$ denotes the expectation value. We then write 
\begin{equation} \label{eq:GP}
    h(\bm{x}) \sim \mathcal{GP}(m(\bm{x}), k(\bm{x}, \bm{x}^\prime))
\end{equation}
if for any finite subset $X = \{\bm{x}_1, ..., \bm{x}_n\}$, 
\begin{equation} \label{eq:GPPrior}
    \bm{h}(X) \sim \mathcal{N}(\bm{m}(X), K(X, X)).
\end{equation}
Following standard notation \cite{rasmussen_gaussian_2006}, we here chose the capital letter $K$ to denote the \textit{kernel matrix} or \textit{Gram matrix} resulting from evaluating the kernel function $k$ on every combination of input points. In general, a kernel is a symmetric, positive definite function $k: \Omega \times \Omega \to \mathbb{R}$ with $k(\bm{x},\bm{x})=1$. It can be understood as a similarity measure between pairs of input points $\bm{x}, \bm{x}^\prime \in \Omega$. The symmetry requirement implies that computing $K(X,X)$ reduces to evaluating the kernel on the $n(n-1)/2$ independent index combinations $(i,j)$, $i \in \{1,...,n\}, j \in \{i,...,n\}$. For two distinct datasets $X$ and $X^\prime$ with $n$ and $n^\prime$ data points, however, constructing $K(X,X^\prime)$ requires the full $n \cdot n^\prime$ calls to $k$. The mean is often set to be zero from the outset, $m(\bm{x})\equiv 0$ or $\bm{m}(X)\equiv 0$, making the kernel the crucial quantity \cite{rasmussen_gaussian_2006}. To foster expressivity, the input points are usually first transformed according to a non-linear \textit{feature map} $\bm{\phi}: \Omega \to \mathcal{F}$, such that a linear model can be employed in this feature space $\mathcal{F}$ \cite{rasmussen_gaussian_2006}. The kernel then computes some inner product on $\mathcal{F}$: 
\begin{equation} \label{eq:GPKernelInnerProduct}
    k(\bm{x}, \bm{x}^\prime) = \langle \bm{\phi}(\bm{x}), \bm{\phi}(\bm{x}^\prime)\rangle_\mathcal{F}.
\end{equation}
A prominent example is the \textit{radial basis function (RBF) kernel}
\begin{equation}\label{eq:RBFKernel}
    k^{\text{RBF}}_s(\bm{x},\bm{x}^\prime) = \exp \left(-\frac{\norm{\bm{x} - \bm{x}^\prime}^2}{2s^2}\right)
\end{equation}
with Euclidean norm $\norm{\cdot}$ and free parameter $s \in \mathbb{R}$.

Drawing random functions from the prior distribution \eqref{eq:GPPrior}, short the \textit{prior}, is usually of secondary interest. In realistic scenarios, one usually has access to a number of \textit{training points} $\{\bm{x}_1, ..., \bm{x}_n \} = X$ and associated observations $(y_1, ..., y_n) = \bm{y}$, with the latter corresponding to noisy versions of the wanted function values, 
\begin{equation} \label{eq:NoisyObservations}
    y_i = h(\bm{x}_i) + \epsilon,
\end{equation}
for some error $\epsilon$. With zero mean in \cref{eq:GPPrior}, the prior on the noisy measurements becomes 
\begin{equation} \label{eq:GPPriorObservations}
    \bm{y} \sim \mathcal{N}(0, K(\bm{y}, \bm{y})).\footnote{Recall our notation $K(X,X) \equiv K(\bm{h}(X), \bm{h}(X))$.}
\end{equation}
Under the  conventional assumption of independent (additive) noise that follows a zero-mean normal distribution with variance $\sigma^2$ \cite{rasmussen_gaussian_2006}, \cref{eq:NoisyObservations} implies that the covariance function of the transformed prior \eqref{eq:GPPriorObservations} simply evaluates to
\begin{equation*}
    k(y_i, y_j) = k(\bm{x}_i, \bm{x}_j) + \sigma^2 \delta_{ij} \, ,
\end{equation*}
or, equivalently,
\begin{equation} \label{eq:GPPriorObservationsCovariance}
    K(\bm{y},\bm{y}) = K(X,X) + \sigma^2 \, \mathds{1}.
\end{equation}
It is, in fact, not unusual for kernel functions to have one or even multiple free parameters like $s$ in the RBF kernel \eqref{eq:RBFKernel}. Determining a (near) optimal choice for them is commonly referred to as \textit{training} a Gaussian process. The quality of a parameter value is assessed based on the marginal likelihood
\begin{equation*}
    p(\bm{y} \, | \, X) = \int p(\bm{y} \, | \, \{\bm{h}, X\})  \, p(\bm{h} \, | \, X) \mathrm{d}\bm{h}.
\end{equation*}
In our zero-mean setting $\bm{h} \, | \, X \sim \mathcal{N}(0,K)$, with shorthand notation $K:=K(X,X)$, we find 
\begin{equation*}
    \log p(\bm{h} \, | \, X) = - \frac{1}{2} \bm{h}^T K^{-1} \bm{h} - \frac{1}{2} \log (\det K) - \frac{n}{2} \log(2\pi).
\end{equation*}
In \cite{rasmussen_gaussian_2006}, it is shown that this yields the \textit{log marginal likelihood}
\begin{align} \label{eq:LogMarginalLikelihood}
    \log p(\bm{y} \, | \, X) = &- \frac{1}{2} \bm{y}^T \left(K + \sigma^2\, \mathds{1}\right)^{-1} \bm{y} \\
    &- \frac{1}{2} \log\left(\det \left(K + \sigma^2 \, \mathds{1}\right)\right) - \frac{n}{2} \log(2\pi) \, , \nonumber
\end{align}
which also follows from combining \cref{eq:GPPriorObservations,eq:GPPriorObservationsCovariance}. Depending on the number of free parameters and the optimization landscape, maximizing \cref{eq:LogMarginalLikelihood} is a more or less complex task.

Once the GP is trained, the main task of interest is to make predictions based on a set of test points $\{\bm{x}^{\ast}_1, ..., \bm{x}^{\ast}_n \} =: X^\ast$. To this end, we need the posterior distribution, short the \textit{posterior}, which can be understood as the updated distribution that results from incorporating the information about the observations in the prior, and feeding it with the test points. More specifically, the posterior may be written as the conditional Gaussian 
\begin{equation} \label{eq:GPPosterior}
    h(X^\ast) \, | \, \{X,\bm{y}\} \sim \mathcal{N} (\bm{m}^\ast(X^\ast), \text{Cov}^\ast(X^\ast, X^\ast)).
\end{equation}
Let us additionally define 
\begin{equation*}
    K^\ast:=K(X^\ast,X), K_\ast:=K(X,X^\ast), K^\ast_\ast:=K(X^\ast,X^\ast).
\end{equation*}
Then, starting from \cref{eq:GPPriorObservationsCovariance}, the mean and covariance of the posterior \eqref{eq:GPPosterior} are given by
\begin{subequations}
\begin{align}
    \bm{m}^\ast(X^\ast) &:= \expectation [h(X^\ast) \, | \, \{X, Y\}] \nonumber \\
    &= K^\ast \left[K + \sigma^2 \mathds{1}\right]^{-1} \, \bm{y} \, , \\
    \text{Cov}^\ast(X^\ast, X^\ast) &= K^\ast_\ast - K^\ast \left[ K + \sigma^2 \mathds{1} \right]^{-1} \, K_\ast \, , \label{eq:GPPosteriorCovariance}
\end{align}
\end{subequations}
as derived in \cite{rasmussen_gaussian_2006}. It is a characteristic property of Gaussian processes that the predictive covariance \eqref{eq:GPPosteriorCovariance} is entirely governed by the prior kernel, meaning that it is independent of the observed targets. Note that the variance of the Gaussian measurement error in \cref{eq:NoisyObservations} carries over to both the covariance \eqref{eq:GPPriorObservationsCovariance} of the transformed prior \eqref{eq:GPPriorObservations} and the covariance \eqref{eq:GPPosteriorCovariance} of the posterior \eqref{eq:GPPosterior}. Effectively, it acts as a regularization, often promoting numerical stability \cite{rapp_quantum_2024}.

Up to this point, we described only standard Gaussian processes with a single output dimension. \Cref{eq:GP} can be extended straightforwardly to multi-output GPs, which generalize 1D GPs in the same fashion as multivariate Gaussian distributions generalize 1D normal distributions. In our setting, we will exclusively work with these multi-output GPs.

%% file: SECTIONS/02_Preliminaries/03_History-Matching.tex
\subsection{History Matching} \label{subsec:HistoryMatching}

\textit{History matching} (HM) is one possible approach for reducing the large amount of subjectivity inherent in the manual schemes that are still common in tuning climate models. It goes back to the simulation of oil sources \cite{thomas_nonlinear_1972,craig_bayes_1996,pievatolo_bayes_2013}, can however also be found in distinct fields like the formation of galaxies in the early universe \cite{vernon_galaxy_2010}, before expanding to (more or less complex) climate models \cite{edwards_precalibrating_2011,williamson_history_2013,williamson_identifying_2015,lguensat_semi-automatic_2023,bonnet_tuning_2024}.

Roughly speaking, history matching is a routine that approaches the optimal values of initially unknown parameters in a given model in a reversed direction: Instead of successively drawing individual samples $\parameters$ from the parameter space $\Omega$ and improving the corresponding value of the black-box function, $g(\parameters)$, HM iteratively rules out implausible regions of $\Omega$ to narrow down the pool of potential candidates for the true solution. Starting from a dense initial sample ensemble $\Theta$ of size $\nosamplepoints$, it shrinks the space of the not-ruled-out-yet parameter configurations, short the \textit{NROY space}, in each of the so-called \textit{waves}. The exclusion of samples is based on a suitably chosen implausibility metric that usually scales with the discrepancy between observations and model output. If a specified threshold is exceeded, the parameter configuration in question is deemed implausible.

In a Bayesian optimization setting where the computationally expensive black-box function $g(\parameters)$ is substituted by an emulator $f(\parameters)$, HM yields an approximate solution to the optimization problem 
\begin{equation}\label{eq:OptimizationProblem}
    \parameters^\ast = \underset{\parameters \in \Omega}{\arg \min} \norm{\observations - f(\parameters)}_M
\end{equation}
where $\norm{\cdot}_M$ is the \textit{Mahalanobis distance}  
\begin{align}\label{eq:MahalanobisDistance}
\begin{split}
    \norm{&\observations - f(\parameters)}_M = \\
    &\sqrt{[\observations - f(\parameters)]^T \variance[\observations - f(\parameters)]^{-1} [\observations - f(\parameters)]}
\end{split}
\end{align}
with variance $\variance[\cdot]$. The implausibility is then defined \cite{lguensat_semi-automatic_2023,williamson_tuning_2017} as 
\begin{equation}\label{eq:HMGeneralImplausibility}
    I_f(\observations, \parameters) = \norm{\observations - \expectation [f(\parameters)]}_M.
\end{equation}
Taking into account potential inaccuracies in the measured observations $\observations$ (e.g., due to instrumental uncertainties) compared to the true observations $\trueobservations$ under ideal settings, we may rewrite the variance implicitly included in \cref{eq:HMGeneralImplausibility} as 
\begin{align}\label{eq:VarianceInImplausibility}
    \variance &\left[\observations - \expectation[f(\parameters)]\right] = \nonumber \\
    &\variance\left[(\observations - \trueobservations) + (\trueobservations - f(\parameters)) + f(\parameters) - \expectation[f(\parameters)]\right] \nonumber \\
    &=: V_e + V_{\eta} + \variance[f(\parameters)]
\end{align}
where $\observationalvariance$ and $\emulatorvariance$ denote the error variances related to the observations themselves and the limited emulator accuracy, respectively, following the notation in \cite{lguensat_semi-automatic_2023,williamson_tuning_2017}. The last equation in \cref{eq:VarianceInImplausibility} assumes that $\observationalvariance$ and $\emulatorvariance$ are statistically independent and uses $\variance[\expectation[\cdot]] \equiv 0$. If the implausibility is evaluated on a single-component level as suggested in \cite{lguensat_semi-automatic_2023}, combining \cref{eq:HMGeneralImplausibility} with \cref{eq:MahalanobisDistance,eq:VarianceInImplausibility} yields 
\begin{equation}\label{eq:HMImplausibility}
    I_f(\observations, \parameters)_z = \frac{\left|(\observations)_z - \expectation[f(\parameters)]_z\right|}{\sqrt{\variance[f(\parameters)]_z + \observationalvariance + \emulatorvariance}}.
\end{equation}
for $z \in [Z:=|\observations|]$.\footnote{Note that some authors define the implausibility via the reduced form $I_f(\observations, \parameters)_z=((\observations)_z - \expectation[f(\parameters)]) / \sqrt{\variance[f(\parameters)]_z}$ \cite{williamson_history_2013,williamson_identifying_2015,bonnet_tuning_2024}.} 

For each component $z$, it is then checked whether $I_f(\observations, \parameters)_z > \implausibilitythreshold$. Typically, $\implausibilitythreshold = 3$ is set according to the 3-sigma rule of \textcite{pukelsheim_three_1994}, stating that at least 95\% of the probability mass corresponding to a unimodal distribution is contained within a distance of three standard deviations from its mean. A sample $\parameters \in \nroy$ is then discarded if the maximum permitted number of implausible components is exceeded, i.e., if
\begin{equation}\label{eq:HMImplausibilityCheck}
    \left|\{z \in [Z]: I_f(\observations, \parameters)_z > \implausibilitythreshold\}\right| > \maxnoimplausibilities.
\end{equation}
Next to the implausibility threshold, this number $\maxnoimplausibilities$ is a hyperparameter of the HM algorithm and needs to be specified by the user.

The property of Gaussian processes (GPs) to directly output the associated variance makes them a favorable emulator choice in the context of BO-based history matching. Moreover, GPs are efficient to evaluate due to the limited number of fitting parameters. In each wave, a GP is then trained for a fixed amount of fresh \textit{design points}, with targets given by the observations. This number remains invariant throughout the full HM procedure; $\nodesignpoints=10d$ design points for $d$ model parameters has become established in practice \cite{chapman_arctic_1994,jones_efficient_1998,loeppky_choosing_2009}. Drawing the new training points from the NROY space is a non-trivial task and can still be considered an open research problem \cite{lguensat_semi-automatic_2023,andrianakis_bayesian_2015,garbuno-inigo_history_2020}. Its difficulty arises from the characteristic that the NROY space can, in principle, be highly disconnected. Also, there is no uniform behavior that it evolves according to, meaning that its shape may change completely from one wave to the next. The general rationale here is to find a reasonable balance between exploring the parameter space exhaustively on the one hand and exploiting the information about the structure of the NROY space gained in previous waves on the other. 

History matching in general should be understood as a high-level recipe for tuning the parameters of a given black-box model, which needs to be tailored to the specific situation at hand. This also includes the definition of a break criterion, a question we will come back to later (see \cref{subsec:SemiAutomaticTuningofL96,subsec:AutomaticTuningOfL96}). The core GP-based process is outlined in \cref{alg:HistoryMatching}.

\begin{algorithm}
    \caption{HM$(\observations, \Omega, g, d, \nosamplepoints, \implausibilitythreshold, \maxnoimplausibilities, \texttt{BC})$}
    \label{alg:HistoryMatching}
        Generate a dense ensemble $\Theta$ of $\nosamplepoints$ samples $\parameters$ from the parameter space $\Omega$ \label{line:SamplingOfParameterSpace}\\
        Choose an emulator $f$ for $g$, e.g., a GP architecture \\
        Initialize $\nroy := \Theta$ \\
        Initialize a wave number counter $w := 0$ \\
        \While{\texttt{BC} not satisfied}{
            Increment wave number counter: $w \gets w + 1$ \\
            Draw $\nodesignpoints=10d$ design points $\designpoints$ \label{line:DrawDeisgnPoints} \\
            Compute targets $\targets=f(\designpoints)$ \label{line:ComputeTargets}\\
            Train $\gp_w=\gp(\designpoints, \targets)$ \\
            Evaluate the implausibility $I_w(\parameters)_z := I_{\text{GP}_w}(\observations, \parameters)_z$ for each component $z \in [Z]$ and every point $\parameters \in \nroy$ via \cref{eq:HMImplausibility} \\
            Update the NROY space: $\nroy \gets \nroy \setminus \{\parameters \in \nroy: \text{\cref{eq:HMImplausibilityCheck} satisfied}\}$ 
        }
        \If{$\nroy \neq \emptyset$}{
            Determine candidate solutions $\candidates$ from $\nroy$ \label{line:DetermineCandidates} \\
            Evaluate the implausibilities $I_w(\candidate)_z$ for each candidate $\candidate \in \candidates$ with $z \in [Z]$ via \cref{eq:HMImplausibility} \\
            Update the set of candidates: $\qquad \qquad \qquad \qquad \candidates \gets \candidates \setminus \{\candidate \in \candidates: \text{\eqref{eq:HMImplausibilityCheck} satisfied}\}$ \\
            \If{$\candidates \neq \emptyset$}{
                \Return $\text{argmin}_{\candidate \in \candidates} \; \frac{1}{Z} \sum_{z=1}^Z I_w(\candidate)_z$
            }
            \Else{
                \Return ``History matching failed!" \label{line:HMFailedForNoFeasibleCandidates}
            }
        }
        \Else{
            \Return ``History matching failed!" \label{line:HMFailedForEmptyNROY}
        }
        
\end{algorithm}

It should be stressed that \cref{line:HMFailedForNoFeasibleCandidates,line:HMFailedForEmptyNROY} in \cref{alg:HistoryMatching} reveal the distinguishing HM property of being able to fail, which may indicate that the employed model is not suitable (or expressive enough) for explaining the observed data.

%% file: SECTIONS/03_Previous-Work/01_Semi-Automatic-Tuning-of-L96.tex
\subsection{Semi-Automatic Tuning of the Lorenz-96 Model} \label{subsec:SemiAutomaticTuningofL96}

\textcite{lguensat_semi-automatic_2023} configured the raw HM scheme sketched in \cref{alg:HistoryMatching} for the Lorenz-96 model, beginning with the definition of a parameter space:
\begin{equation}\label{eq:L96ParameterSpace}
    \parameterspacelorenz := [-20,20]_F \times [-2, 2]_h \times [0,20]_c \times [-20,20]_b
\end{equation}
The physical motivation behind the interval boundaries is rather small; upper and lower bounds are simply chosen as twice the respective true value from \cref{eq:L96ParameterTruth}, differing only in sign. The parameter $c$ forms an exception, because it corresponds to a time-related quantity, which cannot take negative values.

In contrast to tuning an advanced climate model, we do not have actual observations of the Earth system. Instead, we need ``observations" of L96 to mimic the mature climate case. In \cite{lguensat_semi-automatic_2023}, they are created by evolving the Lorenz-96 model based on the parameter truth \eqref{eq:L96ParameterTruth}. More specifically, a first short simulation consisting of 10 \textit{model time units} (MTUs) starts from the initial state\footnote{Only a slight perturbation in one (arbitrary) component is added in the initial state \eqref{eq:L96InitialState} to not let the system start in its fixpoint $\bm{X}=(F,...,F) \in \mathbb{R}^{36} , \bm{Y}=(0,...,0) \in \mathbb{R}^{360}$ where all derivatives in \cref{eq:Lorenz96} vanish.} 
\begin{align}\label{eq:L96InitialState}
\begin{split}
    X_k = F &\quad \forall k \in [36] \setminus \{19\} \, , \quad X_{19}=F + 0.01 \\ 
    &\text{and} \quad Y_{j,k}=0 \quad \forall j\in [10], k \in [36]
\end{split}
\end{align}
and, by design, terminates in the L96 attractor (see, e.g., \cite{van_kekem_dynamics_2018} for an extensive study of the dynamics L96 undergoes). Based on this state, a larger 100-MTU simulation gives a trajectory that is considered the ground truth. The physical observables in real climate models are here replaced by first- and second-order momenta of both variables, 
\begin{equation}\label{eq:L96Metrics}
    \left(\langle \bm{X} \rangle_{10}^{110}, \langle \bar{\bm{Y}} \rangle_{10}^{110}, \langle \bm{X}^2 \rangle_{10}^{110}, \langle \bm{X} \bar{\bm{Y}}\rangle_{10}^{110}, \langle \bar{\bm{Y}}^2 \rangle_{10}^{110} \right)
\end{equation}
where 
\begin{equation*}
    \langle A\rangle_{t_i}^{t_f} := \frac{1}{t_f - t_i} \int_{t_i}^{t_f} A(t)\mathrm{d}t
\end{equation*}
is the temporal average of $A(t)$ over the interval $[t_i, t_f]$. The 180-dimensional statistical quantities vector \eqref{eq:L96Metrics} serves as a metric to compare the model outcomes for different parameter configurations. The large dimensionality of \eqref{eq:L96Metrics} poses a challenge for fitting a full multivariate GP (four inputs, 180 outputs). Instead of following the naive approach of replacing the high-dimensional GP by a multitude of single-output GPs \cite{alvarez_kernels_2012,wackernagel_multivariate_2003,boyle_dependent_2004,duvenaud_additive_2011}, a \textit{principal component analysis} (PCA) is used in \cite{lguensat_semi-automatic_2023}, which is designed to cover at least 99\% of the total variance. This allows to avoid losing track of correlations and the resulting loss of information. In the context of \cite{lguensat_semi-automatic_2023}, the PCA reduces the dimension of the target metrics from 180 to eight.\footnote{The feasibility of such a dimensionality reduction ansatz was confirmed in, e.g., \cite{wilkinson_bayesian_2010}, where the author demonstrates that fitting a GP in the PCA-reduced space can achieve a similar accuracy compared to the full multivariate emulator, especially for large training data sets.} Then, the standard RBF kernel \eqref{eq:RBFKernel} is used to perform the reduced 4-to-8 GP regressions.

Next to the canonical choice $\implausibilitythreshold = 3$ (cf. \cref{subsec:HistoryMatching}), \textcite{lguensat_semi-automatic_2023} are very restrictive in prohibiting any implausible component, meaning $\maxnoimplausibilities = 0$.

The first algorithmic component to be configured is the initial dense sampling of the parameter space \eqref{eq:L96ParameterSpace} in \cref{line:SamplingOfParameterSpace} of \cref{alg:HistoryMatching}. A common approach to achieve as much uniformity as possible is the \textit{Latin hypercube sampling} (LHS) \cite{mckay_comparison_1979}, which is employed in \cite{lguensat_semi-automatic_2023} and also \cite{williamson_tuning_2017}, here yielding $\nosamplepoints = 10^6$ sample points.

For the L96 model with $d=4$, each wave starts off by drawing $\nodesignpointsmodel=10d=40$ design points. LHS can, however, not be reused for this task, because -- as explained in \cref{subsec:HistoryMatching} -- the NROY space will in general not be a hypercube (or rather a hyper-rectangle) anymore. Inspired by \cite{bower_galaxy_2010,boukouvalas_bayesian_2014}, \textcite{lguensat_semi-automatic_2023} work with a simple rejection sampling method: In every wave, the procedure of sampling from the full parameter space via LHS and refusing samples according to the feasibility check \eqref{eq:HMImplausibilityCheck} is repeated until the desired $40$ design points are reached. More specifically, each sample is tested against all GPs that were trained in previous waves. With an increasing amount of waves and shrinking NROY space, it becomes more challenging to find feasible parameter configurations. To mitigate this issue, the authors upscale the LHS size throughout the HM run as $\nosamplepointsforwave = \lceil \nodesignpointsmodel / r_{w-1} \rceil$ for $w \geq 1$ where $r_w:=|\nroy_w|/\nosamplepoints$ measures the sample space reduction ratio in wave $w$.

In \cite{lguensat_semi-automatic_2023}, there is no sharply defined break criterion that makes \cref{alg:HistoryMatching} terminate. Because the roll-out of new waves is manually triggered, their approach should be considered semi-automatic. The conducted numerical study is stopped after six waves, based on observing a low NROY ratio $r_6=0.02\%$. Following \cite{williamson_tuning_2017}, this decision is driven by an economic consideration, estimating that additional waves would not reduce the NROY space significantly any further.

Lastly, for the open problem of determining a number of candidate solutions in \cref{line:DetermineCandidates} of \cref{alg:HistoryMatching}, \textcite{lguensat_semi-automatic_2023} suggest to partition the final NROY space into $k$ clusters via the \textit{$k$-means} algorithm \cite{lloyd_least_1982}, whose centroids are then used as candidates. As it cannot be guaranteed that these cluster centers are contained in the remaining NROY space, they must be tested for implausibility according to \cref{eq:HMImplausibilityCheck} -- and potentially be discarded as well. In order to decide on an amount $k$ of clusters to fit, it is further proposed to utilize the \textit{silhouette score} \cite{rousseeuw_silhouettes_1987}: Among a handful of examined values for $k$, the one with the largest silhouette score is selected for computing the ultimate candidates, from which the HM solution is in turn deduced via minimal mean implausibility.

Note that the authors investigate the effect of incorporating domain expert knowledge into their HM framework. In particular, for them, it boils down to restricting the parameter space \eqref{eq:L96ParameterSpace} further from the outset. However, in this work, we stick to $\parameterspacelorenz$ as defined in \cref{eq:L96ParameterSpace} without any physical prior.

%% file: SECTIONS/03_Previous-Work/02_QGP-Regression-for-Bayesian-Optimization.tex
\subsection{Quantum Gaussian Process Regression for Bayesian Optimization} \label{subsec:QuantumGPRegressionForBayesionOptimization}

The definition of a GP kernel via the inner product on some feature space in \cref{eq:GPKernelInnerProduct} lends itself to a straightforward extension to Hilbert spaces. In full generality, where quantum states are then described by density matrices $\rho(\bm{x})$, the \textit{quantum kernel} is defined via the Hilbert-Schmidt inner product \cite{schuld_supervised_2021},
\begin{equation} \label{eq:QuantumKernelFidelity}
    k(\bm{x}, \bm{x}^\prime) = \text{Tr}[\rho(\bm{x}) \rho(\bm{x}^\prime)].
\end{equation}
For pure states $\rho(\bm{x})=\ket{\psi(\bm{x})} \bra{\psi(\bm{x})}$, the fidelity \eqref{eq:QuantumKernelFidelity} reduces to the overlap
\begin{equation} \label{eq:QuantumKernelOverlap}
    k(\bm{x}, \bm{x}^\prime) = \left|\braket{\psi(\bm{x}) | \psi(\bm{x}^\prime)}\right|^2.
\end{equation}
Both the general \cref{eq:QuantumKernelFidelity} and the more specific \cref{eq:QuantumKernelOverlap} obey the requirements of symmetry and positive definiteness discussed in \cref{subsec:GPs}, making them proper kernel functions that can readily be used in the GP framework. In their work, \textcite{rapp_quantum_2024} propose to create \textit{quantum Gaussian processes (QGPs)} by replacing the classical covariance function \eqref{eq:GPKernelInnerProduct} in favor of the quantum kernel \eqref{eq:QuantumKernelOverlap} and using these QGPs for regression tasks. The crucial part in the transition from classical to quantum is the specification of a quantum feature map $U: \Omega \to \mathcal{U}(\mathcal{H})$, which is used instead of its classical counterpart $\bm{\phi}: \Omega \to \mathcal{F}$ to encode the classical data in a quantum system. Here, $\mathcal{U}(\mathcal{H})$ denotes the group of unitary operators acting on Hilbert space $\mathcal{H}$. Usually, $\mathcal{H}=\left(\mathbb{C}^2\right)^{\otimes N}$ and the encoding is achieved by wrapping the input points as parameters of a parameterized quantum circuit (PQC), such that the pure states in \cref{eq:QuantumKernelOverlap} can be described by
\begin{equation} \label{eq:QuantumFeatureMapEncoding}
    \ket{\psi(\bm{x})} = U(\bm{x}) \ket{\bm{0}}
\end{equation}
where $\ket{\bm{0}} \equiv \ket{0}^{\otimes N}$ abbreviates the all-zero state on $N$ qubits. For the sake of completeness, the density matrices in \cref{eq:QuantumKernelFidelity} are then given by
\begin{equation*}
    \rho(\bm{x}) = U(\bm{x}) \ket{\bm{0}}\bra{\bm{0}} U^\dagger(\bm{x}).
\end{equation*}
Despite the decisive role of the encoding, there is no general rule for how to optimally construct the actual PQC, which is why one must mostly resort to heuristic approaches \cite{schuld_supervised_2021,rapp_quantum_2024}. Unlike in the classical case, specifying an encoding is, however, not sufficient to fully determine a quantum kernel: While the results from evaluating a classical covariance function may be directly post-processed, accessing the information contained in a quantum circuit is a nontrivial task in its own right. Specifically, the number of measurements required to read out all amplitudes of the corresponding quantum state scales exponentially with the number of qubits, which becomes particularly challenging for larger qubit requirements. 

Similar to their classical analogs (cf. \cref{subsec:GPs}), quantum feature maps often come with a (relatively small) number of free parameters. Optimizing the parameters of a PQC is still an active area of research, especially regarding Barren plateaus, see, e.g., \cite{mcclean_barren_2018,sack_avoiding_2022,sannia_engineered_2024,nadori_batched_2025}. As this optimization is outsourced to a classical device, QGPs in the sense of \cite{rapp_quantum_2024} should rather be considered hybrid instruments, just like most QML techniques.

%% file: SECTIONS/04_Methods/01_Automatic-Tuning-of-L96.tex
\subsection{Automatic Tuning of the Lorenz-96 Model} \label{subsec:AutomaticTuningOfL96}

We improve the L96-HM configuration by \textcite{lguensat_semi-automatic_2023} in three aspects, which shall be outlined according to the following order: (i) a more sophisticated sampling technique reduces the number of GP evaluations required for drawing design points; (ii) a quantitative convergence criterion turns the partly-manual tuning procedure into a fully automated process; (iii) a refined candidate selection heuristic ensures that the HM does not fail from the absence of feasible candidate solutions.

(i) As described in \cref{subsec:SemiAutomaticTuningofL96}, the sample size of the naive rejection sampling used in \cite{lguensat_semi-automatic_2023} scales inversely with the cardinality of the NROY space. Additionally, there is an extra GP for each preceding wave against which the drawn samples must be tested for implausibility. Hence, the number of emulator evaluations (counted once per input point) grows as $\mathcal{O}(w/|\nroy|)$. Although this can surely be considered suboptimal, simplicity wins over efficiency in the classical case, as the HM runtime is heavily dominated by computing the targets for the new design points (running the Lorenz-96 model, computing the metrics, and reducing them via the PCA; cf. \cref{line:ComputeTargets} in \cref{alg:HistoryMatching}). While we find this to still hold when using statevector simulation (with proper code optimization) for QGPs (see \cref{subsec:HPOviaOptuna}), the quantum emulator clearly becomes the bottleneck of our quantum HM\footnote{Note that we call it \textit{quantum HM} for the same reason we call it \textit{quantum GP}, knowing that both concepts arguably consist of at least as many classical as quantum parts.} when making the step to real quantum hardware (cf. \cref{subsec:TowardsRealQuantumHardware}). Instead of the expensive rejection sampling, we employ a noisy maximin ansatz to draw the design points. More specifically, after picking $\nodesignpointsmodel$ points from the current NROY space according to a maximin heuristic, our method adds some uniformly distributed noise to them and performs the implausibility check \eqref{eq:HMImplausibilityCheck} against all previously fitted (Q)GPs; the set of surviving parameter configurations is then extended by the same procedure until the desired number $\nodesignpointsmodel$ of feasible design points is reached. In each of these iterations, the samples are selected successively according to the largest minimum distance to any of the already selected points (hence the name \textit{maximin}); the initial sample is drawn uniformly at random. Our sampling strategy clearly shifts the exploration-exploitation trade-off (cf. \cref{subsec:HistoryMatching}) towards exploiting the NROY structure and can really be described by sampling from the NROY space. The noise is added artificially to allow finding favorable parameter configurations that have possibly been overlooked in previous waves. In summary, replacing the rejection sampling in \cite{lguensat_semi-automatic_2023} by our noisy maximin sampling can be understood as shifting the complexity from many emulator evaluations to computing pairwise distances between all NROY points and the current design points. In particular, it eliminates the NROY cardinality dependence, reducing the emulator-call scaling to $\mathcal{O}(w)$.

(ii) When working with a true climate model, the target observations of the Earth system are always associated with an uncertainty that comes from averaging over multiple measurements at different times (or from various instruments), or from the intrinsic uncertainty corresponding to the measurement instruments themselves. Similar to generating ``observations" of the Lorenz-96 model in the first place (cf. \cref{subsec:Lorenz96Model}), we aim to mimic the real-world case by artificially creating observational uncertainties, which will ultimately give rise to a convergence criterion for our HM tuning scheme. To this end, we draw a set of 300 random parameter configurations from the parameter space \eqref{eq:L96ParameterSpace}, calculate the full metrics \eqref{eq:L96Metrics} and apply a 99\%-coverage PCA as outlined in \cref{subsec:Lorenz96Model}. For each principal component, 5\% of the range of obtained values is set as the uncertainty corresponding to the observation of that component. At the beginning of each wave, the PCA-transformed metrics (targets in \cref{line:ComputeTargets}, \cref{alg:HistoryMatching}) are then compared against the defined uncertainties. A sample is considered close to the parameter truth \eqref{eq:L96ParameterTruth} if all components of the new targets are contained within the uncertainty hyperellipsoids spanned around the observations. With this, we consider the HM procedure converged if more than a given ratio $\convergencethresholdmetrics$ of the $\nodesignpointsmodel$ design points is close to the parameter truth.\footnote{Naively, we would expect proper convergence to only be reached once \textit{all} new points lie within the uncertainty (i.e., their PCA-reduced metrics do), meaning that no sample can be distinguished in terms of quality compared to the observations anymore. However, the break criterion is subject to a trade-off between computational resources and accuracy, especially because \cref{alg:HistoryMatching} is not promised to converge to the exact solution (if only given enough waves). In practice, it is often sufficient to have a certain amount of points being close to the observations to obtain good solutions, see \cref{subsec:HPOviaOptuna}.} This convergence threshold is introduced as a hyperparameter in our HM framework (see \cref{subsec:HPOviaOptuna}). The final convergence criterion establishes a uniform measure for assessing and comparing the performances of different kernel architectures when tuning L96 via HM, applying to both classical and quantum approaches. Ultimately, it is responsible for making the transition to a fully automatic workflow that is less prone to subjective biases.

(iii) Assuming convergence has been reached, the k-means clustering might not be the ideal choice for determining candidate solutions, because it is specifically tailored to spherical clusters. For more general shapes like ellipsoids or disconnected regions, the \textit{Gaussian mixture model (GMM)} \cite{pearson_contributions_1894} is better suited thanks to more flexibility. The GMM is called a \textit{soft clustering}, as it does not assign each point to a single cluster\footnote{In contrast to the k-means algorithm, which is therefore called \textit{hard clustering}.}; instead, clusters are modeled as Gaussian distributions and every point is assigned $k$ \textit{responsibilities} (one for each cluster), describing the probability of belonging to the respective cluster. Next to the mean, every cluster thus comes with an individual covariance. Additionally, the different distributions (clusters) are related via the \textit{mixture weights}, which encompass their overall contribution to the GMM. Like k-means, an optimization routine is underlying the GMM: The \textit{expectation-maximization (EM)} \cite{dempster_maximum_1977} maximizes the conditional likelihood of the data given the values for means, covariances, weights, and the thereby induced responsibilities. In our framework, the GMM is then fitted for a number of clusters that ranges between one and a specified upper bound $\noclusters$. This maximum number of clusters will be handled as another hyperparameter of our HM algorithm (see \cref{subsec:HPOviaOptuna}). In total, this leads to 
\begin{equation}
    \sum_{k=1}^{\noclusters} k = \frac{\noclusters \left(\noclusters + 1\right)}{2}
\end{equation}
solution candidates. As the GMM means will, in general, not be part of the clusters themselves, similar to the k-means centroids, they need to subsequently be checked for implausibility according to \cref{eq:HMImplausibilityCheck} (cf. \cref{subsec:SemiAutomaticTuningofL96}). In the event of an empty list of surviving feasible candidates, we decide to fall back to the \textit{k-medoids} algorithm \cite{kaufman_clustering_1987} for the same series of cluster numbers to fit. The cluster medoids (which then form the candidates) are those cluster representatives that minimize the sum of distances to the other points in the same cluster. This prevents the HM from failing, as the medoids are, by design, real members of their respective cluster.\footnote{Note that the k-medoids clustering is more resource demanding than k-means or GMM, which is why it should not be applied to large final NROY spaces. Accordingly, it is only used as a fallback.}


%% file: SECTIONS/04_Methods/02_Quantum-Kernel-Evaluation-Methods.tex
\subsection{Quantum Kernel Evaluation Methods} \label{subsec:QuantumKernelEvaluationMethods}

\subsubsection{Inversion Test} \label{subsubsec:InversionTest}

Given a feature map $U(\bm{x}): \Omega \to \mathcal{U}(\mathcal{H})$ in the sense of \cref{subsec:QuantumGPRegressionForBayesionOptimization}, one of the most widely-used methods to compute the overlap \eqref{eq:QuantumKernelOverlap} is the \textit{inversion test (IT)} \cite{peters_machine_2021,havlicek_supervised_2019}. It is based on simply re-writing the kernel \eqref{eq:QuantumKernelOverlap} via the encoding \eqref{eq:QuantumFeatureMapEncoding} as 
\begin{equation} \label{eq:InversionTest}
    k(\bm{x}_i, \bm{x}_j) = \left| \bra{\bm{0}} U(\bm{x}_i)^{\dagger} \, U(\bm{x}_j) \ket{\bm{0}}  \right|^2.
\end{equation}
\Cref{eq:InversionTest} suggests the straightforward strategy to execute both feature map PQCs, perform a measurement, and repeat this procedure to estimate the probability of the all-zero state. We will generally refer to the routine of repeatedly executing and measuring a quantum circuit as \textit{(quantum) circuit sampling}. It leads to an approximation $\itkernel(\bm{x}_i,\bm{x}_j) \approx k(\bm{x}_i, \bm{x}_j)$, whose accuracy increases with the number of shots. Specifically, the error scales as $\mathcal{O}(S^{-1/2})$ for $S$ shots. Assuming the noise-free gates and measurements of statevector simulations (or fault-tolerant quantum computers), the inversion test is only exact if the probability of the all-zero state is either exactly zero or one. Especially, this is the case for $\bm{x}_j=\bm{x}_i$, which gives $\itkernel(\bm{x}_i, \bm{x}_i) = 1$ in accordance with \cref{eq:InversionTest}, as desired. In essence, the idea of IT is based on the principle that proximity in the parameter space corresponds to proximity in the feature or Hilbert space. Usually, we are interested in the full kernel matrix for one or two given datasets rather than individual kernel values. When estimating the kernel as described above, each entry of the kernel matrix requires one quantum circuit sampling. This makes $|X|(|X|-1)/2$ circuit samplings to determine $\itkernelmatrix(X,X)$ for a single dataset $X$ and $|X| \cdot |X^\prime|$ for the non-symmetric $\itkernelmatrix(X,X^\prime)$ in case of two datasets $X$ and $X^\prime$ (compare \cref{subsec:GPs}). Hence, the number of IT circuit samplings scales quadratically with the number of input points as $\mathcal{O}(|X|^2)$ or $\mathcal{O}(|X|\cdot|X^\prime|)$, respectively. \Cref{alg:InversionTest} schematically depicts in pseudo-code how to compute the full kernel matrix using the inversion test.

\begin{algorithm}
    \caption{$\itkernelmatrix(X, X^\prime=\text{None})$}
    \label{alg:InversionTest}
        Pick a number of shots $S$ \\
        \For{$i \in [|X|]$}{
            \For{$j \in [|X|]$ \textbf{if} $X^\prime$ is None \textbf{else} $j \in [|X^\prime|]$}{
                Set counter $n_0:=0$ \\
                \For{\_ $ \in [S]$}{
                    Prepare a register $\ket{q} := \ket{\bm{0}} \equiv \ket{0}^{\otimes N}$ \\
                    \If{$X^\prime$ is None}{
                        Transform register $\ket{q} \gets U(\bm{x}_j) \ket{q}$
                    }
                    \Else{
                        Transform register $\ket{q} \gets U(\bm{x}^\prime_j) \ket{q}$
                    }
                    Transform register $\ket{q} \gets U(\bm{x}_i)^{\dagger} \ket{q}$ \\
                    Measure basis state $\ket{b} := \mathcal{M} \ket{q}$  \label{line:Measure0} \\
                    \If{$\ket{b} = \ket{\bm0}$}{
                        $n_0 \gets n_0 + 1$
                    }
                }
                Store kernel value $\itkernelmatrix_{ij} = n_0/S$
            }
        }
        \Return $\itkernelmatrix$
\end{algorithm}

Usually, the number of shots that are necessary to obtain reliable statistics grows exponentially with the number of qubits, $S \in \mathcal{O}(2^N)$. However, for the moderate qubit requirements of our quantum HM, this concern can safely be dismissed.

\subsubsection{Randomized Measurements} \label{subsubsec:RandomizedMeasurements}

An alternative to the canonical inversion test was proposed by \textcite{haug_quantum_2023}, who suggest employing randomized measurements (RM) in order to reduce the complexity from a quadratic to a linear circuit sampling dependence. Their idea goes back to \cite{elben_cross-platform_2020}, where it was proven that the fidelity between two separate quantum systems $A,B$, described by (reduced) density matrices $\rho_A, \rho_B$, can be obtained by applying the same sequence of local uniformly random unitaries and classically cross-correlating the corresponding basis state probabilities as
\begin{align} \label{eq:RandomizedMeasurements}
\begin{split}
    \text{Tr}[\rho_A\rho_B] = \; \lim_{R \, \to \, \infty} &\left[2^N \sum_{v,v^\prime=0}^{2^N-1}(-2)^{\text{Ham}(v,v^\prime)} \right. \\
    &\left.\times \frac{1}{R} \sum_{r=1}^R p_A^{(r)}(v) \, p_B^{(r)}(v^\prime) \right]
\end{split}
\end{align}
where $\text{Ham}(v,v^\prime)$ is the Hamming distance between the computational basis states $v,v^\prime$, $p_{A/B}^{(r)}(v)$ denotes the probability of measuring basis state $v$ in system $A/B$ after applying the $r^{\text{th}}$ set of random unitaries. It is crucial to draw these unitaries according to the Haar measure to achieve a true uniform distribution across $\text{SU}(2)$. 

Combining \cref{eq:QuantumKernelFidelity,eq:QuantumKernelOverlap} with the fidelity identity \eqref{eq:RandomizedMeasurements} and truncating the limit at a finite repetition number $R$ motivates how this purely classical cross-correlation can be used to approximate the kernel value $k(\bm{x}_i,\bm{x}_j)$. More specifically, the critical implication of \cref{eq:RandomizedMeasurements} is that the basis state probabilities can be precomputed once, see \cref{alg:RMProbabilities}, and then simply be coupled classically every time the kernel shall be evaluated. This makes no difference for a single call to $k$ but reduces the amount of quantum circuit samplings significantly if we are interested in a complete kernel matrix.

\begin{algorithm}
    \caption{$\bm{P}(X, \bm{V}^{\text{Haar}}, S)$}
    \label{alg:RMProbabilities}
        Retrieve the number of repetitions $R = \left|\bm{V}^{\text{Haar}}\right|$ \\
        \For{$i \in [|X|]$}{
            \For{$r \in [R]$}{
                \For{$v \in \{0,...,2^N-1\}$}{
                    Set counter $n_v := 0$
                }
                \For{\_ $\in [S]$}{
                    Prepare a register $\ket{q_1 \cdots q_N}\equiv \ket{q} \gets \ket{\bm{0}}$ \\
                    Transform register $\ket{q} \gets U(\bm{x}_i) \ket{q}$\\
                    Transform register $\ket{q} \gets \bigotimes_{j=1}^N V_{r,j}^{\text{Haar}}\ket{q_j}$ \\
                    Measure basis state $\ket{v} := \mathcal{M} \ket{q}$ \\
                    Update $n_v \gets n_v + 1$
                }
                \For{$v \in \{0,...,2^N-1\}$}{
                    Store probability $P_{irv} = p_i^{(r)}(v) = n_v / S$
                }
            }
        }
        \Return $\bm{P} = \left\{P_{irv}: \,  v \in \{0,...,2^N-1\}\right\}_{i \in [|X|]}^{r \in [R]}$
\end{algorithm}

The application of random unitaries in each shot and repetition iteration in \cref{alg:RMProbabilities}, followed by measuring out the probabilities of all computational basis states, motivates the term \textit{randomized measurements} \cite{elben_statistical_2019,elben_cross-platform_2020,zhu_cross-platform_2022}. Based on this, the RM approximation of the kernel matrix is sketched in \cref{alg:RandomizedMeasurements}.

\begin{algorithm}
    \caption{$K^{\text{RM}}(X, X^\prime = \text{None})$}
    \label{alg:RandomizedMeasurements}
        Pick a number of repetitions $R$ and shots $S$ \\
        Draw $R$ Haar-random sets $\bm{V}_r^{\text{Haar}}=\left(V_{r,1}^{\text{Haar}}, ..., V_{r,N}^{\text{Haar}}\right)$ \\
        Compute probabilities $\bm{P}(X,\bm{V}^{\text{Haar}}, S) = \left\{p_i^{(r)}(v) \right\}$ for all combinations via to \cref{alg:RMProbabilities} \\
        \If{$X^\prime$ is not None}{
            Compute $\widetilde{\bm{P}}(X^\prime,\bm{V}^{\text{Haar}}, S) = \left\{\tilde{p}_i^{(r)}(v)\right\}$ for all combinations via \cref{alg:RMProbabilities}
        }
        \For{$i \in [|X|]$}{
            \For{$j \in [|X|]$ \textbf{if} $X^\prime$ is None \textbf{else} $j \in [|X^\prime|]$}{
                \If{$X^\prime$ is None}{
                    Store kernel value $K^{\text{RM}}_{ij} = \bigg[2^N \times $ $\sum_{v,v^\prime=0}^{2^N-1} (-2)^{\text{Ham}(v,v^\prime)} \frac{1}{R} \sum_{r=1}^R p_i^{(r)}(v) \, p_j^{(r)}(v^\prime) \bigg]$
                }
                \Else{
                    Store kernel value $K^{\text{RM}}_{ij} = \bigg[2^N \times $ $\sum_{v,v^\prime=0}^{2^N-1} (-2)^{\text{Ham}(v,v^\prime)} \frac{1}{R} \sum_{r=1}^R p_i^{(r)}(v) \, \tilde{p}_j^{(r)}(v^\prime) \bigg]$
                }
            }
        }
        \Return $\rmkernelmatrix$
        
\end{algorithm}

Due to the precomputation of the probabilities, determining the full kernel matrix $\rmkernelmatrix(X,X)$ or $\rmkernelmatrix(X, X^\prime)$ according to \cref{alg:RandomizedMeasurements} requires $R|X|$ or $R(|X| + |X^\prime|)$ quantum circuit samplings, respectively. This corresponds to a quadratic speedup to $\mathcal{O}(|X|)$ or $\mathcal{O}(|X|,|X^\prime|)$ compared to the inversion test (cf. \cref{subsubsec:InversionTest}).

Both theoretical and numerical arguments in \cite{elben_statistical_2019,elben_cross-platform_2020} indicate that the error of estimating a single kernel value $\rmkernel(\bm{x}, \bm{x}) \approx 1$ scales as $\mathcal{O}(S^{-1}R^{-1/2})$. As for IT, the number of required shots increases as $\mathcal{O}(2^N)$ with the number of qubits (compare \cref{subsubsec:InversionTest}). However, in \cite{elben_cross-platform_2020,haug_quantum_2023}, there is evidence that the actual measurement cost can be significantly lower than in conventional techniques like quantum state tomography. As the RM method trades random measurement repetitions for data-induced circuit executions and needs to deliver profound statistics for all computational basis states (instead of only the all-zero state), it turns out particularly beneficial for low qubit requirements and big datasets \cite{haug_quantum_2023}. Moreover, RM shows advantages when it comes to real hardware, as will be further discussed in \cref{subsec:TowardsRealQuantumHardware}.

%% file: SECTIONS/04_Methods/03_Quantum-Kernel-Architectures.tex
\subsection{Quantum Kernel Architectures}\label{subsec:QuantumKernelArchitectures}

Irrespective of which of the three coming quantum feature maps is investigated, we motivate employing $N \geq 4$ qubits, meaning that there is at least one qubit per dimension of the parameter space $\parameterspacelorenz$ defined in \cref{eq:L96ParameterSpace}. This ensures that each of the model parameters $\modelparams = \bm{\theta}$ can be varied over its full range (especially including the value $0$) without affecting the encoding of the others. Moreover, following \cite{lguensat_semi-automatic_2023}, any configured quantum kernel is accompanied by a constant kernel as a multiplicative factor and a white-noise kernel that promotes regularization of the Gram matrix. Both the rescaling strength and the Gaussian noise level will be optimized during a QGP fit.

\subsubsection{Chebyshev Kernel}\label{subsubsec:ChebyshevKernel}

\begin{figure*}[!ht]
    \centering
    \begin{minipage}{0.7\textwidth}
        \resizebox{\textwidth}{!}{
        \begin{quantikz}
            \lstick{$\ket{0}$} & \gate{\ry(\phi_{0,1})} & \gate{\rx(\phi_{l,1} \arccos{\theta_1})} \gategroup[wires=4,steps=5,style={dashed,rounded corners,fill=blue!10},background,label style={yshift=0.1cm}]{Repeated $L$ times} & \ctrl{1} & \qw & \qw & \gate{\rz(\phi_{l,8})} & \gate{\ry(\phi_{L+1,1})} & \qw \\
            \lstick{$\ket{0}$} & \gate{\ry(\phi_{0,2})} & \gate{\rx(\phi_{l,2} \arccos{\theta_2})} & \gate{\rz(\phi_{l,5})} & \ctrl{1} & \qw & \qw & \gate{\ry(\phi_{L+1,2})} & \qw \\
            \lstick{$\ket{0}$} & \gate{\ry(\phi_{0,3})} & \gate{\rx(\phi_{l,3} \arccos{\theta_3})} & \qw & \gate{\rz(\phi_{l,6})} & \ctrl{1} & \qw & \gate{\ry(\phi_{L+1,3})} & \qw \\
            \lstick{$\ket{0}$} & \gate{\ry(\phi_{0,4})} & \gate{\rx(\phi_{l,4} \arccos{\theta_4})} & \qw & \qw & \gate{\rz(\phi_{l,7})} & \ctrl{-3} & \gate{\ry(\phi_{L+1,4})} & \qw
        \end{quantikz}
    }
    \end{minipage}\hfill
    \begin{minipage}{0.25\textwidth}
        \vspace{0.2cm}
        \caption{Entangling Chebyshev feature map for $N=4$ qubits. In total, the quantum circuit is parameterized by $8(L+1)$ angles $\phi_{i,k} \in [0,2\pi)$, as well as the L96 model parameters $(\theta_1,\theta_2,\theta_3,\theta_4)=\modelparams$.}
        \label{fig:ChebysevFeatureMap}
    \end{minipage}
\end{figure*}
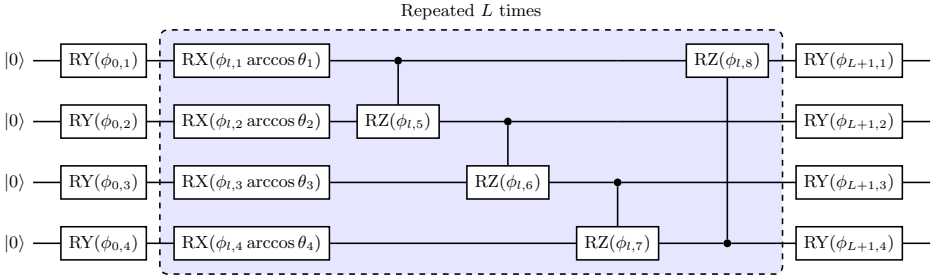

Our \textit{Chebyshev kernel} is based on the \textit{Chebyshev feature map} originally proposed in \cite{kyriienko_solving_2021}. The latter is a non-linear encoding scheme that processes a classical data point $x$ from a one-dimensional domain via the depth-1 unitary 
\begin{equation}\label{eq:ChebyshevFeatureMap}
    U_{\bm{\phi}}(x) = \bigotimes_{j=1}^N \ry_{j} (\phi_j \arccos{x})
\end{equation}
with a number of parameters $\bm{\phi}=(\phi_1,...,\phi_N)$, one assigned to each qubit. Despite its purely linear operations, \eqref{eq:ChebyshevFeatureMap} is said to be a nonlinear encoding due to the nonlinear dependence on the input $x$. Choosing $\phi_j =:a \in \N_0$ as in \cite{kyriienko_solving_2021} and writing out a single rotation in \cref{eq:ChebyshevFeatureMap} via Euler's formula reveals the origin of this feature map's name:
\begin{align*}
    \ry_j(a\arccos{x}) &= \exp\left({-i \frac{a \arccos{x}}{2} \y_j}\right) \\
    &= \cos\left(a \arccos{x}\right) \idty_j - i \sin\left(a \arccos{x}\right) Y_j \\
    &= T_a(x) \idty_j + \sqrt{1 - x^2} \, U_{a-1}(x) \x_j \z_j \, ,
\end{align*}
where $T_d$ and $U_d$ are the degree-$d$ Chebyshev polynomials of the first and second kind, respectively. As any smooth function can be represented by an infinite series of (weighted) Chebyshev polynomials, the encoding \eqref{eq:ChebyshevFeatureMap} promises high expressivity. To access the full Hilbert space, we add $L$ layers of an entangling block and shift the dependence on the input inside. Wrapping these layers by initial and final $\ry$-rotations gives what should best be called the \textit{entangled Chebyshev feature map}, see \cref{fig:ChebysevFeatureMap}.
This way, it is made of $\gatecount_{\text{Chebyshev}} = 2N(L+1)$ single- and two-qubit gates, each associated with one of the $[0,2\pi)$-rotation angles $(\phi_{0,j},\phi_{l,j},\phi_{l,N+j},\phi_{L+1,j})$ for $l \in \{1,...,L\}$ and $j\in\{1,...,N\}$. This general formulation of the entangled Chebyshev feature map goes back to \cite{kreplin_reduction_2023}. Depending on the choice of these circuit parameters, different versions of the entangled Chebyshev feature map can be created. For example, fixing every angle by, e.g., drawing them uniformly at random from $[0,2\pi)$, gives what we call the \textit{static Chebyshev feature map}. We are, however, more interested in a tunable version: Setting $\phi_{0,j}=\phi_{l,j}= \phi_{L+1,j}=:\phi_j$ and $\phi_{l,N+j}=:\phi_{N+j}$ yields the \textit{trainable Chebyshev feature map} where each of the remaining $2N$ independent angles can be employed as a tunable parameter in the QGP regression.\footnote{Note that our implementation also allows every possible subset to be tunable, accompanied by a static complement. This can help saving computational resources.} The trainable Chebyshev feature map with $2N$ free parameters was originally used in \cite{rapp_quantum_2024}. 

In contrast to \cref{eq:ChebyshevFeatureMap}, our domain is 4-dimensional. Note that it is up to the developer how exactly the four parameters $\modelparams$ are processed in case that $N>4$. In our implementation, we stick to the following rule: Repeat $(\theta_1,\theta_2,\theta_3,\theta_4)=\modelparams$ cyclically $\lceil N L/4\rceil$ times, fill up the circuit successively, and cut the sequence after the last position $NL$. This is, we proceed with the next model parameter at the beginning of each layer instead of starting the order from the beginning. Incorporating this, our version of the Chebyshev feature map can be written as
\begin{align}
\begin{split}
    U^{\text{Cheb}}_{\bm{\phi}} &(\bm{\theta}) = \left[\bigotimes_{j=1}^N \ry\left(\phi_j\right)\right] \\
    & \times \left[\prod_{l=1}^L U^{\text{Cheb}}_{\bm{\phi},l}(\bm{\theta}) \right] \times \left[\bigotimes_{j=1}^N \ry\left(\phi_j\right) \right]
\end{split}
\end{align}
with layer unitaries
\begin{align}
\begin{split}
    U&^{\text{Cheb}}_{\bm{\phi},l}(\bm{\theta}) = \left[\prod_{j=1}^N \controlled_j\rz_{j+1 \, \text{mod} \, N}(\phi_{N+j}) \right] \\
    &\times \left[\bigotimes_{j=1}^N \rx_j(\phi_j \arccos\theta_{[(l-1)N+j-1] \, \text{mod} \, 4 + 1})\right].
\end{split}
\end{align}
Assuming suitable parallelization capabilities, we find that the entangled Chebyshev feature map has a circuit depth of $\depth_{\text{Chebyshev}}=L(N+1)+2$, taking into account that the entangling gates have to be executed consecutively in each layer.\footnote{Given these parallelization capabilities, disjoint gates acting on different qubits constitute a single cycle on a QPU. Provided hardware-specific gate execution times, the circuit depth can thus be used as a proxy for the runtime.}

\subsubsection{NPQC Kernel} \label{subsubsec:NPQCKernel}

In \cite{haug_optimal_2021}, Haug and Kim argue that for small distances $\bm{\varepsilon}$ in parameter space, the fidelity \eqref{eq:QuantumKernelFidelity} between the parameterized quantum states $\ket{\psi(\bm{\theta})}, \ket{\psi(\bm{\theta}+\bm{\varepsilon})}$ can be approximated by
\begin{equation} \label{eq:QuantumFidelityLocallyRBF}
    K(\bm{\theta},\bm{\theta}+\bm{\varepsilon}) \approx \exp \left(-\frac{1}{4}\bm{\varepsilon}^T \mathcal{F}(\bm{\theta}) \bm{\varepsilon}\right)
\end{equation}
where $\mathcal{F}(\bm{\theta})$ is the quantum Fisher information metric (QFIM). The QFIM is a measure for how the parameter space geometry is related to the geometry of the feature space. \Cref{eq:QuantumFidelityLocallyRBF} means that a PQC-induced quantum kernel behaves locally like the classical Gaussian RBF kernel \eqref{eq:RBFKernel} for $\sigma=\sqrt{2}$, with the QFIM as weight matrix.
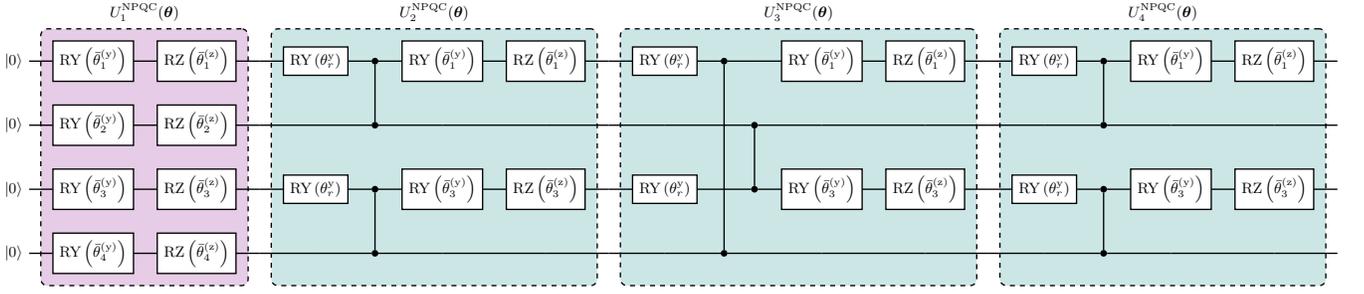
\begin{figure*}[!ht]
    \centering
    \resizebox{\textwidth}{!}{
    \begin{quantikz}
        \lstick{$\ket{0}$} & \gate{\ry\left(\bar{\theta}_1^{(\text{y})}\right)} \gategroup[wires=4,steps=2,style={dashed,rounded corners,fill=violet!20},background,label style={yshift=0.1cm}]{$U_1^{\text{NPQC}}(\bm{\theta})$} & \gate{\rz\left(\bar{\theta}_1^{(\text{z})}\right)} & \qw & \gate{\ry\left(\npqcreferenceparamy\right)} \gategroup[wires=4,steps=4,style={dashed,rounded corners,fill=teal!20},background,label style={yshift=0.1cm}]{$U_2^{\text{NPQC}}(\bm{\theta})$} & \ctrl{1} & \gate{\ry\left(\bar{\theta}_1^{(\text{y})}\right)} & \gate{\rz\left(\bar{\theta}_1^{(\text{z})}\right)} & \qw & \gate{\ry\left(\npqcreferenceparamy\right)} \gategroup[wires=4,steps=5,style={dashed,rounded corners,fill=teal!20},background,label style={yshift=0.1cm}]{$U_3^{\text{NPQC}}(\bm{\theta})$} & \ctrl{3} & \qw & \gate{\ry\left(\bar{\theta}_1^{(\text{y})}\right)} & \gate{\rz\left(\bar{\theta}_1^{(\text{z})}\right)} & \qw & \gate{\ry\left(\npqcreferenceparamy\right)} \gategroup[wires=4,steps=4,style={dashed,rounded corners,fill=teal!20},background,label style={yshift=0.1cm}]{$U_4^{\text{NPQC}}(\bm{\theta})$} & \ctrl{1} & \gate{\ry\left(\bar{\theta}_1^{(\text{y})}\right)} & \gate{\rz\left(\bar{\theta}_1^{(\text{z})}\right)} & \qw \\
        \lstick{$\ket{0}$} & \gate{\ry\left(\bar{\theta}_2^{(\text{y})}\right)} & \gate{\rz\left(\bar{\theta}_2^{(\text{z})}\right)} & \qw & \qw & \control{} & \qw & \qw & \qw & \qw & \qw & \ctrl{1} & \qw & \qw & \qw & \qw & \control{} & \qw & \qw & \qw \\
        \lstick{$\ket{0}$} & \gate{\ry\left(\bar{\theta}_3^{(\text{y})}\right)} & \gate{\rz\left(\bar{\theta}_3^{(\text{z})}\right)} & \qw & \gate{\ry\left(\npqcreferenceparamy\right)} & \ctrl{1} & \gate{\ry\left(\bar{\theta}_3^{(\text{y})}\right)} & \gate{\rz\left(\bar{\theta}_3^{(\text{z})}\right)} & \qw & \gate{\ry\left(\npqcreferenceparamy\right)} & \qw & \control{} & \gate{\ry\left(\bar{\theta}_3^{(\text{y})}\right)} & \gate{\rz\left(\bar{\theta}_3^{(\text{z})}\right)} & \qw & \gate{\ry\left(\npqcreferenceparamy\right)} & \ctrl{1} & \gate{\ry\left(\bar{\theta}_3^{(\text{y})}\right)} & \gate{\rz\left(\bar{\theta}_3^{(\text{z})}\right)} & \qw \\
        \lstick{$\ket{0}$} & \gate{\ry\left(\bar{\theta}_4^{(\text{y})}\right)} & \gate{\rz\left(\bar{\theta}_4^{(\text{z})}\right)} & \qw & \qw & \control{} & \qw & \qw & \qw & \qw & \control{} & \qw & \qw & \qw & \qw & \qw & \control{} & \qw & \qw & \qw
    \end{quantikz}
    }
    \caption{NPQC feature map for $N=4$ qubits and the maximum number of $L=2^{N/2}=4$ layers. Rotation gates are parameterized by $\left(\bar{\theta}_1^{(\text{y/z})},\bar{\theta}_2^{(\text{y/z})},\bar{\theta}_3^{(\text{y/z})},\bar{\theta}_4^{(\text{y/z})}\right) = \bm{\theta}_r^{(\text{y/z})} + c \, \modelparams$ according to \cref{eq:NPQCEncoding} with reference parameter values \eqref{eq:NPQCReferenceParameter}. The shift factors evaluate to $a_2=0, a_3=1,a_4=0$; see \cref{alg:NPQCShiftFactor}.}
    \label{fig:NPQC}
\end{figure*}

While the QFIM corresponding to a given PQC is generally not known a priori, the \textit{natural parameterized quantum circuit} (NPQC) \cite{haug_natural_2022,haug_quantum_2023} is characterized by featuring $\mathcal{F}(\bm{\theta}_r)=\idty$ for a reference parameter $\bm{\theta}_r$ with values 
\begin{equation}\label{eq:NPQCReferenceParameter}
    \theta_{r,l,j}^{(\text{y})} = \frac{\pi}{2} =: \theta_r^{(\text{y})} \quad \text{and} \quad \theta_{r,l,j}^{(\text{z})}=0 =: \theta_r^{(\text{z})}
\end{equation}
where the superscripts $(\text{y})$ and $(\text{z})$ indicate the axes around which the rotations are applied, and the subscripts $l,j$ the layer and qubit on which they act, respectively. The NPQC is then based on the linear encoding
\begin{equation}\label{eq:NPQCEncoding}
    \bar{\bm{\theta}} = \bm{\theta}_r + c \, \bm{\theta}
\end{equation}
where $c$ is a scaling constant. The full circuit is composed of an initial cascade of accordingly parameterized single-qubit rotations, followed by $L-1$ mixed layers
\begin{align}\label{eq:NPQCLayer}
\begin{split}
    U_l&^{\text{NPQC}}(\bm{\theta}) = \left[\bigotimes_{j=1}^{N/2} \rz_{2j-1}\left(\bar{\theta}_{(j-1)\modulo 4 + 1}^{(\text{z})}\right)\right] \\
    & \times \left[\bigotimes_{j=1}^{N/2} \ry_{2j-1} \left(\bar{\theta}_{(j-1)\modulo 4 + 1}^{(\text{y})}\right)\right] \\
    & \times \left[\bigotimes_{j=1}^{N/2} \cz^{2j-1}_{2(j+a_l)} \right] \times \left[\bigotimes_{j=1}^{N/2} \ry_{2j-1}\left(\npqcreferenceparamy \right) \right]
\end{split}
\end{align}
where $a_l \in \left\{0,...,\frac{N}{2} - 1\right\}$ is a recursively defined shift factor, which yields entanglement among different pairs of qubits, depending on the layer $l > 1$. The explicit routine for determining the factors $a_l$ may be found in \cref{appendix:NPQCShiftFactors}. \Cref{eq:NPQCLayer} implies that the NPQC is only well-defined for even qubit numbers $N \, \text{mod} \, 2 = 0$. In total, this gives 
\begin{align}\label{eq:NPQCFeatureMap}
\begin{split}
    U^{\text{NPQC}}(\bm{\theta}) = \; &\left[\prod_{l=0}^{L-2} U_{L-l}^{\text{NPQC}}(\bm{\theta}) \right] \\
    & \times \left[\bigotimes_{j=1}^N\rz_j\left(\bar{\theta}_{(j-1)\modulo 4 + 1}^{(\text{z})}\right) \right] \\ 
    & \times \left[\bigotimes_{j=1}^N \ry_j\left(\bar{\theta}_{(j-1)\modulo 4 + 1}^{(\text{y})}\right)\right].
\end{split}
\end{align}
By construction \cite{haug_natural_2022}, the number of layers is bounded from above by $L \leq 2^{N/2}$. \Cref{fig:NPQC} illustrates what \cref{eq:NPQCFeatureMap} amounts to for $N=4$ qubits and the maximum number of $L=4$ layers. In contrast to the strategy for the trainable/static Chebyshev feature map, here every layer is constructed to start over again from the first parameter $\bar{\theta}_1$ used on the first qubit, independent of the number of qubits employed.

By simple counting, the NPQC \eqref{eq:NPQCFeatureMap} consists, in total, of 
\begin{equation}\label{eq:NPQCGateCount}
    \gatecount_{\text{NPQC}} = 2N + (L - 1) \, 4 \, \frac{N}{2} = 2NL \leq N \,2^{N/2+1}
\end{equation}
gates. The shift factors are defined such that every qubit is participating in exactly one entangling process per layer. Hence, they are disjoint and can therefore, in principle, be executed in parallel. \Cref{eq:NPQCLayer,eq:NPQCFeatureMap} then imply an NPQC-depth of 
\begin{equation}\label{eq:NPQCDepth}
    \depth_{\text{NPQC}} = 2 + (L-1) \, 4 = 4L - 2 \leq 2^{N/2+2} - 2. 
\end{equation}
Note that \eqref{eq:NPQCDepth} is only upper-bounded by an expression in $N$, despite the lack of a direct dependency.

The only trainable parameter of the NPQC, which can be optimized during a QGP fit, is the scaling constant $c$ in \cref{eq:NPQCEncoding}. Together with the prescribed model parameter ranges \eqref{eq:L96ParameterSpace}, it influences how much the encoded parameters $\bar{\bm{\theta}}$ can deviate from the reference parameter $\bm{\theta}_r$. Since we expect the QFIM to remain close to the identity in a small region around $\bm{\theta}_r$, the chosen value of $c$ thus determines how close the NPQC kernel is to the RBF kernel. In any case, it can fairly be asserted the quantum analog of our classical competitor.

\subsubsection{YZ-CX Kernel}\label{subsubsec:YZCXKernel}

Thirdly, we employ another PQC that was also studied in \cite{haug_quantum_2023} and which we will accordingly call \textit{YZ-CX feature map}. It is based on the same linear encoding \eqref{eq:NPQCEncoding} as the NPQC; however, the reference parameters $\bm{\theta}_r$ here are drawn uniformly at random from the interval $[0,2\pi)$, with no difference being made between y- and z-rotations.\footnote{Note that the YZ-CX feature map comes with a non-trivial QFIM $\mathcal{F}(\bm{\theta}_r) \neq \idty$.} Also, there is no restriction on the number of qubits $N$ and, in particular, the number of layers $L$ in our implementation. The latter is different to the version in \cite{haug_quantum_2023}. The NPQC feature map is characterized by alternating layers of equal single-qubit parameterized rotation blocks and shifted CNOT gates:
\begin{align}\label{eq:YZCXFeatureMap}
\begin{split}
    U^{\text{YZ-CX}}_l(\bm{\theta}) = \; &\left[\bigotimes_{j=1}^{\left\lfloor \frac{N-(l-1)\modulo 2}{2} \right\rfloor} \cnot^{2j+(l-1)\modulo 2-1}_{2j +(l-1)\modulo 2} \right] \\
    & \times \left[\bigotimes_{j=1}^N \rz_j\left(\bar{\theta}_{(j-1)\modulo 4 + 1}\right) \right] \\
    & \times \left[\bigotimes_{j=1}^N \ry_j\left(\bar{\theta}_{(j-1)\modulo 4 + 1}\right) \right],
\end{split}
\end{align}
which differ for even and odd values of $l \in \{1,...,L\}$.\footnote{$\cnot^{i}_k$ denotes a $\cnot$-gate with control qubit $i$ and target qubit $k$.} 
\begin{figure}[!ht]
    \centering
    \resizebox{\columnwidth}{!}{
        \begin{quantikz}
            \lstick{$\ket{0}$} & \gate{\ry\left(\bar{\theta}_1\right)} \gategroup[wires=4,steps=3,style={dashed,rounded corners,fill=violet!20},background,label style={yshift=0.1cm}]{Applied $\lceil L/2 \rceil$ times} & \gate{\rz\left(\bar{\theta}_1\right)} & \ctrl{1} & \qw & \gate{\ry\left(\bar{\theta}_1\right)} \gategroup[wires=4,steps=3,style={dashed,rounded corners,fill=teal!20},background,label style={yshift=0.1cm}]{Applied $\lfloor L/2\rfloor$ times} & \gate{\rz\left(\bar{\theta}_1\right)} & \qw & \qw \\
            \lstick{$\ket{0}$} & \gate{\ry\left(\bar{\theta}_2\right)} & \gate{\rz\left(\bar{\theta}_2\right)} & \targ{} & \qw &  \gate{\ry\left(\bar{\theta}_2\right)} & \gate{\rz\left(\bar{\theta}_2\right)} & \ctrl{1} & \qw \\
            \lstick{$\ket{0}$} & \gate{\ry\left(\bar{\theta}_3\right)} & \gate{\rz\left(\bar{\theta}_3\right)} & \ctrl{1} & \qw & \gate{\ry\left(\bar{\theta}_3\right)} & \gate{\rz\left(\bar{\theta}_3\right)} & \targ{} & \qw \\
            \lstick{$\ket{0}$} & \gate{\ry\left(\bar{\theta}_4\right)} & \gate{\rz\left(\bar{\theta}_4\right)} & \targ{} & \qw & \gate{\ry\left(\bar{\theta}_4\right)} & \gate{\rz\left(\bar{\theta}_4\right)} & \qw & \qw
        \end{quantikz}
    }
    \caption{YZ-CX feature map for $N=4$ qubits. Rotation gates are parameterized by $(\bar{\theta}_1, \bar{\theta}_2, \bar{\theta}_3, \bar{\theta}_4) = \bm{\theta}_r + c \modelparams$ according to \cref{eq:NPQCEncoding} with reference parameters $\bm{\theta}_r$ drawn uniformly at random from $[0,2\pi)$.}
    \label{fig:YZCXFeatureMap}
\end{figure}
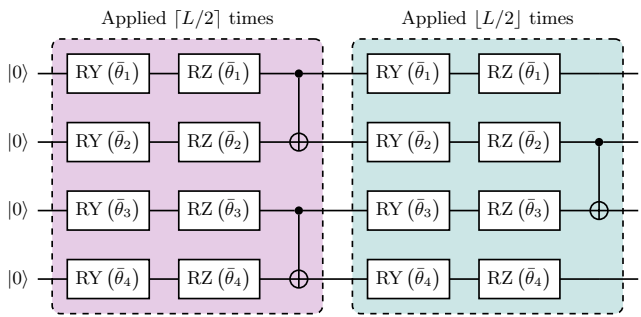

For $N=4$, \cref{eq:YZCXFeatureMap} yields the quantum circuit shown in \cref{fig:YZCXFeatureMap}. Unlike the static/trainable Chebyshev feature map and the NPQC, \eqref{eq:YZCXFeatureMap} entangles only neighboring qubits via $\cnot$s, making the YZ-CX feature map particularly hardware-efficient. The full unitary $U^{\text{YZ-CX}}(\bm{\theta})=\prod_{l=1}^L U^{\text{YZ-CX}}_l(\bm{\theta})$ is composed of 
\begin{align}   
\begin{split}
    \gatecount_{\text{YZ-CX}} &= \left\lceil \frac{L}{2} \right\rceil \left(2N + \left\lfloor \frac{N}{2} \right\rfloor\right) \\
    &\quad + \left\lfloor \frac{L}{2} \right\rfloor \left(2N + \left\lfloor \frac{N-1}{2} \right\rfloor\right) \\
    &= 2NL + \left\lceil \frac{L}{2} \right\rceil \left\lfloor \frac{N}{2} \right\rfloor + \left\lfloor \frac{L}{2} \right\rfloor \left\lfloor \frac{N-1}{2} \right\rfloor
\end{split}
\end{align}
gates. Under the assumption of parallelizability, this reduces to an elementary depth of $\depth_{\text{YZ-CX}}=3L$, which is -- as \cref{eq:NPQCDepth} for the NPQC -- independent of $N$. Another similarity to the NPQC is the characteristic that the scaling constant $c$ used for the input data encoding \eqref{eq:NPQCEncoding} is the only free parameter when training a QGP based on the YZ-CX feature map.

%% file: SECTIONS/05_Results/01_HPO-via-Optuna.tex
\subsection{Hyperparameter Optimization via \texttt{Optuna}} \label{subsec:HPOviaOptuna}

In \cref{subsec:SemiAutomaticTuningofL96}, we outlined how \textcite{lguensat_semi-automatic_2023} configured the HM \cref{alg:HistoryMatching} for the Lorenz-96 model. On the one hand, this comprises implementations of the generic methods in \cref{alg:HistoryMatching} -- our extensions and enhancements of them were discussed in \cref{subsec:AutomaticTuningOfL96}. On the other hand, this means the specific values that were set for the inputs of \cref{alg:HistoryMatching}, e.g., the observations, the parameter space, etc. To avoid confusion with the parameters of the feature maps presented in \cref{subsec:QuantumKernelArchitectures}, we refer to these ``outer" parameters as \textit{hyperparameters} of the full HM algorithm. Due to the different architectures, it is not reasonable to expect the corresponding quantum kernels to have the same optimal hyperparameter values for which they deliver the best results. For sound benchmarking, we should employ the quantum kernels in their optimal setting. Hence, we perform an extensive hyperparameter optimization (HPO) using \texttt{Optuna} \cite{akiba_optuna_2019} for each of the architectures in \cref{subsec:QuantumKernelArchitectures}. Adding a \textit{study} for our classical RBF opponent \eqref{eq:RBFKernel} allows us to make a fair comparison at the end. The resulting four \texttt{Optuna} studies explore $300-500$ hyperparameter configurations each. These \textit{trials} are guided by two objectives for assessing their quality: Most naturally, we choose the Euclidean distance of the final solution returned by \cref{alg:HistoryMatching} to the parameter truth \eqref{eq:L96ParameterTruth}, after rescaling each dimension of the parameter space \eqref{eq:L96ParameterSpace} to the interval $[0,1]$ to weight them equally. On the other hand, the number of waves is used as a measure for the spent computational resources. In combination, minimizing both metrics, equally weighted, can be understood as optimizing the price-performance ratio associated with the trials.

Since our implementation of the inversion test in \texttt{JAX} \cite{bradbury_jax_2018} turns out to be faster than the randomized measurements approach (compare \cref{subsubsec:InversionTest,subsubsec:RandomizedMeasurements}), the HPO is performed with IT-based kernel evaluation.\footnote{Despite the expected quadratic reduction in quantum circuit samplings for big datasets from IT to RM, we find the inversion test to be more amenable to a significant \texttt{JAX}-induced speedup. This especially comes into play when computing the gradient matrix, i.e., the derivative of the kernel matrix with respect to a PQC parameter. In this case, the full probability calculation in \cref{alg:RMProbabilities} must be differentiated in one go for RM, while it can be done element-wise when using IT.} Moreover, until this point, our idea of employing QGPs in tuning L96 via HM should be considered a \textit{quantum-inspired} ansatz, as all quantum circuit executions are emulated using statevector simulation on a purely classical machine. Also, there is no need for a proper sampling in \cref{alg:InversionTest} when all amplitudes of the respective quantum states are directly accessible.

An overview of all HM hyperparameters together with the optimal combinations found by our \texttt{Optuna} studies and the corresponding objective values is given in \cref{tab:HPO}. Note that only a subset of these hyperparameters is actually trained by \texttt{Optuna}. In the following, we will introduce those hyperparameters that have not been discussed yet:

\begin{table*}[t!]
    \centering
    \begin{tabular}{@{}c@{}c@{}c@{}c@{}c@{}c@{}c@{}} \toprule
        \textbf{Symbol} & \textbf{Interpretation} & \textbf{Range} & \multicolumn{4}{c}{\textbf{Optimal values}} \\ \midrule
        
         &  &  & $\;$ Chebyshev $\;$ & $\;$ NPQC $\;$ & $\;$ YZ-CX $\;$ & $\;$ RBF $\;$ \\ \midrule
         
         $N$ & Number of qubits & $\{4, 6, 8\}$ & $8$ & $6$ & $6$ & - \\
         
        \rowcolor{gray!10} $L$ & Number of layers & $\left\{1,2,3,...,2^{N/2}\right\}$ & $1$ & $2$ & $6$ & - \\ 
        
        $\nosamplepoints$ & Number of initial sample points & $\left\{1 \times 10^4, 5 \times 10^4, 1 \times 10^5\right\}$ & $1 \times 10^4$ & $1 \times 10^4$ & $5 \times 10^4$ & $1 \times 10^4$ \\
        
        \rowcolor{gray!10} $\trainkernelonlyonce$ & Boolean for whether & $\{0, 1\}$ & - & $0$ & $1$ & $1$ \\
        \rowcolor{gray!10} & kernel is trained only once &  &  &  &  & \\
        
        $\implausibilitythresholdmax$ & Maximum implausibility threshold & $3$ & - & - & - & - \\
        
        \rowcolor{gray!10} $\implausibilitythresholdmin$ & Minimum implausibility threshold & $[0.1 , 1.5]$ & $1.419$ & $1.096$ & $0.814$ & $0.381$ \\
        
        $\implausibilitythresholdmindecayfactor$ & Minimum implausibility & $[0, 0.5]$ & $0.387$ & $0.382$ & $0.403$ & $0.354$ \\
         & threshold decay factor &  &  &  &  & \\
         
        \rowcolor{gray!10} $\maxnoimplausibilities$ & Maximum number of permitted & $\{0, 1\}$ & $0$ & $0$ & $0$ & $1$ \\
        
        \rowcolor{gray!10} & implausible principal components &  &  &  &  & \\
        
        $\convergencethresholdmetrics$ & Convergence threshold of new design & $\{0.1, 0.15, 0.2, ..., 1.0\}$ & $0.2$ & $0.25$ & $0.45$ & $0.95$ \\
         & points ratio close to parameter truth &  &  &  &  & \\
         
        \rowcolor{gray!10} $\noclusters$ & Maximum number of clusters & $4$ & - & - & - & - \\
        
        $\maxnowaves$ & Maximum number of waves & $30$ & - & - & - & - \\
        
        \rowcolor{gray!10} $\maxruntime$ & Maximum HM runtime & $5 \times 60 \times 60 \, \text{s} = 5 \, \text{h}$ & - & - & - & - \\
        
        $\nohmrepetitionsmin$ & Minimum number of HM repetitions & $2$ & - & - & - & - \\
        
        \rowcolor{gray!10} $\nohmrepetitionsmax$ & Maximum number of HM repetitions & $5$ & - & - & - & - \\
        
        $\randomnessseed$ & Randomness seed & $\{42, 43, 44, 45, 46\}$ & - & - & - & - \\ \midrule
        
        \rowcolor{gray!10} $\meandistancerescaled$ & Mean rescaled distance & $[0, 2]$ & $0.318$ & $0.175$ & $0.107$ & $0.191$ \\
        
        $\meannowaves$ & Mean number of waves & $[0, 30]$ & $5.8$ & $4.667$ & $6.4$ & $9.333$ \\ \midrule 
        
        \rowcolor{gray!10} $\optimaldistancerescaled$ & Smallest rescaled distance & $[0,2]$ & $0.038$ & $0.054$ & $0.016$ & $0.075$ \\
        
        $\optimalnowaves$ & Number of waves corresponding to optimum & $\{0,1,2,...,30\}$ & $5$ & $4$ & $7$ & $9$ \\ \bottomrule
    \end{tabular}
    \caption{Hyperparameters and metrics of our history matching for tuning the Lorenz-96 model. For each kernel architecture ((trainable) Chebyshev, NPQC, YZ-CX, and the classical RBF), an \texttt{Optuna} study is used to optimize certain hyperparameters by minimizing the two objectives given in the penultimate row block. A hyperparameter is``trained" for an architecture if an optimal value from the respective range is given in the corresponding subcolumn. The quantum kernels are evaluated via the inversion test (cf. \cref{subsubsec:InversionTest}) and all quantum circuits are emulated using statevector simulation. Although the restriction to even qubit numbers $N$ is only necessary for the NPQC kernel (compare \cref{subsec:QuantumKernelArchitectures}), for the sake of simplicity, we use the same discrete set of possible choices for the other two quantum feature maps as well. The same reasoning is applied to limit the number of layers $L$ to $2^{N/2}$ (cf. \cref{subsubsec:NPQCKernel}). Thereby, the layer number is the only hyperparameter that depends on the choice of another hyperparameter. To account for the more expensive training of $2N$ free parameters in the trainable Chebyshev compared to the single parameter in NPQC and YZ-CX (cf. \cref{subsec:QuantumKernelArchitectures}), the Chebyshev kernel is trained only once per full HM run, meaning $\trainkernelonlyonce \equiv 1$.}
    \label{tab:HPO}
\end{table*}

The hyperparameter $\trainkernelonlyonce$ controls whether a kernel is trained only once at the beginning of an HM run, or whether a new optimization is performed in every wave when fitting a fresh GP to the newly drawn design points.

In order to accelerate the process of shrinking the NROY space, and thereby potentially save computational resources due to faster convergence, we relax the 3-sigma rule \cite{pukelsheim_three_1994} when assessing the implausibility of single principal components (cf. \cref{subsec:HistoryMatching}). More specifically, while we always start with the canonical value $\implausibilitythresholdmax = 3$ as the largest implausibility threshold, it is uniformly reduced from one wave to the next wave until convergence, with a lower bound given by $\implausibilitythresholdmin \in [0.1, 1.5]$. The additive decay factor is denoted by $\implausibilitythresholdmindecayfactor \in [0, 0.5]$. Note that $\implausibilitythresholdmindecayfactor$ can also take a value of zero, corresponding to no reduction at all. For the same objective of improving the efficiency of our HM, we loosen the strict choice $\maxnoimplausibilities = 0$ made by \textcite{lguensat_semi-automatic_2023} (cf. \cref{subsec:SemiAutomaticTuningofL96}) and instead allow a maximum of one implausible component in the feasibility check \eqref{eq:HMImplausibilityCheck}.

For the tractability of performing four mature \texttt{Optuna} studies, we implement some safeguards on the runtime of a single HM execution. In particular, we limit the number of possible waves to $\maxnowaves = 30$ and, in case this is not sufficient, interrupt a trial strictly after exceeding a computing time of $\maxruntime = 5 \, \text{h}$. If either of the two upper bounds is met, we revert to the last successfully performed wave and determine a solution from the associated NROY space as usual.

Some parts of our algorithm are still subject to randomness, especially the initial LHS (cf. \cref{subsec:SemiAutomaticTuningofL96}) of the parameter space and the drawing of design points via our maximin heuristic (cf. \cref{subsec:AutomaticTuningOfL96}). To mitigate this effect, we perform multiple repetitions of the full HM for each \texttt{Optuna} trial. This is, for every hyperparameter configuration from \cref{tab:HPO}, a minimum of $\nohmrepetitionsmin = 2$ repetitions is executed. Rolling averages are then used to decide whether or not to top up to $\nohmrepetitionsmax = 5$ repetitions: Only if the rolling average of the rescaled distance deviates not more than $20\%$ from the best value found so far, the current trial is considered promising and new repetitions are granted; otherwise, the trial is terminated prematurely. Also, a trial is pruned in the (unlikely) case of two failing repetitions. To create different results in the simulation, every repetition is equipped with a different randomness seed. For comparability among different trials, the initial seed is always set to $\randomnessseed = 42$. In each successive repetition, it is then incremented by one. Consequently, \texttt{Optuna} compares the mean rescaled distance and the mean number of waves, which replace the respective single-valued objectives:
\begin{subequations}
\begin{align}
    \meandistancerescaled &= \frac{1}{\nohmrepetitions} \sum_{r =1}^{\nohmrepetitions} \distancerescaled \; , \label{eq:MeanDistanceRescaled} \\
    \meannowaves &= \frac{1}{\nohmrepetitions} \sum_{r =1}^{\nohmrepetitions} \nowaves \; , \label{eq:MeanNumberOfWaves}
\end{align}
\end{subequations}
where $\nohmrepetitions$ and $\nowaves$ denote the number of repetitions and the number of waves actually taken in the corresponding trial, respectively. The rescaled distance is defined as 
\begin{equation} \label{eq:DistanceRescaled}
    \distancerescaled (\bm{\theta}_{\text{sol}}) = \sqrt{\sum_{p \in \{h,F,c,b\}} \left(\frac{p_{\text{truth}} - p_{\text{sol}}}{p_{\text{max}} - p_{\text{min}}}\right)^2} \quad ,
\end{equation}
where $p_{\text{truth}}$ is the parameter value of the truth \eqref{eq:L96ParameterTruth} and $p_{\text{max}}$, $p_{\text{min}}$ denote the upper and lower bounds of the respective domain interval from the parameter space \eqref{eq:L96ParameterSpace}. Naturally, the averaged quantities \eqref{eq:MeanDistanceRescaled} and \eqref{eq:MeanNumberOfWaves} are upper bounded by the upper bounds of the individual objectives. More specifically, $\meannowaves \leq 30$ due to our set limit of at most $\maxnowaves =30$ waves per HM run. On the other hand, the rescaling of each parameter interval in \cref{eq:DistanceRescaled} implies that $\meandistancerescaled \leq \sqrt{4}=2$. 

Finally, the best 20 trials are filtered for each architecture based on weighing both objectives equally. Out of these, the hyperparameter configurations with the lowest single rescaled distances $\optimaldistancerescaled$ are ultimately selected as the optima found by \texttt{Optuna}. The numbers of waves corresponding to these individual runs are denoted by $\optimalnowaves$.\footnote{Note that these are, by definition, not necessarily the smallest observed wave numbers.} This strategy respects our HM's statistical nature on the one hand, and shows the peak performance capabilities of the different kernels on the other.

More details on the results of the four \texttt{Optuna} studies can be found in \cref{appendix:HPO}.

%% file: SECTIONS/05_Results/02_Performance-Comparison.tex
\subsection{Performance Comparison} \label{subsec:PerformanceComparison}

\begin{figure}[!t]
    \centering
    \includegraphics[width=\columnwidth]{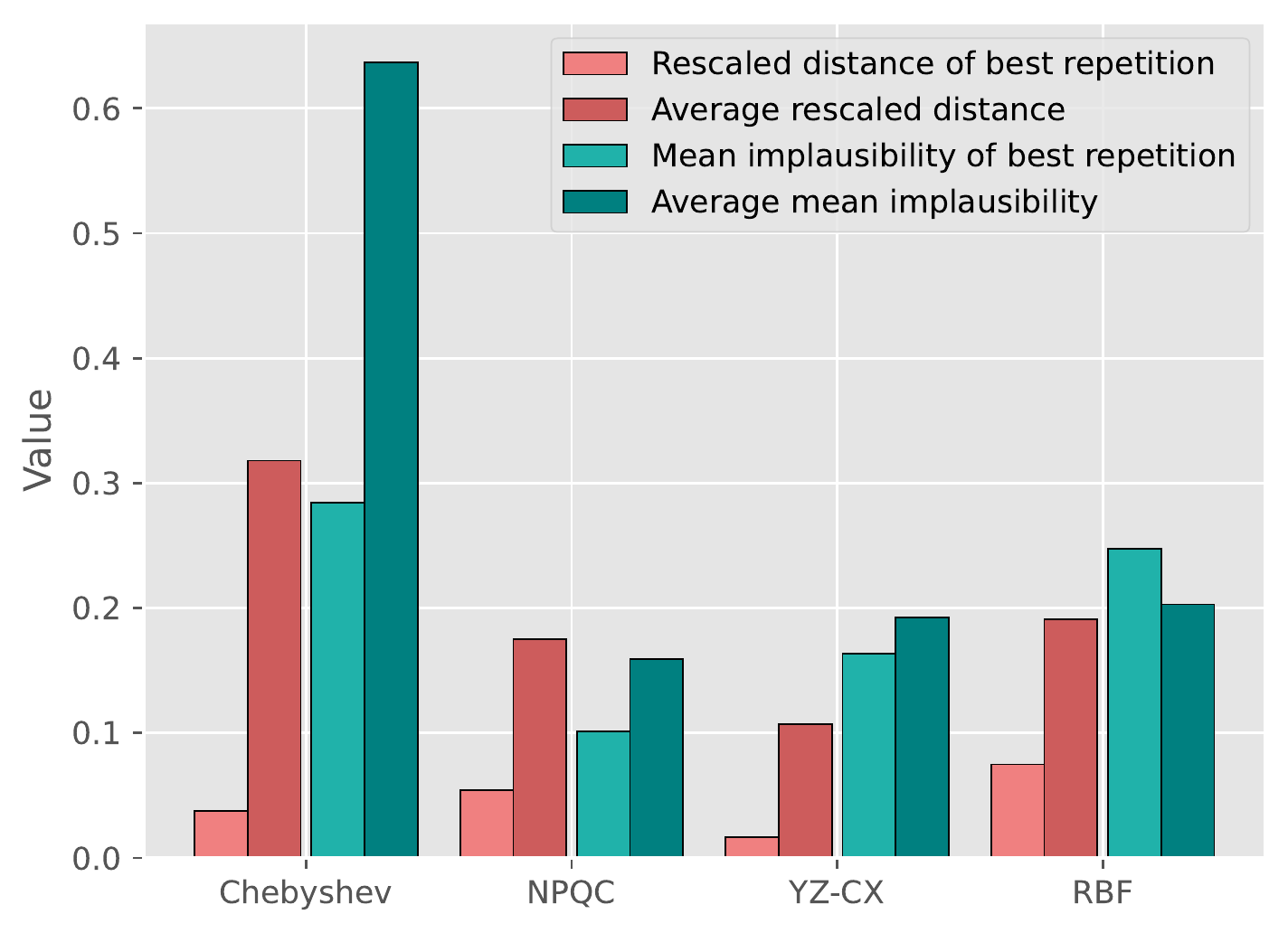}
    \caption{Comparison of the results for the ideal repetition vs. the average values in the best trial for all kernel architectures. Investigated is the smallest rescaled distance in relation to the mean as well as the corresponding implausibility (averaged over the principal components). The optimal and the mean distances are equal to the values in \cref{tab:HPO}. For NPQC, YZ-CX and RBF, the results are in a similar regime. Only the trainable Chebyshev kernel shows qualitative differences.}
    \label{fig:DistancesAndImplausibilities}
\end{figure}

The distances to the true solution and the corresponding numbers of waves of the best \texttt{Optuna} trials in \cref{tab:HPO} indicate already that the explored quantum kernels can achieve similar or even better results than the classical workhorse, namely the RBF kernel. Here, we provide more numerical evidence supporting this observation. To achieve a peak performance comparison, we restrict benchmarking to the best architecture trials only. From \cref{tab:HPO}, we learn that the YZ-CX kernel returned the best single result according to our strategy to assess the study trials.\footnote{Recall that the results in \cref{tab:HPO} only represent the smallest-distance repetitions among the best 20 trials, with a 50/50 weighting applied to determine them. Hence, they do not have to be the global study optima as well.} More details on the best repetitions -- and how they relate to the average values of the trial -- can be found in \cref{fig:DistancesAndImplausibilities}. For NPQC, YZ-CX, and RBF, the distance and the implausibility of the best run, as well as the mean values, are in a similar regime. Only the trainable Chebyshev kernel shows qualitative differences, especially in the average implausibility (which itself is a mean over the principal components, see \cref{alg:HistoryMatching,subsec:SemiAutomaticTuningofL96}). Specifically, it is larger than the competing values by more than a factor of three. However, the mean implausibility of the best Chebyshev repetition is only half the size, representing the largest discrepancy across all architectures. Only the classical RBF kernel has a mean implausibility exceeding the ideal run. This shows that the candidate closest to the parameter truth \eqref{eq:L96ParameterTruth} is not always the one that has the smallest (mean) implausibility score.\footnote{Consequently, the implausibility-driven HM solutions returned by \cref{alg:HistoryMatching} might even be suboptimal in the first place. However, note that this is a retrospective analysis, only possible for our toy model, for which we know the parameter truth in advance. In a mature climate model, the implausibility is the only available guidance.} Most importantly, \cref{fig:DistancesAndImplausibilities} implies that NPQC and YZ-CX outperform the RBF kernel in all displayed metrics. Despite the lower average and best implausibility scores of NPQC, YZ-CX achieves better solutions, both in peak and on average.

\begin{figure*}[!ht]
    \centering
    \begin{minipage}[t]{0.48\textwidth}
        \includegraphics[width=\textwidth]{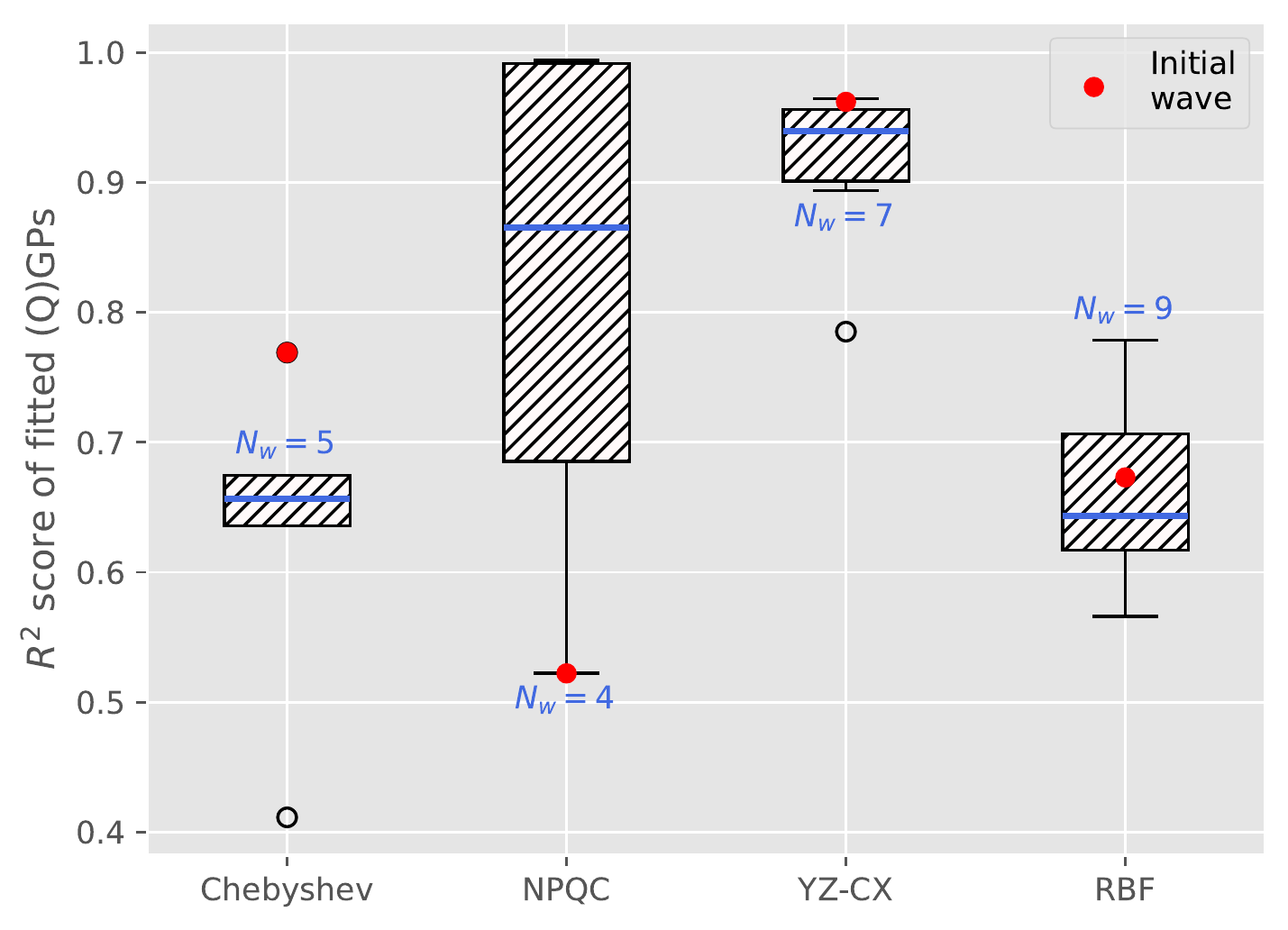}
        \par\vspace{0.5ex}
        \raggedright
        \footnotesize (a) $R^2$ scores of the (quantum) GPs corresponding to the best repetitions. The kernel architectures exhibit different behavior, both qualitatively and quantitatively. The NPQC kernel comes with the largest deviation. In terms of median $R^2$ values, it is solely beaten by YZ-CX, the only feature map with a central value above 0.9. RBF comes with the lowest median $R^2$ score. Except for one extreme outlier, the trainable Chebyshev kernel shows similar results, but a better initial fit. With a significant gap to RBF, the worst initial fit is, however, found at NPQC.
    \end{minipage}\hfill
    \begin{minipage}[t]{0.48\textwidth}
        \includegraphics[width=\textwidth]{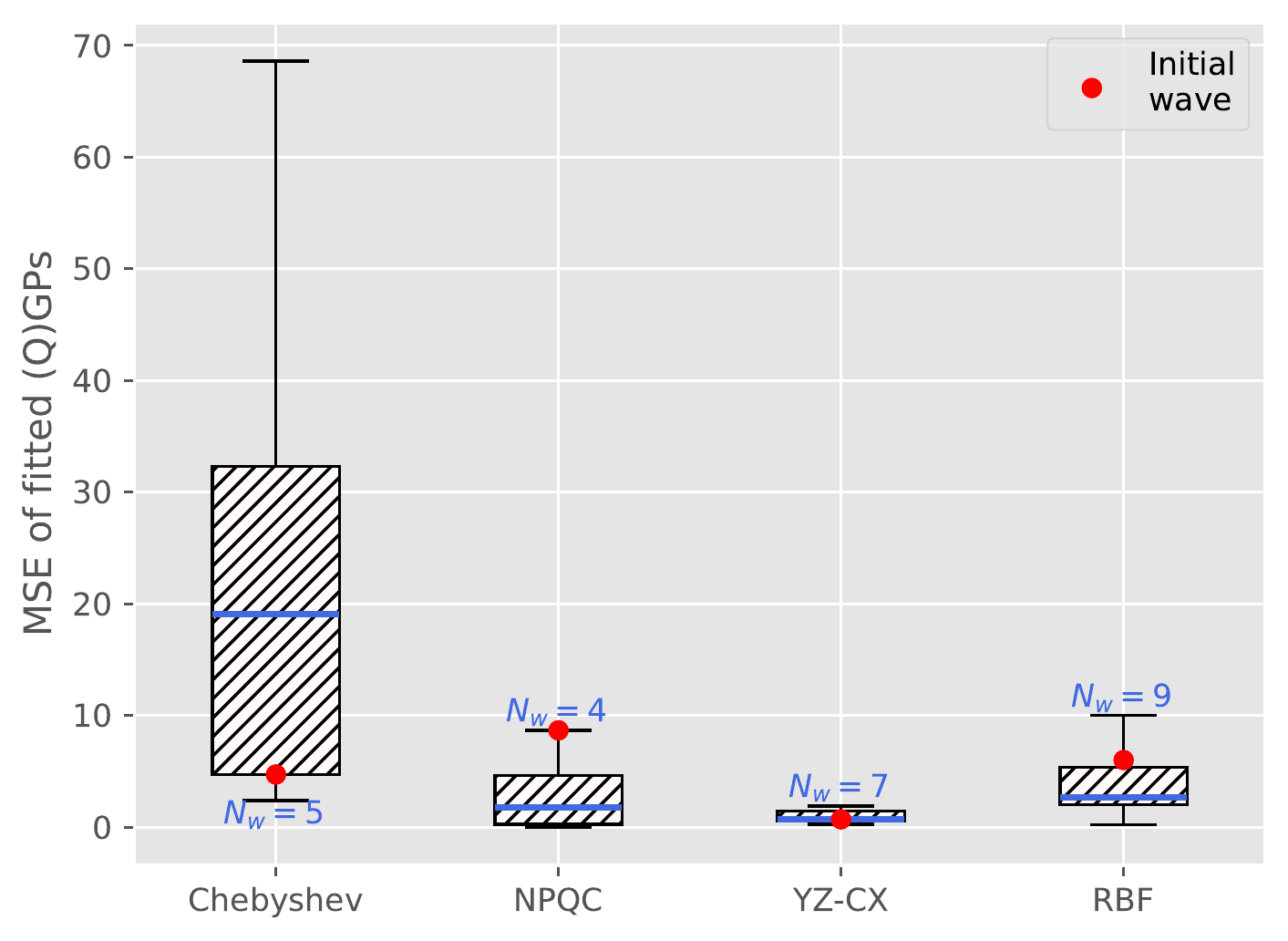}
        \par\vspace{0.5ex}
        \raggedright
        \footnotesize (b) MSE scores of the (quantum) GPs corresponding to the best repetitions. NPQC, YZ-CX, and RBF exhibit comparable behavior, with values smaller than 10. The large displayed range is necessitated by the trainable Chebyshev kernel, reaching a mean squared error of almost 70 at maximum. On this scale, the spread of YZ-CX values can hardly be resolved. Also, YZ-CX achieves the lowest MSE for the initial wave, followed by Chebyshev and RBF with comparable values around 5. RBF and NPQC show similar results in general.
    \end{minipage}
    \caption{$R^2$ and MSE (mean squared error) scores of the (quantum) GPs fitted during the best HM run of each kernel type. Every box is marked with the number of waves, $N_W$, the respective run performed (and hence (Q)GP fits), accounting for the different number of waves needed to reach convergence. Additionally, the first (initial) scores are colored in red to acknowledge the improved comparability due to fixed input points.}
    \label{fig:R2AndMSEScores}
\end{figure*}

So far, we have only investigated the final outcomes of our history matching. \Cref{fig:R2AndMSEScores} provides a first closer look at quantities evaluated during the best HM runs. For regression tasks, the $R^2$ score and the \textit{mean squared error (MSE)} are commonly used metrics to assess how well the observations (targets) are described by the fit. Their distributions over all rolled-out waves for the various kernels in \cref{fig:R2AndMSEScores} confirm the above findings of YZ-CX outperforming the other architectures, including the classical RBF. It returns the highest $R^2$ score at the first wave and the largest median, combined with an ordinary deviation around it.\footnote{The first wave is of special interest, because the initial set of input parameter configurations is fixed. This implies an exceptional comparability across different HM runs and even varying feature map designs.} Remarkably, the worst YZ-CX fit (given by the smallest outlier) is still better than the best RBF fit. Neglecting the $R^2$ score of this single outlier, YZ-CX never significantly falls below 0.9. Moreover, it also asserts itself against the other architectures with regard to the MSE. The trainable Chebyshev kernel spans a drastic range of values, with a worst-case MSE of almost 70. This makes it complicated to properly resolve the spread of YZ-CX values, all smaller than 5. RBF and NPQC exhibit similar MSE distributions, both bounded from above by 10. Nevertheless, their initial MSE scores are worse than the comparably good Chebyshev result. 

\begin{figure}[t!]
    \centering
    \includegraphics[width=\columnwidth]{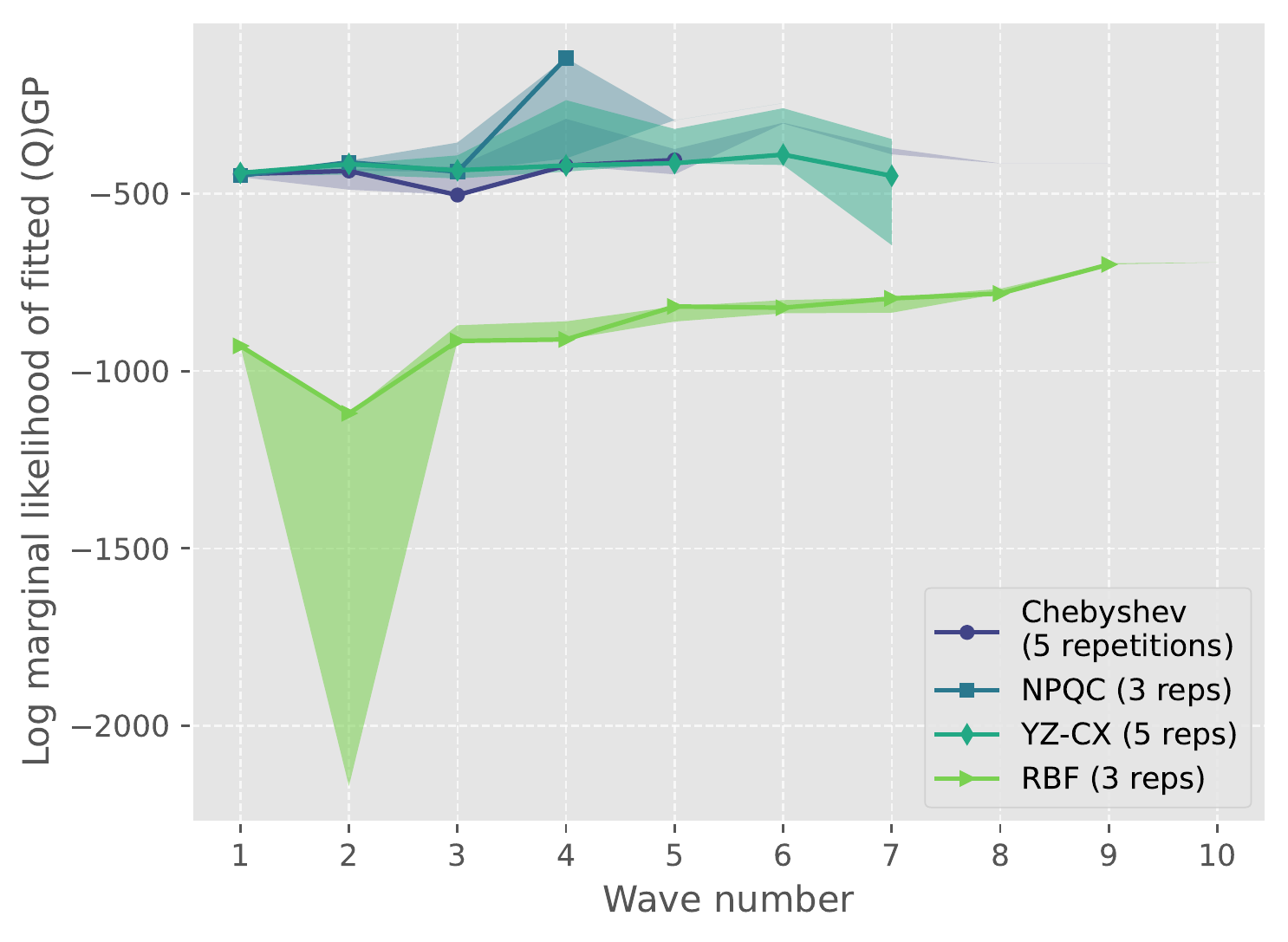}
    \caption{Log marginal likelihood of the fitted (quantum) GP for the different kernel types depending on the wave number. For each architecture, the data associated with the best repetition is drawn as a solid line. The color bands span between the minimum and maximum values, retrieved individually at each wave based on the conducted repetitions. Curves and bands end after convergence has been reached. As different repetitions generally need different numbers of waves to converge, the best trial curves can end earlier than the min-max bands. The likelihood values belonging to quantum kernels are confined in a narrow region between -200 and -600, with a final increase only observable for NPQC. The RBF-induced GPs show significantly worse likelihood values in every wave.}
    \label{fig:LogMarginalLikelihoods}
\end{figure}

However, since the $R^2$ score and the MSE do not take into account the covariance of the prediction, which is decisive for a (Q)GP regression, both generally deliver only limited information. Although a large $R^2$ and a small MSE value typically indicate a good fit, this inference can be corrupted, e.g., by overfitting. As described in \cref{subsec:GPs}, the log marginal likelihood might be better suited as a metric. In \cref{fig:LogMarginalLikelihoods}, we draw the likelihood values over all waves for all repetitions performed by \texttt{Optuna} for the best trials. When defined as in \cref{eq:LogMarginalLikelihood}, it is bounded from above by zero, with larger values (smaller absolute values) representing a better fit. It is impressive that the likelihood values belonging to quantum kernels are confined in a narrow region between -200 and -600. The best Chebyshev and YZ-CX runs stay at an almost constant level throughout the tuning process. For NPQC, we can see a notable increase in the last wave. On the other hand, the curve of the RBF kernel starts off approximately twice as bad. Except for the second wave, it shows a clear upward trend, something that cannot be observed for the quantum feature maps. At the bespoke second wave, the RBF likelihood drops rapidly to values below -2000 at worst. As this is not reflected by the quantum curves and bands, it is unlikely due to a global phenomenon. The continuous gap between quantum and classical results further confirms the feasibility of our quantum-inspired approach and demonstrates that all explored quantum kernels can be superior compared to the classical standard. This supports the hypothesis that the increased expressivity of quantum feature maps over their classical opponents can yield a measurable improvement.

\begin{figure*}[!ht]
    \centering
    \begin{minipage}[t]{0.48\textwidth}
        \includegraphics[width=\textwidth]{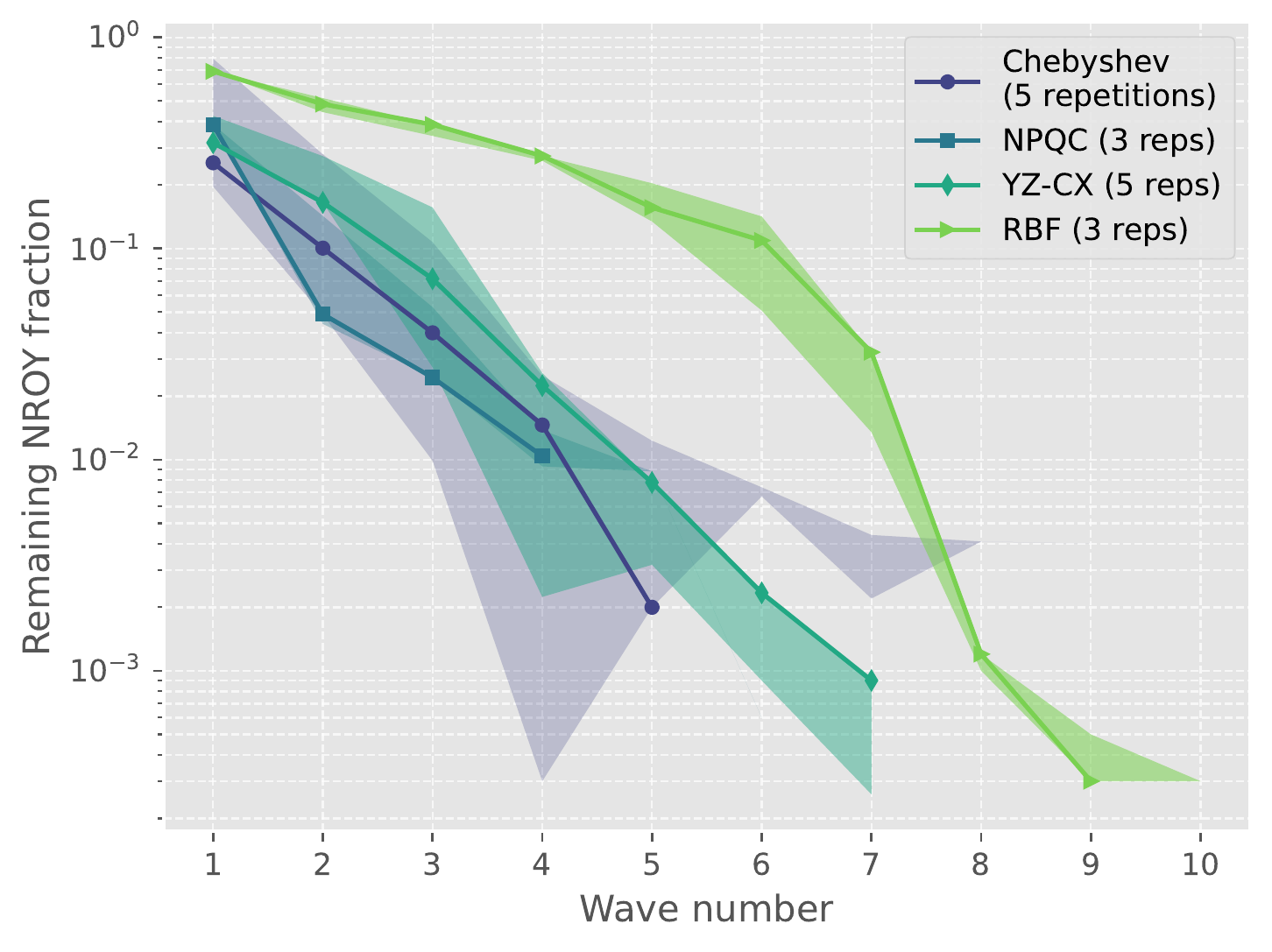}
        \par\vspace{0.5ex}
        \raggedright
        \footnotesize (a) Remaining fraction of the NROY space evolving with the number of waves. This fraction is given by the ratio between the number of points contained in the NROY space at a certain wave and the initial sample size. The quantum kernels collectively feature a smaller NROY fraction compared to the classical RBF at each of their encountered waves. The best RBF trial, however, terminates at a lower value than the quantum challengers (although also achieved by non-optimal repetitions for Chebyshev and YZ-CX). Touchpoints only exist with larger values of the non-ideal Chebyshev repetitions at the initial and the last wave.
    \end{minipage}\hfill
    \begin{minipage}[t]{0.48\textwidth}
        \includegraphics[width=\textwidth]{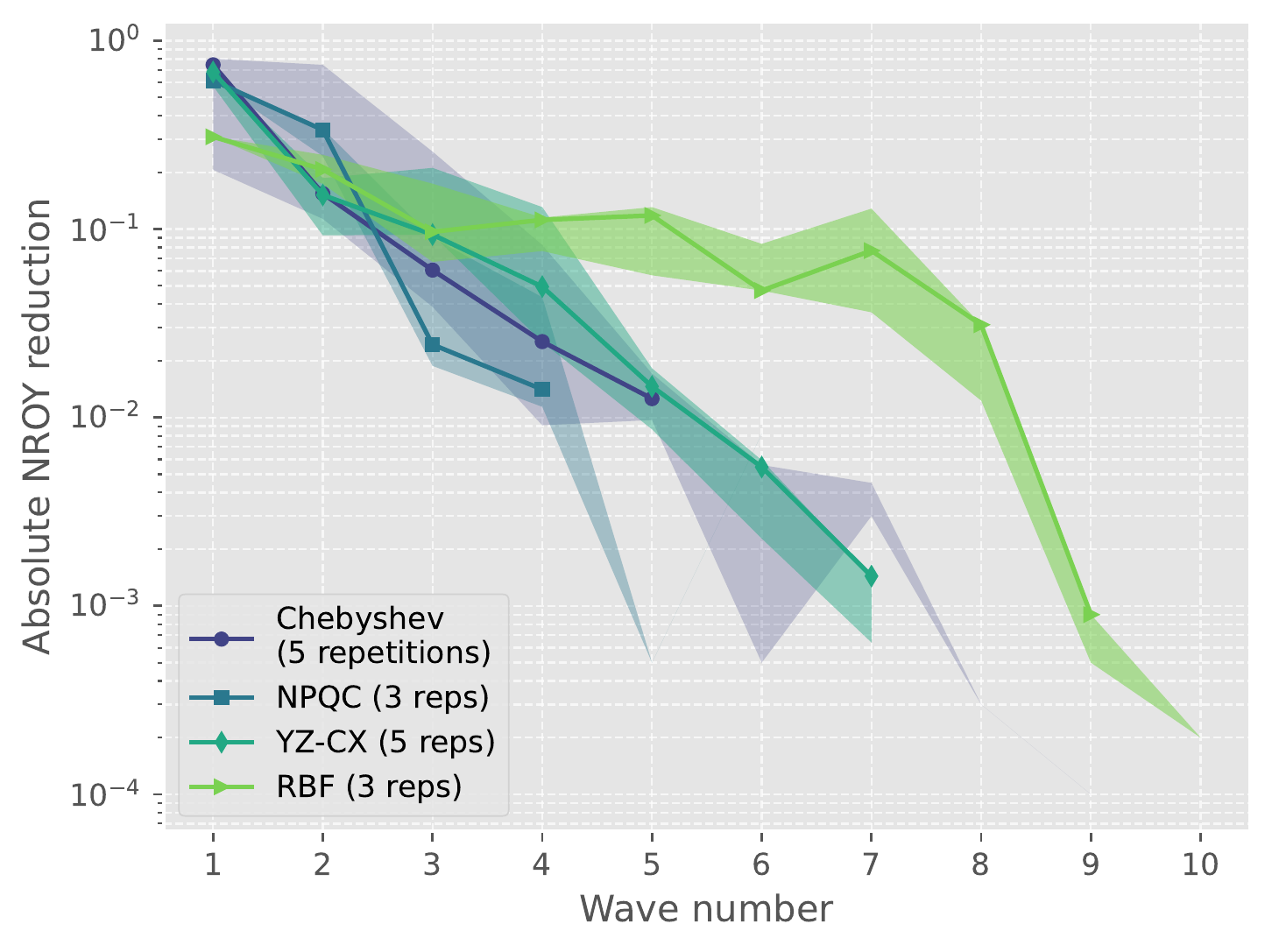}
        \par\vspace{0.5ex}
        \raggedright
        \footnotesize (b) Absolute reduction of the NROY space evolving with the number of waves. This reduction is given by the absolute difference between the remaining NROY fractions of the previous and the current wave. Initially, the remaining fraction equals 1. After a similar evolution until 4 waves, the quantum kernels collectively follow a more or less constant reduction rate. In contrast, the RBF kernel stagnates at a reduction of about 0.1 until wave 7, before it experiences an even stronger decline starting at wave 8. Consequently, the repetitions terminate at a comparable value of $0.3$.
    \end{minipage}
    \caption{Evolution of the NROY space for the different kernel types depending on the wave number. For each architecture, the data associated with the best repetition is drawn as a solid line. The color bands span between the minimum and maximum values, retrieved individually at each wave based on the conducted repetitions. Curves and bands end after convergence has been reached. As different repetitions generally need different numbers of waves to converge, the best trial curves can end earlier than the min-max bands. Across architectures, the remaining fraction and the absolute reduction show similar behavior for the remaining fraction and the reduction of the NROY space.}
    \label{fig:NROYFractionAndReduction}
\end{figure*}

The plotting style in \cref{fig:LogMarginalLikelihoods} can be used to disclose the actual behavior of certain HM-related quantities. The main principle behind \cref{alg:HistoryMatching} is to find good solution approximations by shrinking the NROY space from wave to wave, see \cref{subsec:HistoryMatching}. Accordingly, the evolution of the NROY space with an increasing number of waves, depicted in \cref{fig:NROYFractionAndReduction}, gives lower-level insights into the HM dynamics. For example, across architectures, the remaining fraction of the NROY space (the ratio between the number of NROY points at a certain wave and the initial sample size) and its reduction from one wave to the next (measured in absolute differences between old and new remaining fractions) seem to be subject to the same trends. Especially for the quantum kernels, the evolutions of both quantities show a large overlap. The logarithmic scaling of the ordinate axes in \cref{fig:NROYFractionAndReduction} (a) and (b) indicates an exponential decrease in terms of both the remaining fraction and the absolute reduction. RBF shows a different behavior, with mostly larger NROY fractions and reduction values at each wave. However, the color bands (wrapping all repetitions executed by \texttt{Optuna} for the respective trial, cf. \cref{subsec:HPOviaOptuna}) illustrate that these NROY quantities do not allow for drawing profound conclusions on the overall HM performance. In particular, the repetitions with the best final outcome (drawn as a solid line) do not always have to be the ones with the smallest NROY fraction, not even the smallest fraction remaining in the last wave. This means that the NROY space alone is, despite its central importance, not sufficient for determining convergence of the HM procedure. 

Furthermore, the different shapes and widths of the color bands in \cref{fig:LogMarginalLikelihoods,fig:NROYFractionAndReduction} imply our configuration of \cref{alg:HistoryMatching} is, to a significant degree, still subject to randomness. However, we suspect this to be a general HM characteristic, which can potentially be explained by observing that each wave in \cref{alg:HistoryMatching} depends critically on the choice of the design points in \cref{line:DrawDeisgnPoints}. As the number of design points is typically very small compared to the cardinality of the NROY space, 
\begin{equation*}
    10d = \nodesignpoints \ll |\nroy| \leq \nosamplepoints \, ,
\end{equation*}
any possible selection will struggle to adequately represent the NROY space. Our maximin sampling heuristic, which picks the first design point in each wave uniformly at random (cf. \cref{subsec:AutomaticTuningOfL96}), is not engineered to mitigate this effect.

More detailed insights into an exemplary HM run can be found in \cref{appendix:HM}.

%% file: SECTIONS/05_Results/03_Towards-Real-Quantum-Hardware.tex
\subsection{Towards Real Quantum Hardware} \label{subsec:TowardsRealQuantumHardware}

The results in \cref{subsec:PerformanceComparison} show an improvement of our quantum-inspired approach over the classical RBF-based algorithm, presumably due to the increased expressivity of the quantum feature maps. However, the small qubit requirements and the manageable circuit depths discussed in \cref{subsec:QuantumKernelArchitectures} make the quantum part of our hybrid HM amenable to NISQ hardware, also taking into account the low numbers of trainable parameters associated with the feature map PQCs. We here describe strategies to make the transition from statevector simulation to real quantum hardware. 

Besides the restrictions in terms of spatial and temporal resources, current and near-future quantum hardware is prone to qubit and gate errors, as well as limited coherence times. As the inversion test relies on exact computations, a statistical method like RM might be intrinsically better suited for noisy quantum systems (cf. \cref{subsec:QuantumKernelEvaluationMethods}). More specifically, the usage of random unitaries in \cref{alg:RandomizedMeasurements} could soften the disruptive effect of gate errors. By design, the RM ansatz is able to compensate for unpredictable behavior. Moreover, \textcite{haug_quantum_2023} argue that errors can be mitigated without further measurement cost when using randomized measurements. Assuming depolarizing noise, they derive how to infer the pure kernel values with the aid of the diagonal kernel matrix entries.

Another source of noise that inevitably comes with the transition to real hardware is \textit{shot noise}. When working with an actual quantum computer, we no longer have access to the exact amplitudes of an encoded quantum state. Instead, they need to be estimated via repeated measurements. This sampling procedure was already included in \cref{alg:InversionTest,alg:RMProbabilities}. The deviation of the obtained measurement result from the (unknown) exact value is denoted by shot noise. As described in \cref{subsubsec:InversionTest}, the approximation improves as the measurement error decreases quadratically as $\mathcal{O}(S^{-1/2})$ with growing numbers of shots.

\begin{figure*}[!ht]
    \centering
    \begin{minipage}[t]{0.48\textwidth}
        \includegraphics[width=\textwidth]{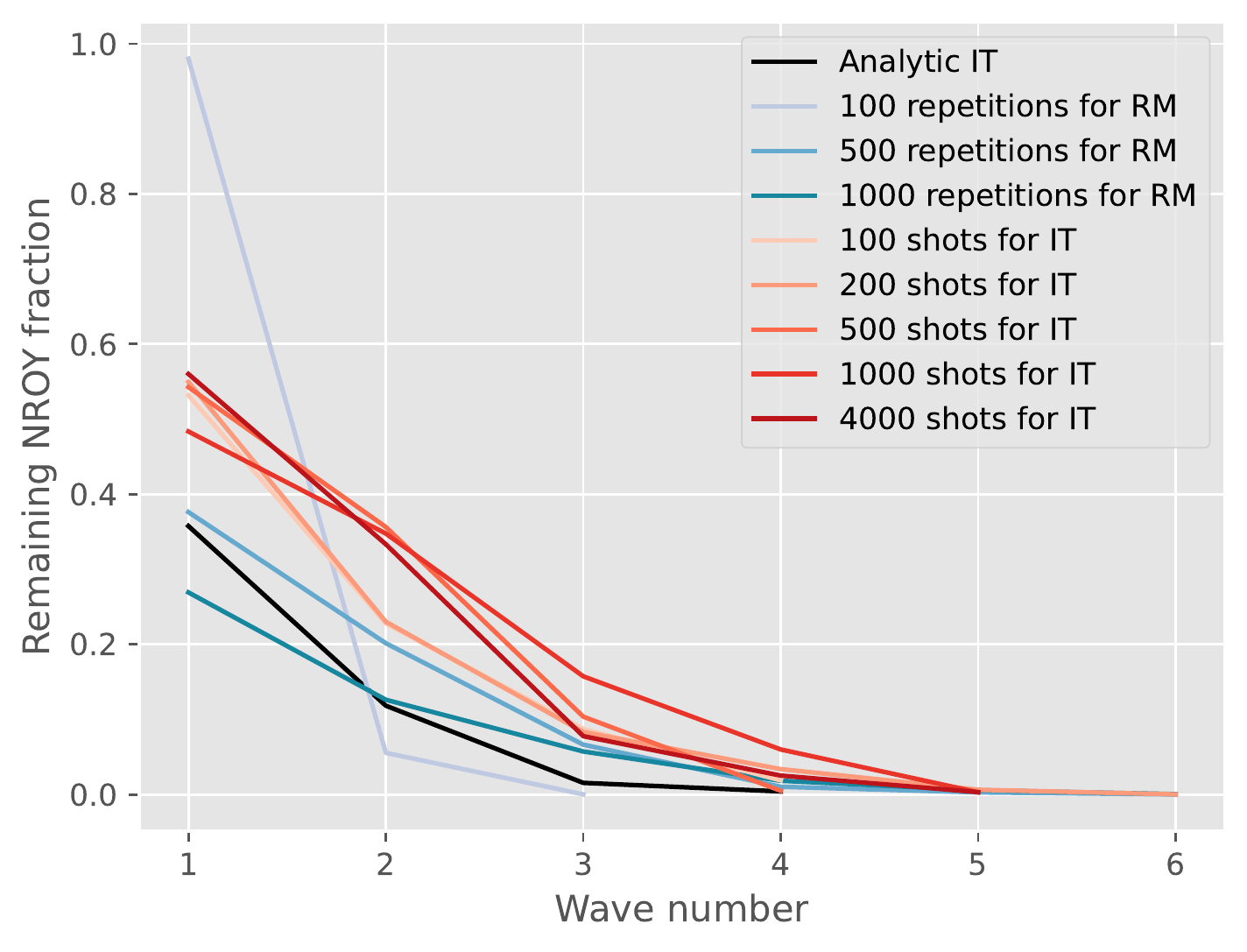}
        \par\vspace{0.5ex}
        \raggedright
        \footnotesize (a) Remaining fraction of the NROY space evolving with the number of waves. The curves show similar behavior, all approaching $0$ until wave $5$ at the latest. Only the $100$-RM-repetitions run stands out by leaving the NROY space mostly untouched in the first wave and emptying it almost completely in the second.
    \end{minipage}\hfill
    \begin{minipage}[t]{0.48\textwidth}
        \includegraphics[width=\textwidth]{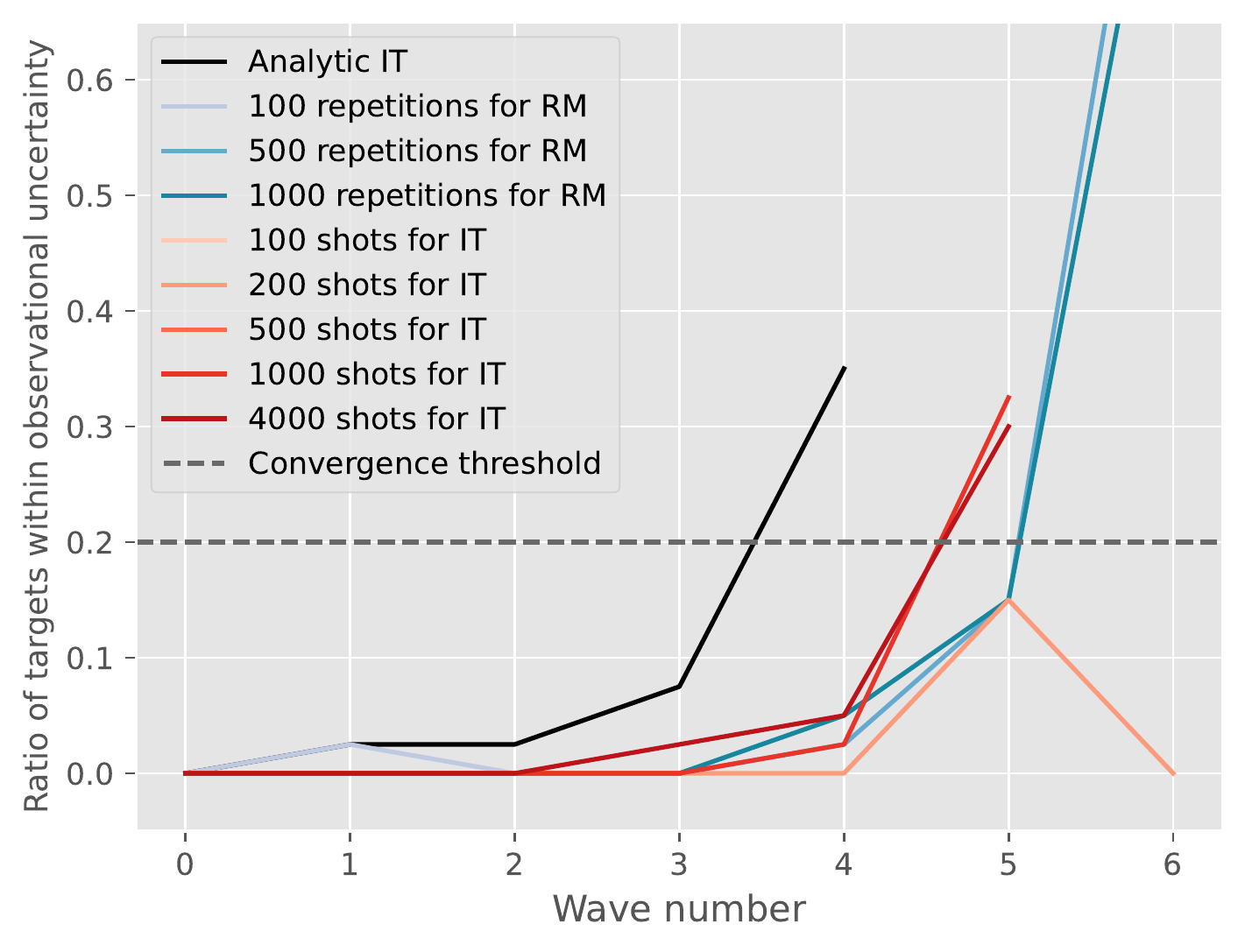}
        \par\vspace{0.5ex}
        \raggedright
        \footnotesize (b) Ratio of targets within observational uncertainty evolving with the number of waves. The convergence threshold $\convergencethresholdmetrics = 0.2$ is indicated via a dashed horizontal line. The analytic IT curve crosses this line first in the transition from wave $4$ to wave $5$. All other runs manage to converge as well, except for $200$ shots. The larger shot numbers resemble the course of the analytic IT curve, shifted by one wave. The range of ratios is limited to $0.65$ to better resolve differences at low values. This only affects RM. Waves start at $0$ because the target distribution is evaluated before the roll-out of a new wave.
    \end{minipage}
    \caption{Evolution of the NROY space and the target distribution with respect to the wave number for different randomized measurements (RM) repetitions and shot numbers. All HM runs are executed based on the YZ-CX kernel and the fixed hyperparameter configuration $(N=8, L=3, \nosamplepoints = 5 \times 10^4, \trainkernelonlyonce = 1, \implausibilitythresholdmin \approx 0.538, \implausibilitythresholdmindecayfactor \approx 0.451, \maxnoimplausibilities = 0, \convergencethresholdmetrics = 0.2, \randomnessseed = 46)$. Values are compared between evaluating the kernel analytically via the inversion test (IT), using $\{100, 500, 1000\}$ repetitions in the RM approach, and estimating the IT result with $\{100, 200, 500, 1000, 4000\}$ shots. All results are generated via statevector simulation. Shot noise is emulated by sampling from the probability distribution of encoded quantum states. The analytic IT outcomes are drawn in black. The RM curves are colored according to a blue-green gradient, from light to dark for growing number of repetitions. A red color gradient is used for the shot-based runs.}
    \label{fig:ITvsRMvsShots}
\end{figure*}

Both aspects -- randomized measurements and shot noise -- shall be investigated separately. To this end, \cref{fig:ITvsRMvsShots} compares the evolutions of two HM quantities for different numbers of RM repetitions in \cref{alg:RandomizedMeasurements} and different numbers of shots in \cref{alg:InversionTest}.
The benchmark is an analytic IT-based execution where the probability of the all-zero state is simply read out as in the \texttt{Optuna} studies (cf. \cref{subsec:HPOviaOptuna}). For comparability among the runs, we use the fixed hyperparameter configuration 
\begin{alignat*}{3}
    N &= 8, \quad L &&= 3, \quad \nosamplepoints &&= 5 \times 10^4, \\
    \trainkernelonlyonce &= 1, \; \implausibilitythresholdmin &&\approx 0.538, \; \implausibilitythresholdmindecayfactor &&\approx 0.451, \\
    \maxnoimplausibilities &= 0, \quad \convergencethresholdmetrics &&= 0.2, \quad \randomnessseed &&= 46,
\end{alignat*}
and the YZ-CX kernel, which appears to be preferable (compare \cref{tab:HPO} and \cref{subsec:PerformanceComparison}). Our test set consists of $\{100, 500, 1000\}$ randomized measurement repetitions and $\{100, 200, 500, 1000, 4000\}$ shots. All results are again generated via statevector simulation. For RM, this means to read out the basis state probabilities in \cref{alg:RandomizedMeasurements} exactly. On the other hand, shot noise is emulated by sampling from the fully-accessible probability distribution of encoded quantum states. We start with the remaining fraction of the NROY space in \cref{fig:ITvsRMvsShots} (a), displayed analogously to \cref{fig:NROYFractionAndReduction} (a). The curves show similar behavior, only the 100-RM-repetitions run stands out and does not fit the pattern. As discussed in \cref{subsec:PerformanceComparison}, the NROY evolution is not a suitable metric for the outcome of the HM. Nevertheless, the large degree of similarity in \cref{fig:ITvsRMvsShots} (a) indicates that our algorithm still works when using randomized measurements or shot noise. This can be inferred equivalently from \cref{fig:ITvsRMvsShots} (b), which plots the ratio of targets lying within the uncertainty ball around the observations. At each wave, the corresponding values are obtained by evolving the L96 model \eqref{eq:Lorenz96} on the newly drawn design points, calculating the metrics \eqref{eq:L96Metrics}, reducing them via the initial PCA, and retrieving the ratio of inputs for which all principal components are contained in the observational uncertainty (cf. \cref{subsec:AutomaticTuningOfL96}). In particular, \cref{fig:ITvsRMvsShots} (b) shows the process towards convergence for the different HM runs. The analytic IT curve exceeds the critical value $\convergencethresholdmetrics = 0.2$ first. All other runs manage to converge as well, except for $200$ shots. Although the larger shot numbers resemble the analytic IT course, shifted by one wave, we cannot observe a clear tendency of approaching the benchmark for increasing numbers of RM repetitions or shots, neither in \cref{fig:ITvsRMvsShots} (b) nor in \cref{fig:ITvsRMvsShots} (a). This is presumably due to the non-negligible effect of randomness in our configuration of \cref{alg:HistoryMatching}, which we already observed in \cref{subsec:PerformanceComparison}.


%% file: SECTIONS/06_Conclusion.tex
In this work, we present a hybrid quantum-classical history matching algorithm for tuning the Lorenz-96 model. As a first step, we refined and extended the classical framework from \textcite{lguensat_semi-automatic_2023}. In particular, we introduced a convergence criterion based on artificially created observational uncertainty, turning the mostly manual tuning procedure into a fully automated process. This allows, in principle, for a straightforward generalization to more advanced climate models, where noise on the observed data is inevitable due to intrinsic measurement uncertainties. L96, on the other hand, was selected as a surrogate model for its simplicity while still exhibiting chaotic behavior.

Inspired by \textcite{rapp_quantum_2024}, we propose to use quantum Gaussian processes as emulators for the (usually expensive) model inside the HM routine. As in the classical case, these QGPs are fully determined by a quantum kernel function, which itself mainly relies on an encoding scheme that maps points from the parameter space to a high-dimensional Hilbert space. We benchmark three such quantum feature maps by first performing an extensive hyperparameter optimization via \texttt{Optuna} for each architecture based on $300$ to $500$ trials. This enables a peak performance comparison on the basis of the best HM hyperparameter configurations. Quantum kernel values are evaluated via the popular inversion test. Using statevector simulation (i.e., all quantum circuit executions are simulated on purely classical machines), our ansatz amounts to a quantum-inspired algorithm. We numerically demonstrate the superiority of the NPQC and the YZ-CX kernel over the canonical classical RBF kernel with respect to all studied metrics (smallest rescaled Euclidean distance to parameter truth, average distance over multiple repetitions, as well as the implausibility scores corresponding to the best and the average). This makes the quantum-inspired approach valid in its own right. A drawback of our algorithm is that it is, to some extent, still driven by randomness, which complicates the search for ideal hyperparameter configurations.

Our quantum feature maps need at least one qubit for each of the four parameters of the Lorenz-96 model. 
With qubit numbers ranging in $\{4,6,8\}$ in our experiments, and NPQC / YZ-CX circuit depths scaling linearly with the moderate number of layers, the quantum routines of our hybrid algorithm are quite manageable in terms of both spatial and temporal resources. Also taking into account that NPQC and YZ-CX have only one trainable circuit parameter each, we infer that our method is particularly NISQ-friendly. We completed our work by discussing two strategies to make the transition from statevector simulation to real quantum hardware. The first consists of evaluating kernel values using randomized measurements instead of the conventional inversion test as suggested by \textcite{haug_quantum_2023}. The statistical ansatz of averaging over randomly sampled unitaries from the Haar measure on SU(2) could potentially compensate for the disruptive effect of gate errors. On the other hand, we take into account shot noise, which is a consequence of approximating probabilities by repeated preparation-measurement cycles. We provide numerical evidence that our algorithm is capable of yielding competitive solutions even with RM or shot noise in place.

We suspect the increased expressivity of the quantum feature maps due to the exponentially larger feature (Hilbert) space to be a crucial key to success. However, so far, we have only benchmarked the quantum kernels against the standard classical choice. Future work should investigate whether the quantum feature maps are also able to outperform more sophisticated classical architectures. Our results for L96 are promising and can be considered motivational for applying a hybrid HM to more complicated models for the Earth system. As an intermediate step on the pathway to a mature climate model, one could tackle the more realistic shallow water equations \cite{de_saint-venant_theorie_1871}. On the classical side, we see the need to reduce the influence of randomness in history matching in general. Coming up with a more advanced design-point sampling technique that is specifically tailored to address this issue could be a viable ansatz in future work.

%% file: APPENDIX/01_NPQC.tex
The shift factors $a_l$, $l \in \{2,...,L\}$,  in the NPQC layers \eqref{eq:NPQCLayer} are used to determine which pair of qubits to entangle via a $\cz$-gate. In line with \cite{haug_natural_2022}, they are, given values for $N$ and $L$, determined by the recursive relation sketched in \cref{alg:NPQCShiftFactor}. 
\begin{algorithm}[!h]
    \caption{NPQC-ShiftFactors$(N,L)$}
    \label{alg:NPQCShiftFactor}
        Set $A:=\left\{0,...,\frac{N}{2}-1\right\}$ \\
        Initialize $\text{factors} = \{\}$ \\
        Initialize $s=1$ \\
        \While{$|\,\text{factors}\,| < L$}{
            Set $a := A[-1] $ \\
            $A \gets A \setminus \{a\}$ \\
            $\text{factors} \gets \text{factors} \cup \{a\}$ \\
            \For{$q \in \{1,...,s-1\}$}{
                \If{$|\,\text{factors}\,| < L$}{
                    $\text{factors} \gets \text{factors} \cup \{\text{factors}[q]\}$
                }
            }
            $s \gets 2s$
        }
        \Return \text{factors}
\end{algorithm}

%% file: APPENDIX/02_HPO.tex
Here we provide some more details on the results of the \texttt{Optuna} HPO discussed in \cref{subsec:HPOviaOptuna}. First, we take a look at the hyperparameter distributions corresponding to the 20 best configurations (trials) of each kernel type, determined via a 50/50 weighting. Next is an investigation of the importance the different hyperparameters have for the two objectives (mean rescaled distance and mean number of waves, see \cref{subsec:HPOviaOptuna}). This is, we compare how stronly each hyperparameter influences the HM outcome. Then, we dive deeper into the \texttt{Optuna} studies for the individual kernels. Specifically, for each of the four architectures this means the following: (i) a Pareto plot that shows the evolution of the \texttt{Optuna} trials with respect to both objectives \eqref{eq:MeanDistanceRescaled} and \eqref{eq:MeanNumberOfWaves}; (ii) a heatmap matrix visualizing the correlation between pairs of hyperparameters; and finally (iii) objective boxplots comparing the spreads of the best 20 trials.

\subsection{Global Comparison}

\begin{figure*}[!ht]
    \centering
    \includegraphics[width=\textwidth]{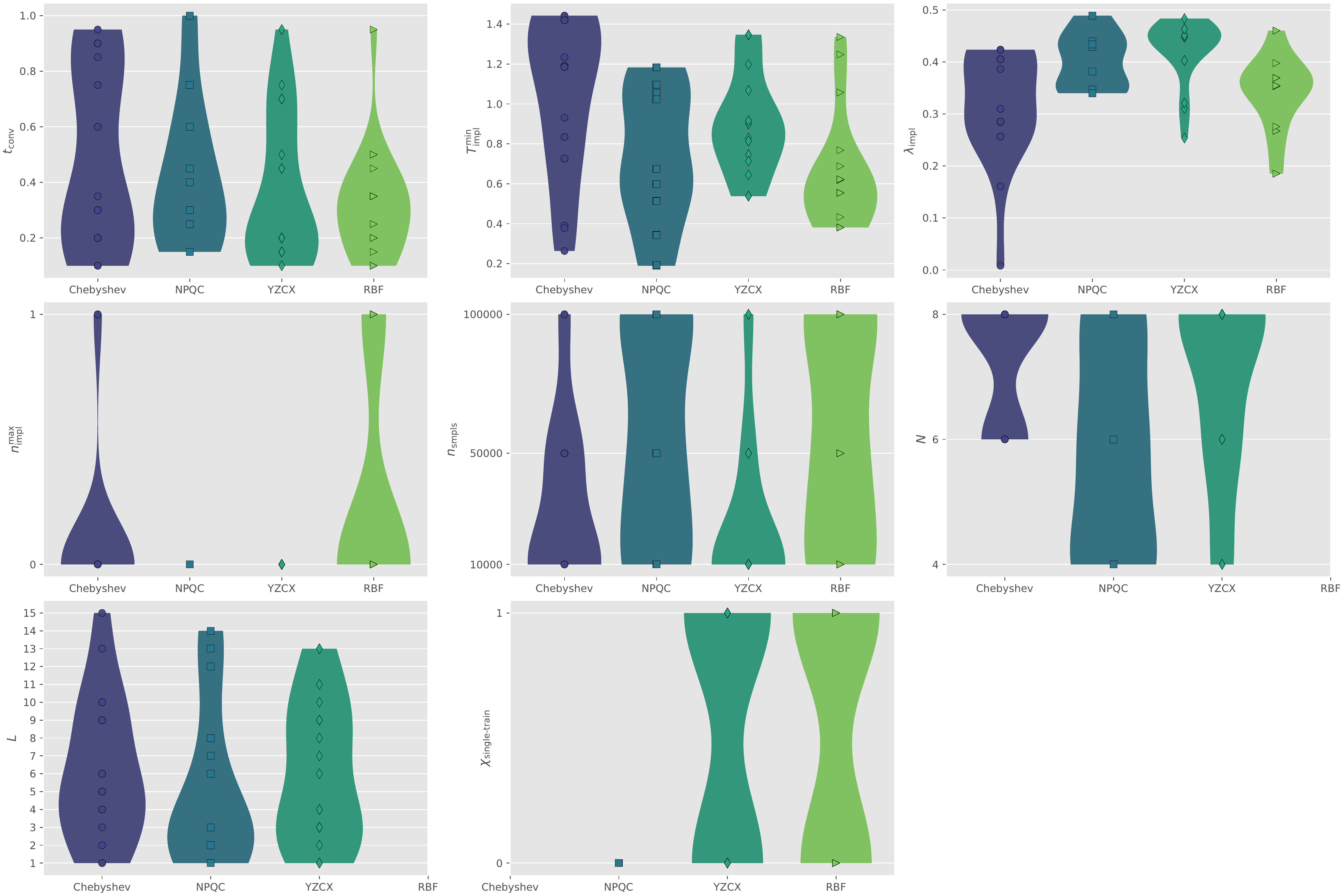}
    \caption{Distributions of hyperparameters corresponding to the 20 best trials for each kernel architecture. For the continuous hyperparameters $\convergencethresholdmetrics, \implausibilitythresholdmin$ and $\implausibilitythresholdmindecayfactor$, the full associated ranges are covered. For $\convergencethresholdmetrics$, coverage is even given by each architecture individually. For $\implausibilitythresholdmindecayfactor$, the vast majority of best trials lie in the upper half of the domain. The best configurations for the trainable Chebyshev kernel use either $6$ or $8$ qubits. The maximum number of layers $L$ decreases from Chebyshev over NPQC to YZ-CX. Concerning both binary hyperparameters $\maxnoimplausibilities, \trainkernelonlyonce$, NPQC only features value $0$. The same holds for YZ-CX and $\maxnoimplausibilities$.}
    \label{fig:HPO20BestHyperparameterDistributions}
\end{figure*}

\begin{figure*}[!ht]
    \centering
    \includegraphics[width=\textwidth]{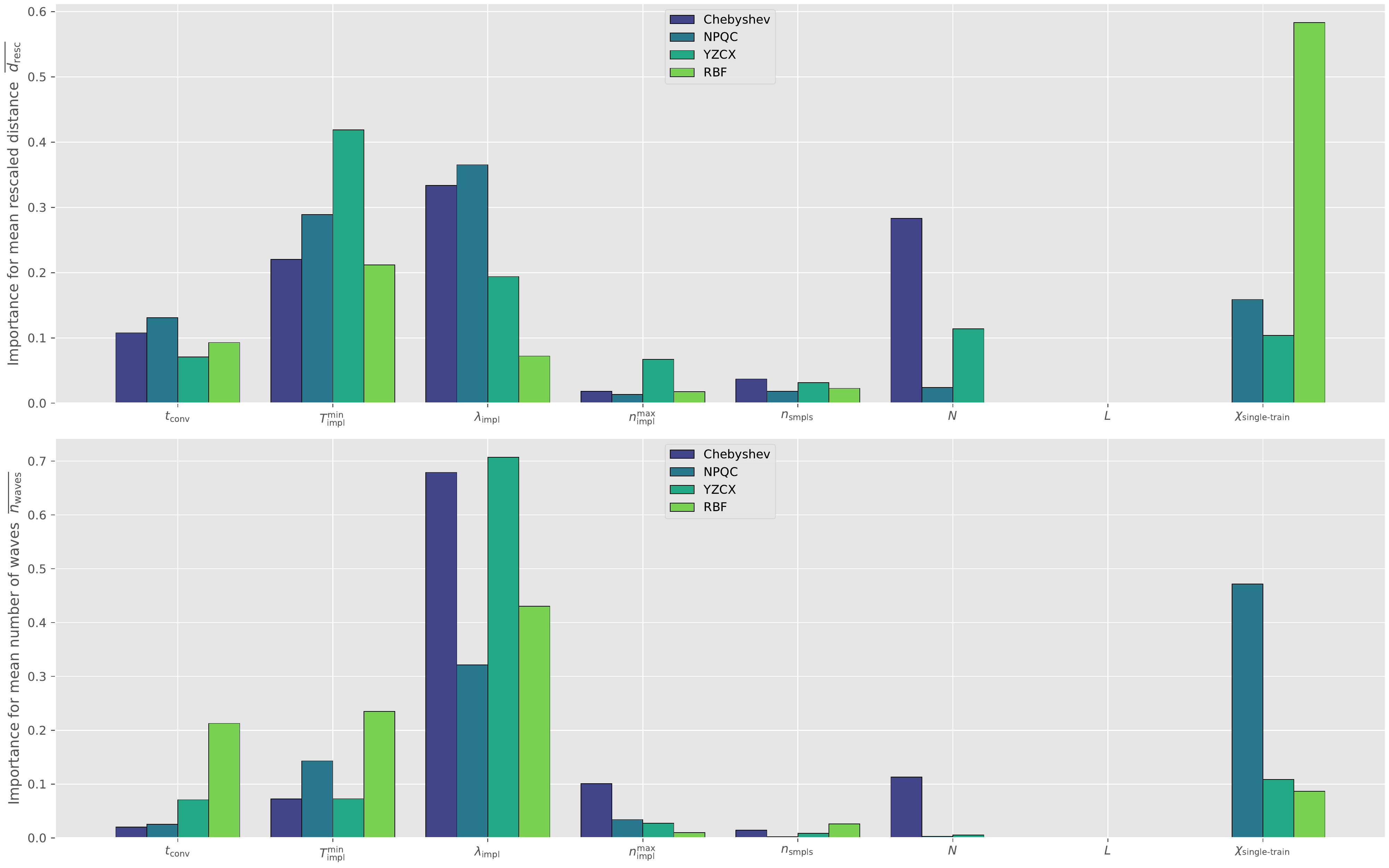}
    \caption{Importance of the hyperparameters for both objectives. On average, $\implausibilitythresholdmin$ and $\implausibilitythresholdmindecayfactor$ have the most impact on the mean rescaled distance. For RBF, $\trainkernelonlyonce$ is by far most influential for $\meandistancerescaled$. In terms of $\meannowaves$, RBF switches roles with NPQC. For the other architectures,  $\implausibilitythresholdmindecayfactor$ is most important for the mean number of waves. The influence of $\implausibilitythresholdmin$ is here reduced globally. For both objectives, $\convergencethresholdmetrics, \maxnoimplausibilities, \nosamplepoints$ and $N$ are of secondary importance. Only for the trainable Chebyshev kernel, the number of qubits shows a medium peak. The number of layers $L$ has no impact at all.}
    \label{fig:HPOHyperparameterImportance}
\end{figure*}

The above-described hyperparameter distributions of the 20 best trials for each kernel architecture can be found in \cref{fig:HPO20BestHyperparameterDistributions}. On the other hand, \cref{fig:HPOHyperparameterImportance} compares the initially described importance of the different hyperparameters.

\subsection{Trainable Chebyshev}

\begin{figure*}[ht!]
    \centering

    \includegraphics[width=\textwidth]{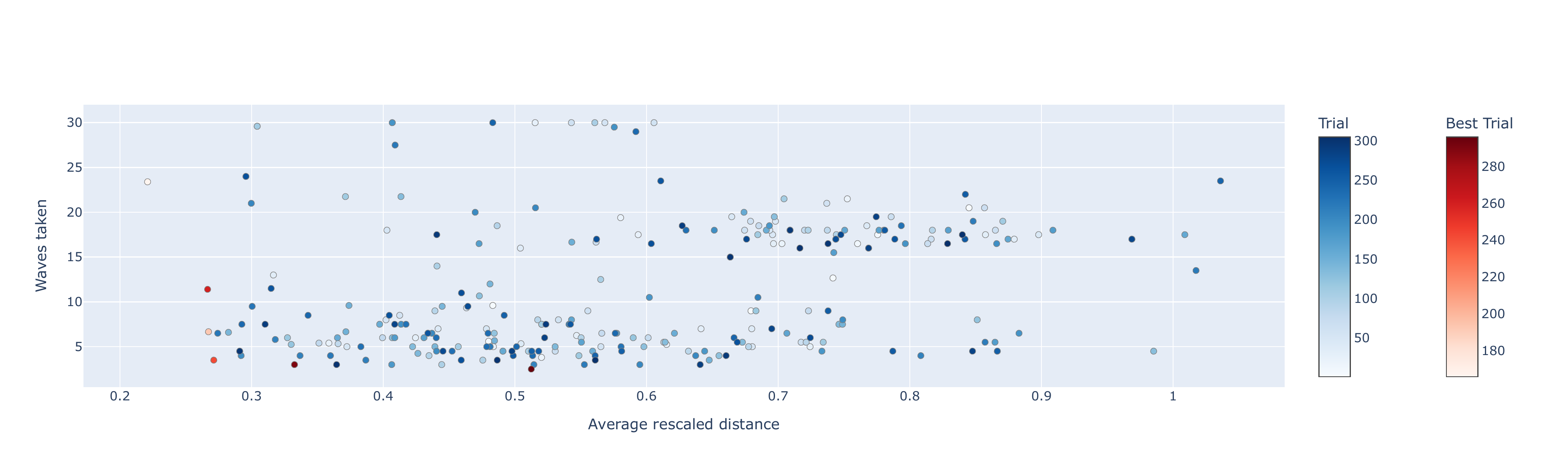}
    \par\vspace{0.5ex}
    \raggedright
    \footnotesize (a) Pareto front of the mean number of waves vs. the average rescaled distance. The five best trials are colored in a shade of red. Among them are the three trials with the smallest mean rescaled distance at around 0.27. The remaining two likewise feature a comparably low distance; more importantly, they correspond to the lowest measured average wave numbers between 2 and 3.

    \par\vspace{1.5ex}

    \begin{minipage}{0.55\textwidth}
        \resizebox{\textwidth}{!}{
            \includegraphics{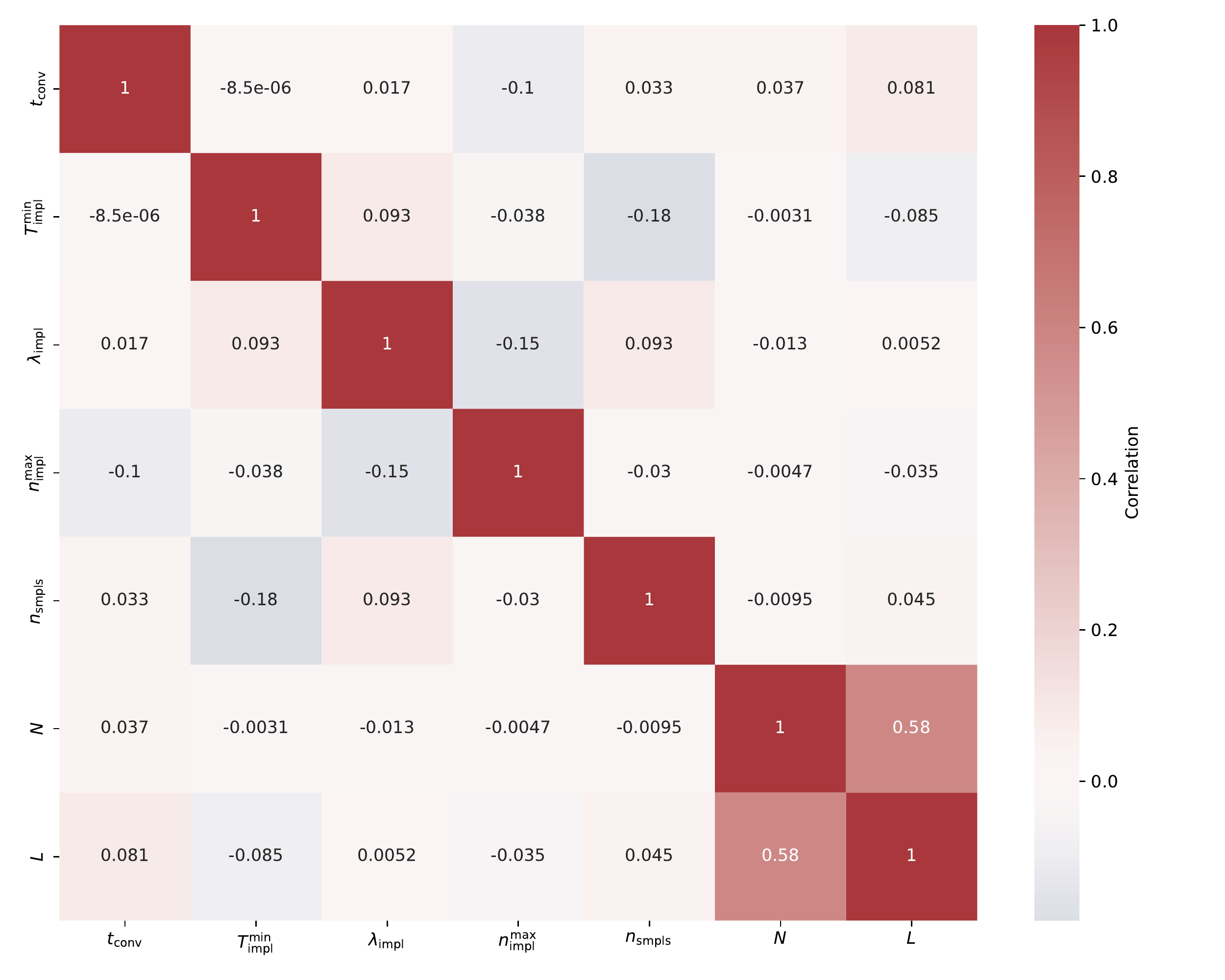}
        }
    \end{minipage}\hfill
    \begin{minipage}{0.4\textwidth}
        \vspace{0.2cm}
        \raggedright
        \footnotesize (b) Heatmap correlation matrix for the optimized hyperparameters. By design, the matrix is symmetric with an identity diagonal. A correlation with an absolute value larger than 0.5 can only be found for the relation between the number of qubits and the number of layers. This observation is in line with the construction of the HPO in \cref{subsec:HPOviaOptuna}: The upper bound for the number of layers, which depends on the number of qubits, represents the only direct relation of hyperparameters from the outset. Other than that, the hyperparameters turn out to be largely uncorrelated.
    \end{minipage}

    \par\vspace{1.5ex}

    \begin{minipage}{0.65\textwidth}
        \resizebox{\textwidth}{!}{
        \includegraphics{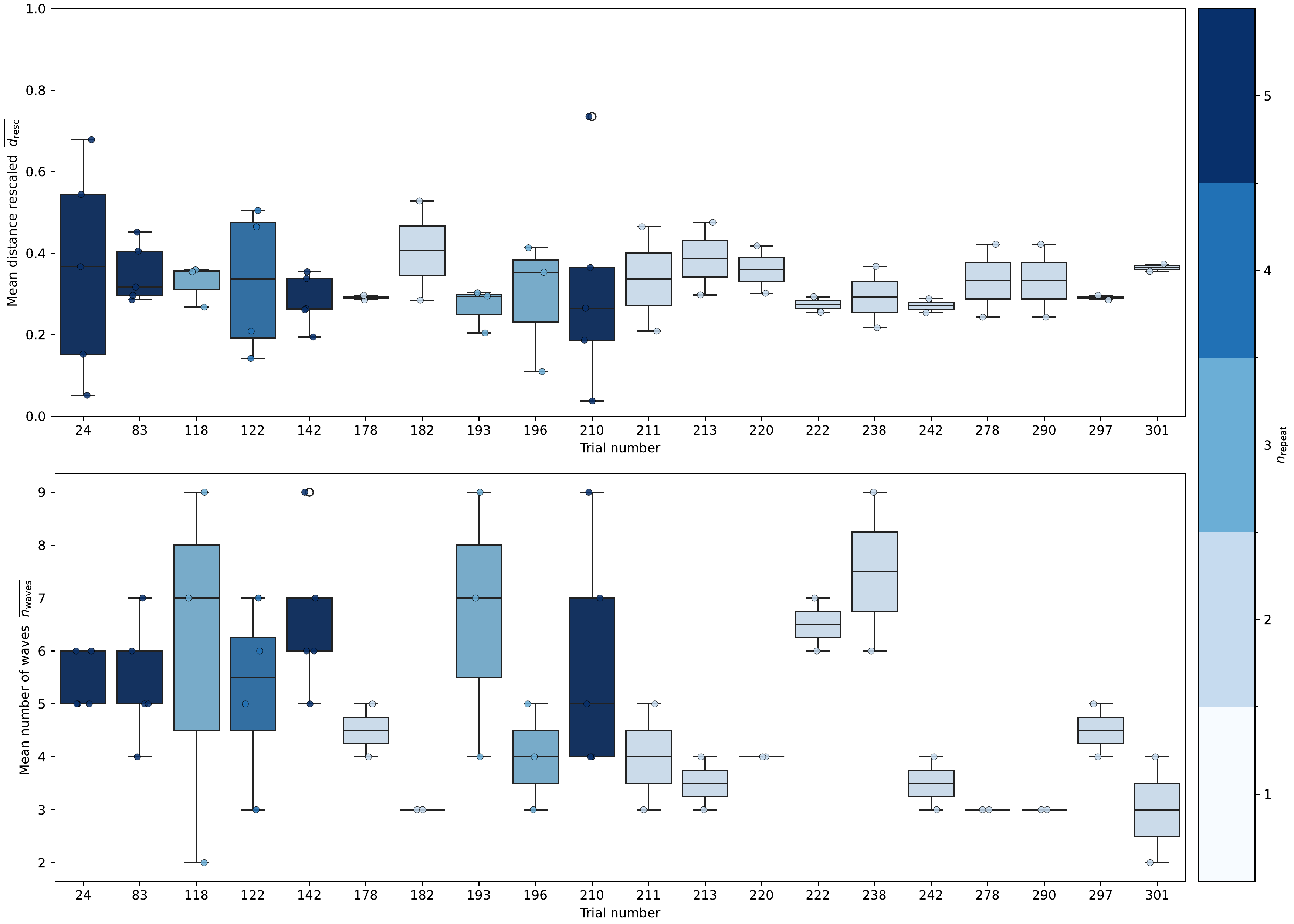}
        }
    \end{minipage}\hfill
    \begin{minipage}{0.3\textwidth}
        \vspace{0.2cm}
        \raggedright
        \footnotesize (c) Spreads of the two HPO objectives corresponding to the 20 best trials according to a uniform weighting. The color indicates the number of repetitions performed by \texttt{Optuna} for the respective trial (cf. \cref{subsec:HPOviaOptuna}). With 12 out of these 20 top trials, 60\% just encountered the minimum number of two repetitions. On the other hand, the full five repetitions were only run for 20\% of the best trials. While the rescaled distances of all repetitions of all trials are mostly contained in a narrow band between 0.3 and 0.4, there are larger fluctuations in the number of waves. For trial 118, the three conducted repetitions even span the full range from 2 to 9 waves.
    \end{minipage}

    \caption{Results of the HPO for the trainable Chebyshev kernel based on 300 \texttt{Optuna} trials.}
    \label{fig:HPOChebyshev}
\end{figure*}

The above-described selection of results from the \texttt{Optuna} HPO for the trainable Chebyshev kernel (cf. \cref{subsubsec:ChebyshevKernel}) can be found in \cref{fig:HPOChebyshev}.

\subsection{NPQC} 

\begin{figure*}[h!]
    \centering

    \includegraphics[width=\textwidth]{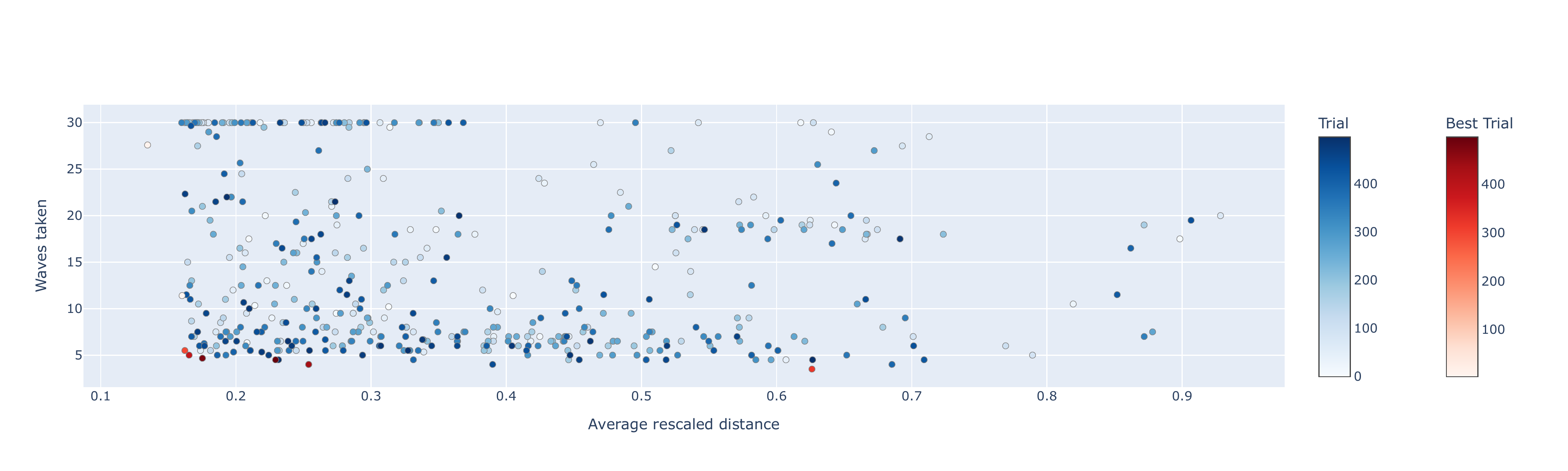}
    \par\vspace{0.5ex}
    \raggedright
    \footnotesize (a) Pareto front of the mean number of waves vs. the average rescaled distance. The six best trials are colored in a shade of red. Among them are the three trials in the lower left corner, where both objectives are small. By far the best mean rescaled distance is achieved in a low-order trial with approximately 28 waves on average. The other half of the top trials corresponds to the lowest measured average wave numbers between ca. 2.5 and 4. With a rescaled distance of about 0.63, the best mean wave number belongs to the upper half of the distance range.

    \par\vspace{1.5ex}

    \begin{minipage}{0.5\textwidth}
        \resizebox{\textwidth}{!}{
            \includegraphics{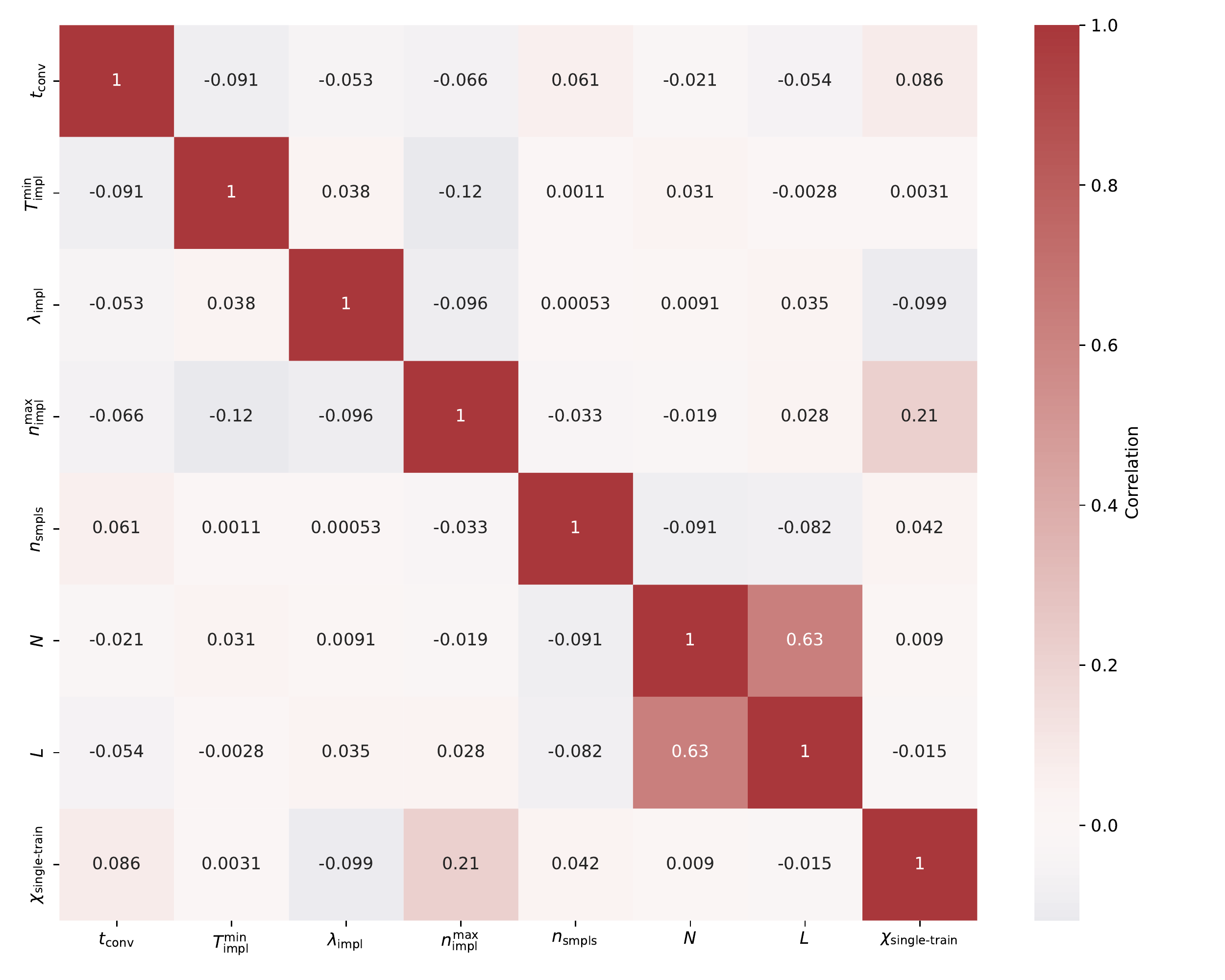}
        }
    \end{minipage}\hfill
    \begin{minipage}{0.4\textwidth}
        \vspace{0.2cm}
        \raggedright
        \footnotesize (b) Heatmap correlation matrix for the optimized hyperparameters. By design, the matrix is symmetric with an identity diagonal. A correlation with an absolute value larger than 0.5 can only be found for the relation between the number of qubits and the number of layers. This observation is in line with the construction of the HPO in \cref{subsec:HPOviaOptuna}: The upper bound for the number of layers, which depends on the number of qubits, represents the only direct relation of hyperparameters from the outset. Other than that, the hyperparameters turn out to be largely uncorrelated.
    \end{minipage}

    \par\vspace{1.5ex}

    \begin{minipage}{0.65\textwidth}
        \resizebox{\textwidth}{!}{
        \includegraphics{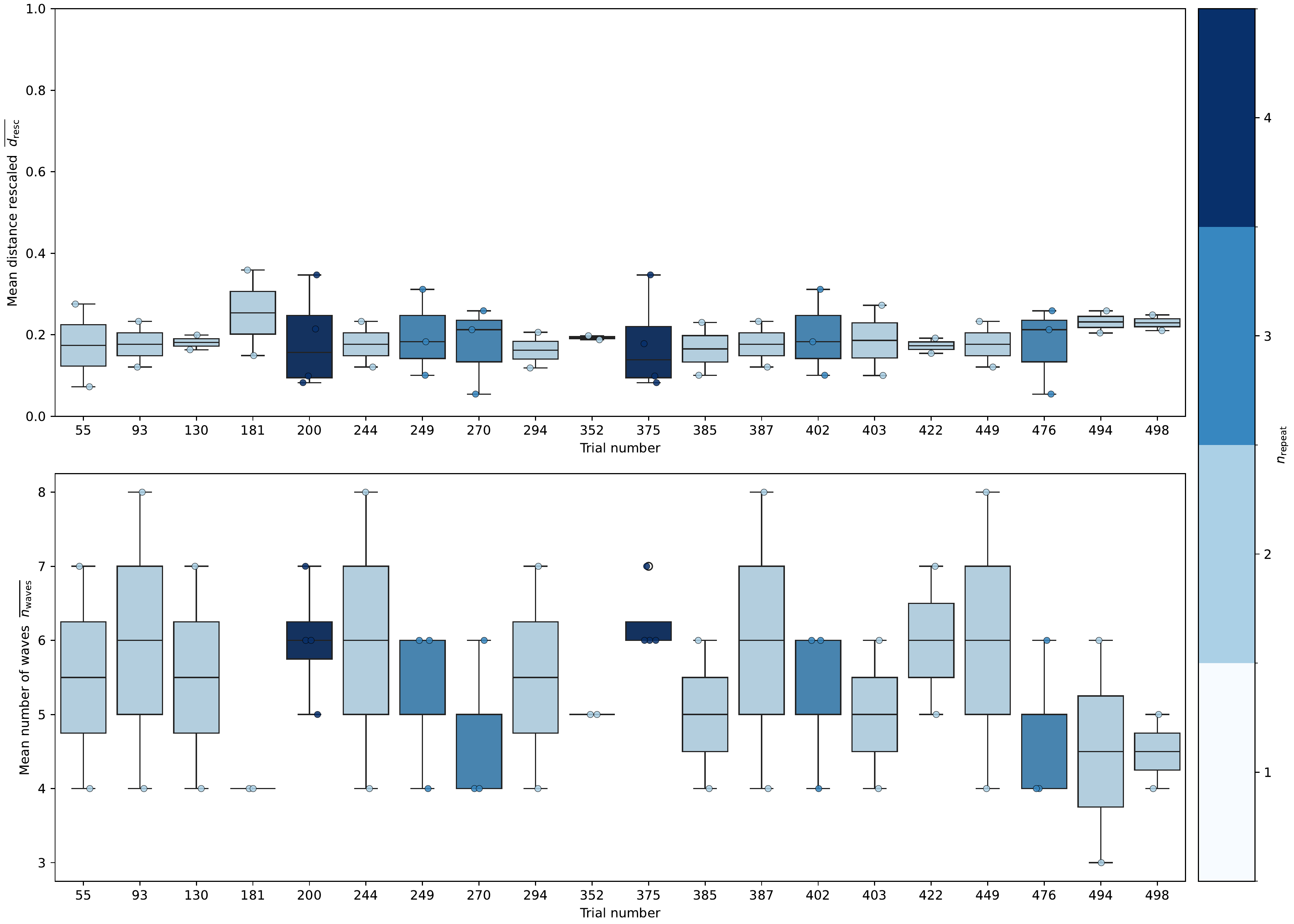}
        }
    \end{minipage}\hfill
    \begin{minipage}{0.3\textwidth}
        \vspace{0.2cm}
        \raggedright
        \footnotesize (c) Spreads of the two HPO objectives corresponding to the 20 best trials according to a uniform weighting. The color indicates the number of repetitions performed by \texttt{Optuna} for the respective trial (cf. \cref{subsec:HPOviaOptuna}). With 14 out of these 20 top trials, 70\% just encountered the minimum number of two repetitions. On the other hand, the full five repetitions were only run for 10\% of the best trials. While the rescaled distances of all repetitions of all trials are mostly contained in a narrow band between 0.2 and 0.3, there are larger fluctuations in the number of waves. Except for trial 494, all values lie within 4 and 8 waves.
    \end{minipage}

    \caption{Results of the HPO for the NPQC kernel based on 500 \texttt{Optuna} trials.}
    \label{fig:HPONPQC}
\end{figure*}

The initially described selection of results from the \texttt{Optuna} HPO for the NPQC kernel (cf. \cref{subsubsec:NPQCKernel}) can be found in \cref{fig:HPONPQC}.

\subsection{YZ-CX} 

\begin{figure*}[h!]
    \centering

    \includegraphics[width=\textwidth]{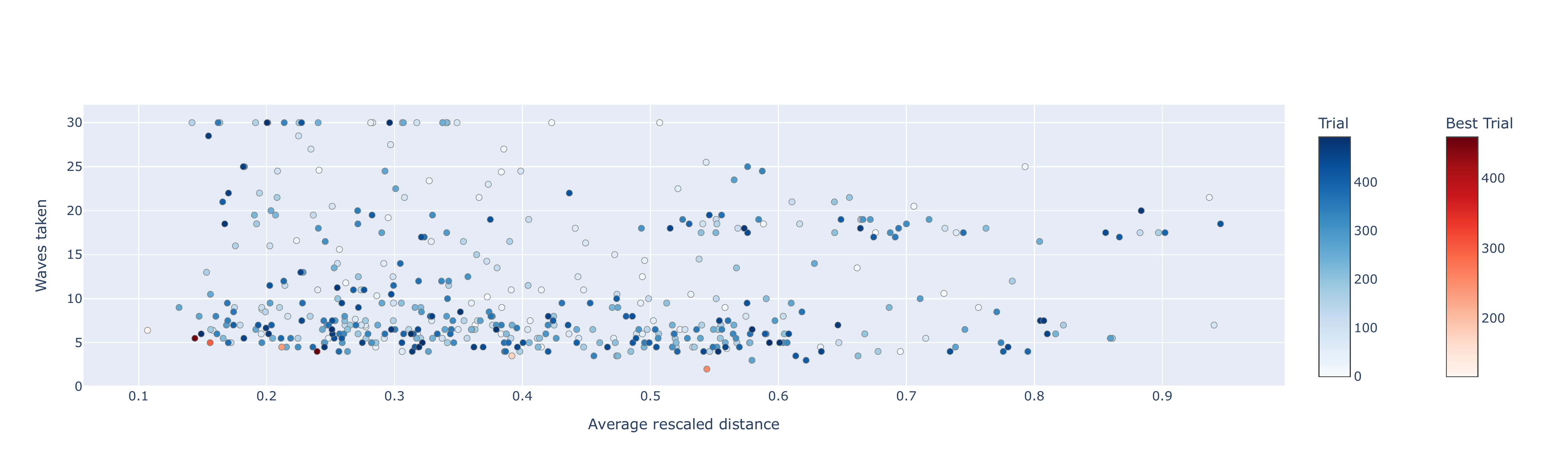}
    \par\vspace{0.5ex}
    \raggedright
    \footnotesize (a) Pareto front of the mean number of waves vs. the average rescaled distance. Six of the best trials are colored in a shade of red. Among them are the two trials in the lower left corner, where both objectives are small. By far the best mean rescaled distance is achieved in a low-order trial with approximately 6 waves on average. Two more are located at the vertical lower border with mean rescaled distances between 0.21 and 0.24. The remaining two top trials stay at small average wave numbers, while the distance is further increased. With a rescaled distance of about 0.55, the best mean wave number belongs to the upper half of the distance range.

    \par\vspace{1.5ex}

    \begin{minipage}{0.5\textwidth}
        \resizebox{\textwidth}{!}{
            \includegraphics{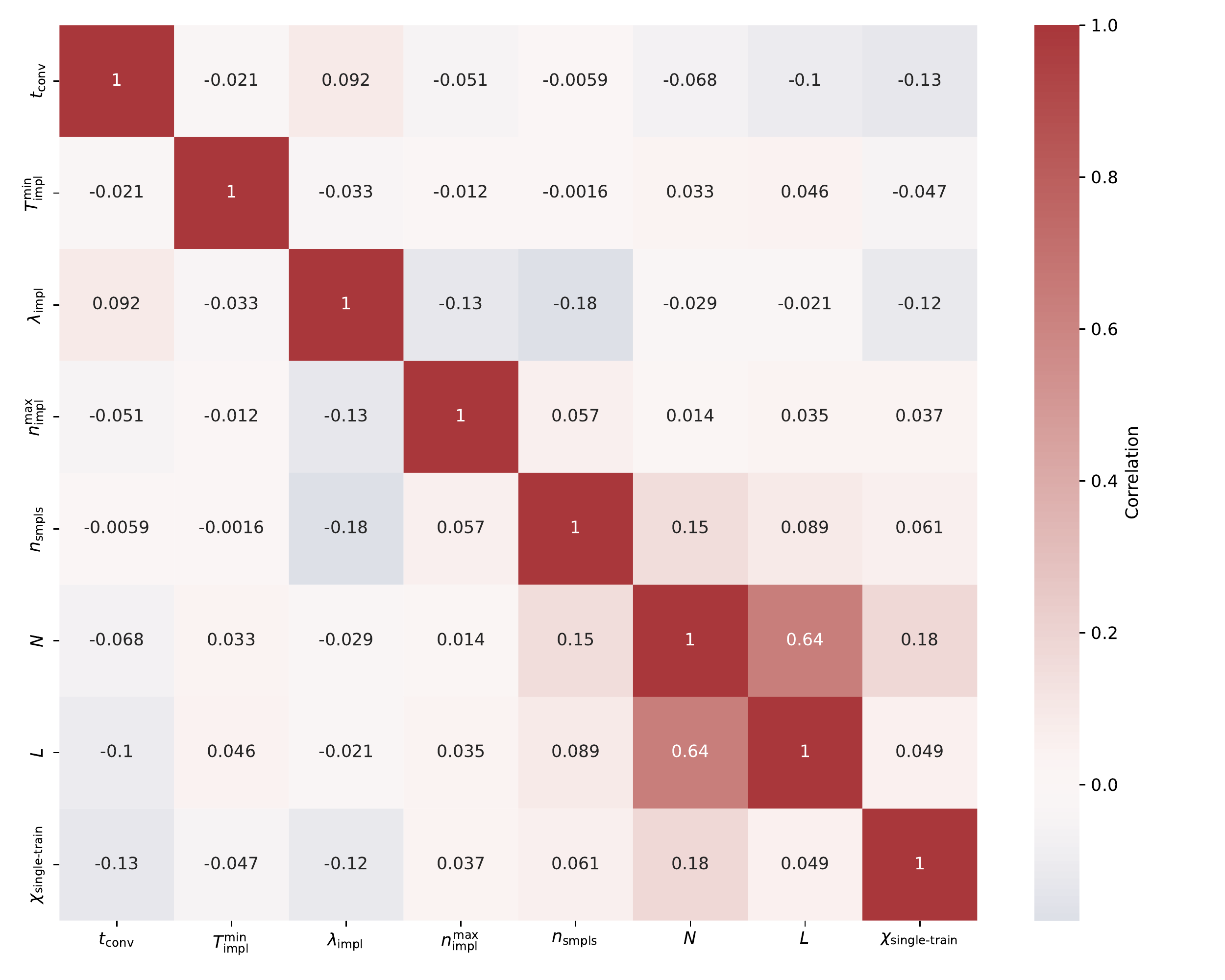}
        }
    \end{minipage}\hfill
    \begin{minipage}{0.4\textwidth}
        \vspace{0.2cm}
        \raggedright
        \footnotesize (b) Heatmap correlation matrix for the optimized hyperparameters. By design, the matrix is symmetric with an identity diagonal. A correlation with an absolute value larger than 0.5 can only be found for the relation between the number of qubits and the number of layers. This observation is in line with the construction of the HPO in \cref{subsec:HPOviaOptuna}: The upper bound for the number of layers, that depends on the number of qubits, represents the only direct relation of hyperparameters from the outset. Other than that, the hyperparameters turn out to be largely uncorrelated.
    \end{minipage}

    \par\vspace{1.5ex}

    \begin{minipage}{0.65\textwidth}
        \resizebox{\textwidth}{!}{
        \includegraphics{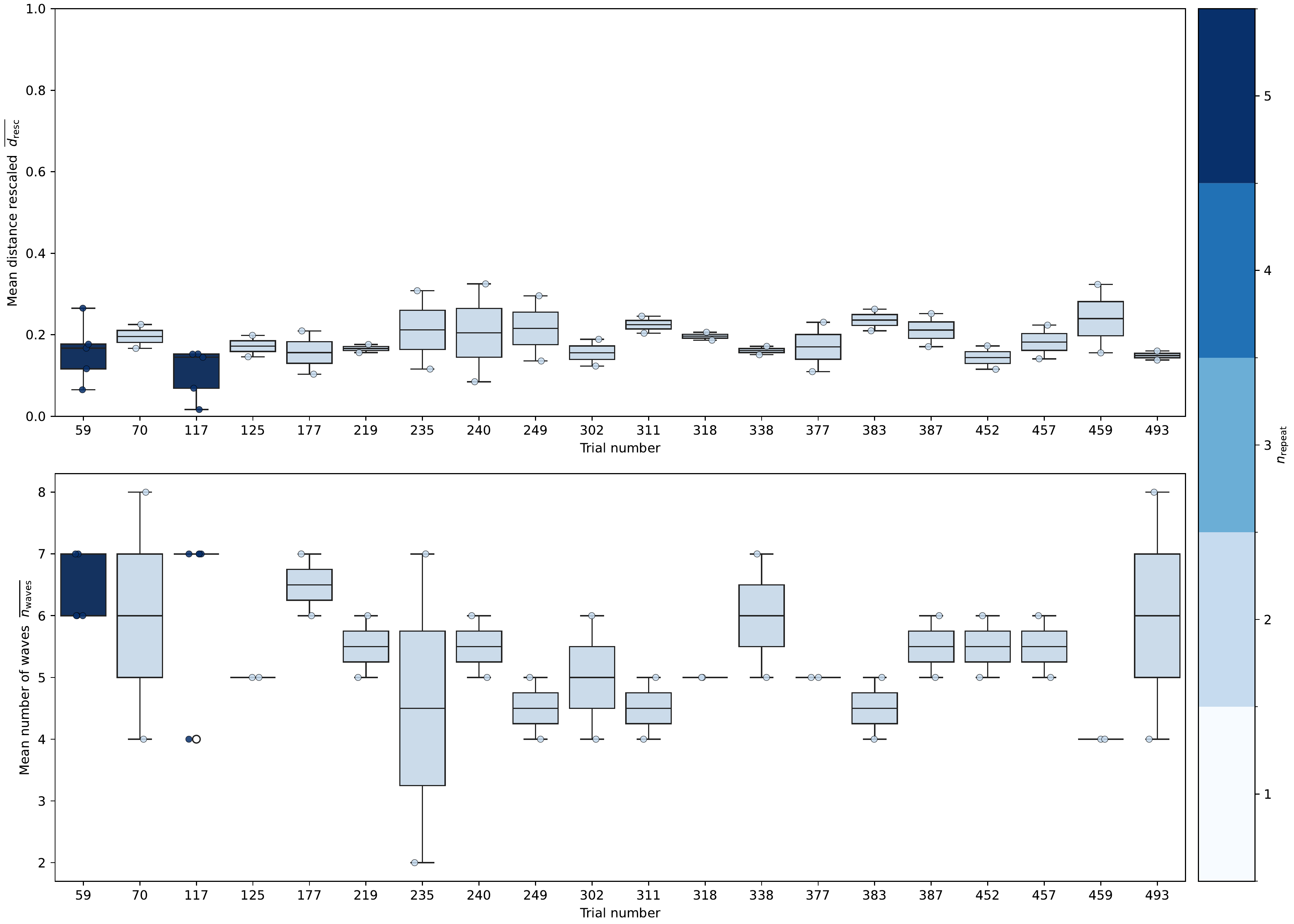}
        }
    \end{minipage}\hfill
    \begin{minipage}{0.3\textwidth}
        \vspace{0.2cm}
        \raggedright
        \footnotesize (c) Spreads of the two HPO objectives corresponding to the 20 best trials according to a uniform weighting. The color indicates the number of repetitions performed by \texttt{Optuna} for the respective trial (cf. \cref{subsec:HPOviaOptuna}). With 18 out of these 20 top trials, 90\% just encountered the minimum number of two repetitions. On the other hand, the full five repetitions were only run for 10\% of the best trials. While the rescaled distances of all repetitions of all trials are mostly contained in a narrow band between 0.1 and 0.2, there are larger fluctuations in the number of waves. However, although the centers for the different trials are more widely distributed in the range from 2 to 8 waves, only three (or four) trials come with a significant spread.
    \end{minipage}

    \caption{Results of the HPO for the YZ-CX kernel based on 500 \texttt{Optuna} trials.}
    \label{fig:HPOYZCX}
\end{figure*}

The initially described selection of results from the \texttt{Optuna} HPO for the YZ-CX kernel (cf. \cref{subsubsec:YZCXKernel}) can be found in \cref{fig:HPOYZCX}.

\subsection{RBF} 

\begin{figure*}[h!]
    \centering

    \includegraphics[width=\textwidth]{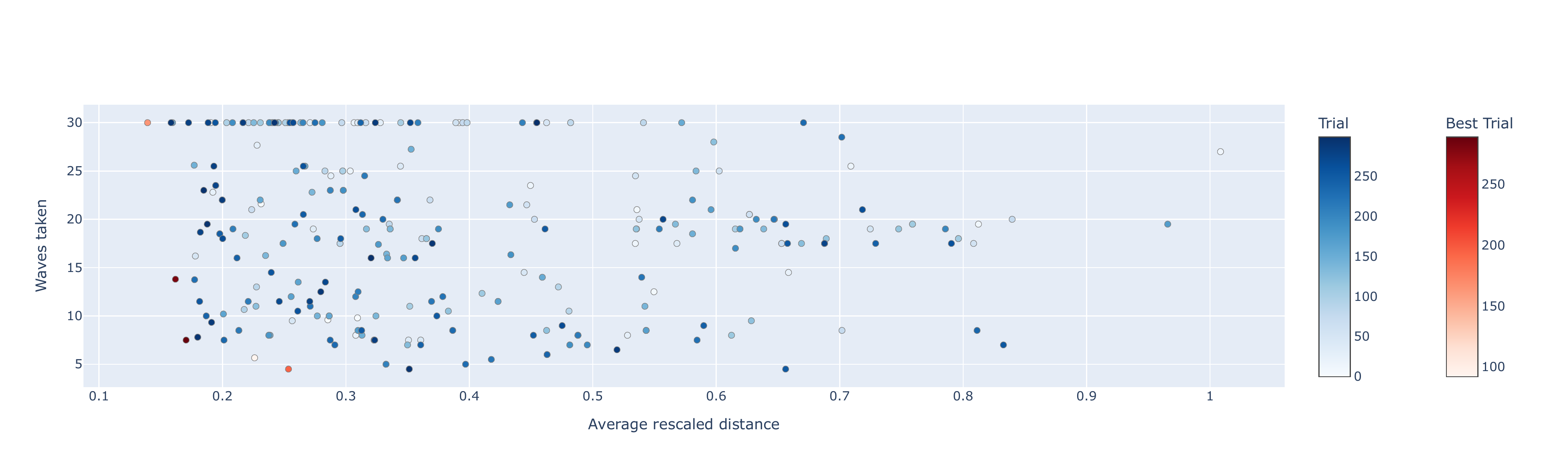}
    \par\vspace{0.5ex}
    \raggedright
    \footnotesize (a) Pareto front of the mean number of waves vs. the average rescaled distance. The four best trials are colored in a shade of red. All of them feature a small mean rescaled distance. Three of the four are located in the lower left corner, where both objectives have good values. The other one represents the trial with the best achieved mean rescaled distance of ca. 0.15. However, the full $30$ permitted waves were needed to reach this optimum.

    \par\vspace{1.5ex}

    \begin{minipage}{0.5\textwidth}
        \resizebox{\textwidth}{!}{
            \includegraphics{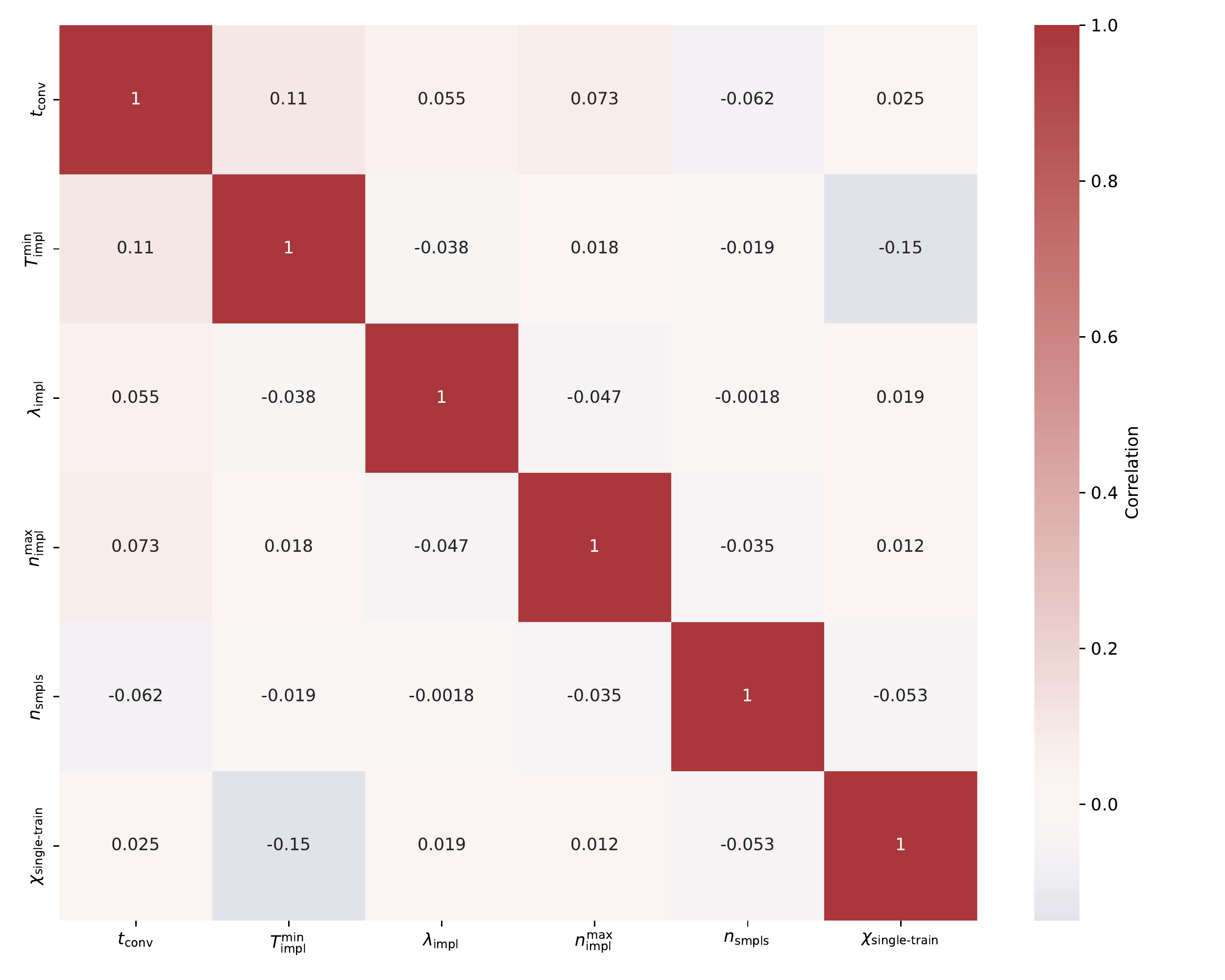}
        }
    \end{minipage}\hfill
    \begin{minipage}{0.4\textwidth}
        \vspace{0.2cm}
        \raggedright
        \footnotesize (b) Heatmap correlation matrix for the optimized hyperparameters. By design, the matrix is symmetric with an identity diagonal. No significant correlation can be found for any pair of distinct hyperparameters. The combination of the convergence threshold $\convergencethresholdmetrics$ and the minimum implausibility threshold $\implausibilitythresholdmin$ is the only one exceeding an absolute value of $0.1$.
    \end{minipage}

    \par\vspace{1.5ex}

    \begin{minipage}{0.65\textwidth}
        \resizebox{\textwidth}{!}{
        \includegraphics{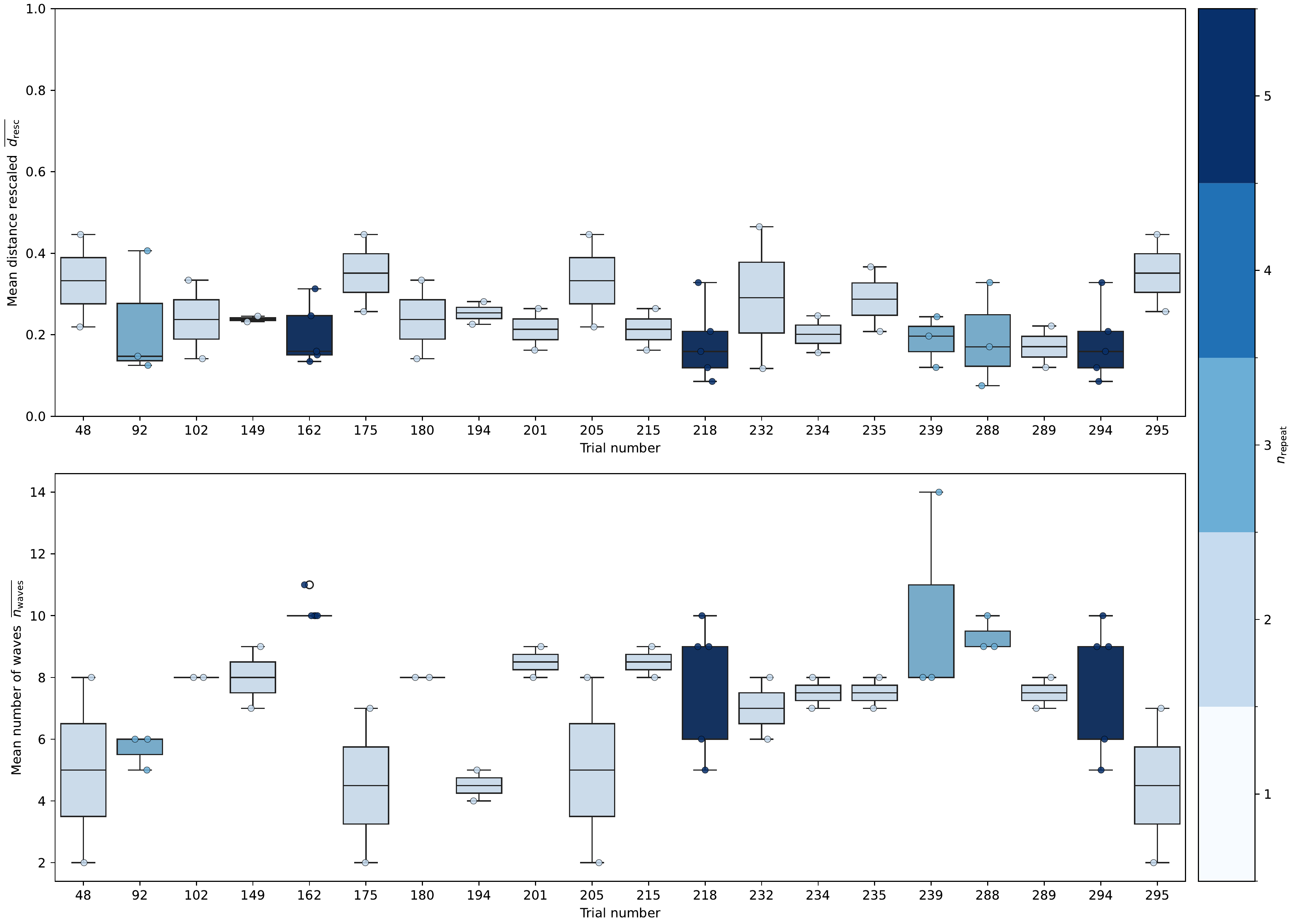}
        }
    \end{minipage}\hfill
    \begin{minipage}{0.3\textwidth}
        \vspace{0.2cm}
        \raggedright
        \footnotesize (c) Spreads of the two HPO objectives corresponding to the 20 best trials according to a uniform weighting. The color indicates the number of repetitions performed by \texttt{Optuna} for the respective trial (cf. \cref{subsec:HPOviaOptuna}). With 14 out of these 20 top trials, 70\% just encountered the minimum number of two repetitions. On the other hand, the full five repetitions were only run for 15\% of the best trials. While the rescaled distances of all repetitions of all trials are mostly contained in a narrow band between 0.1 and 0.4, there are larger fluctuations in the number of waves. However, although the centers for the different trials are more widely distributed in the range from 2 to 14 waves, only seven trials come with a significant spread.
    \end{minipage}

    \caption{Results of the HPO for the RBF kernel based on 300 \texttt{Optuna} trials.}
    \label{fig:HPORBF}
\end{figure*}

The initially described selection of results from the \texttt{Optuna} HPO for the RBF kernel \eqref{eq:RBFKernel} can be found in \cref{fig:HPORBF}.

%% file: APPENDIX/03_HM.tex
\begin{figure*}[!h]
    \centering

    \begin{minipage}[b]{0.48\textwidth}
        \centering
        \includegraphics[width=\textwidth]{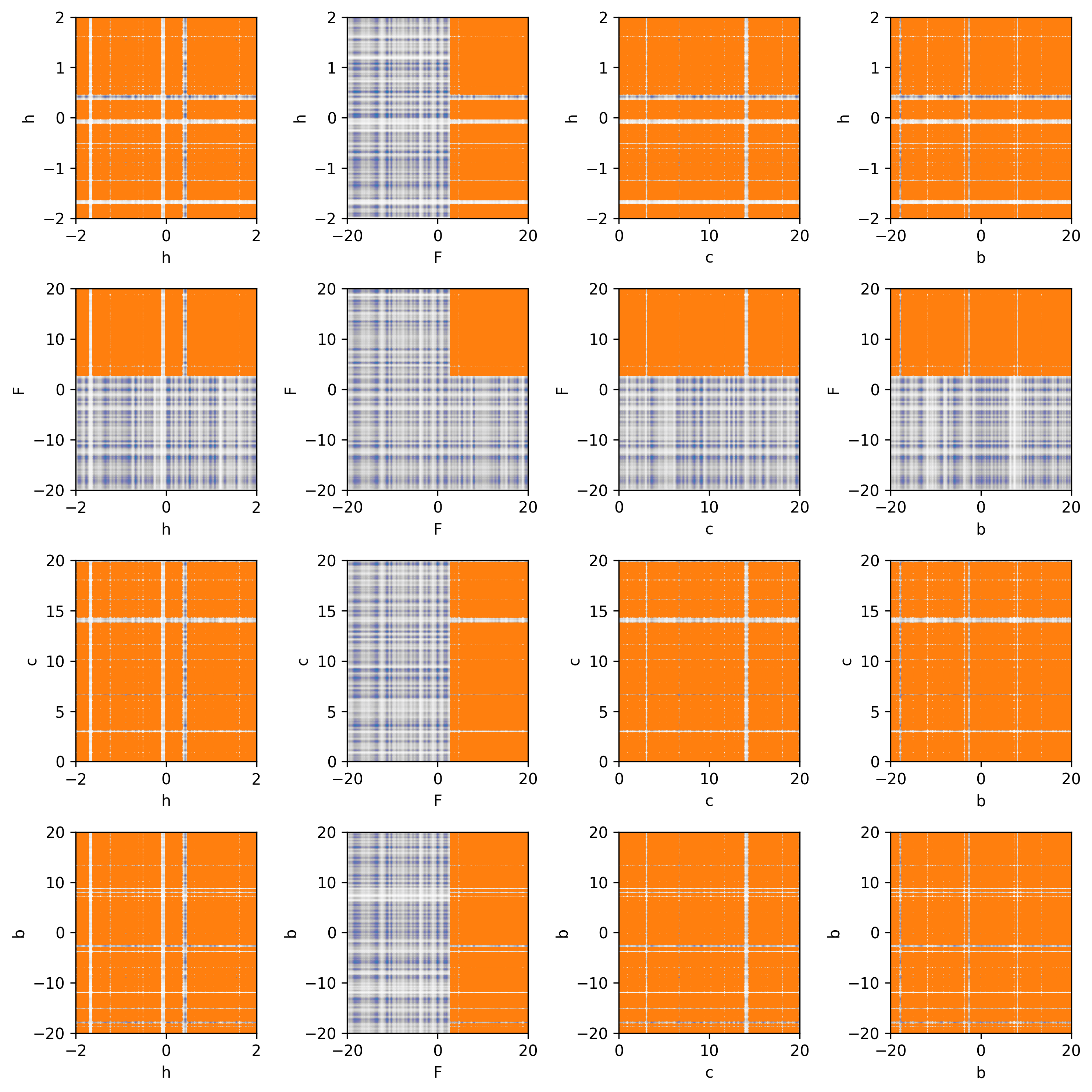}
        \par\vspace{3ex}
        \footnotesize (a) Initial LHS sampling vs. NROY space after the first wave. 
    \end{minipage}
    \hfill
    \begin{minipage}[b]{0.48\textwidth}
        \centering
        \includegraphics[width=\textwidth]{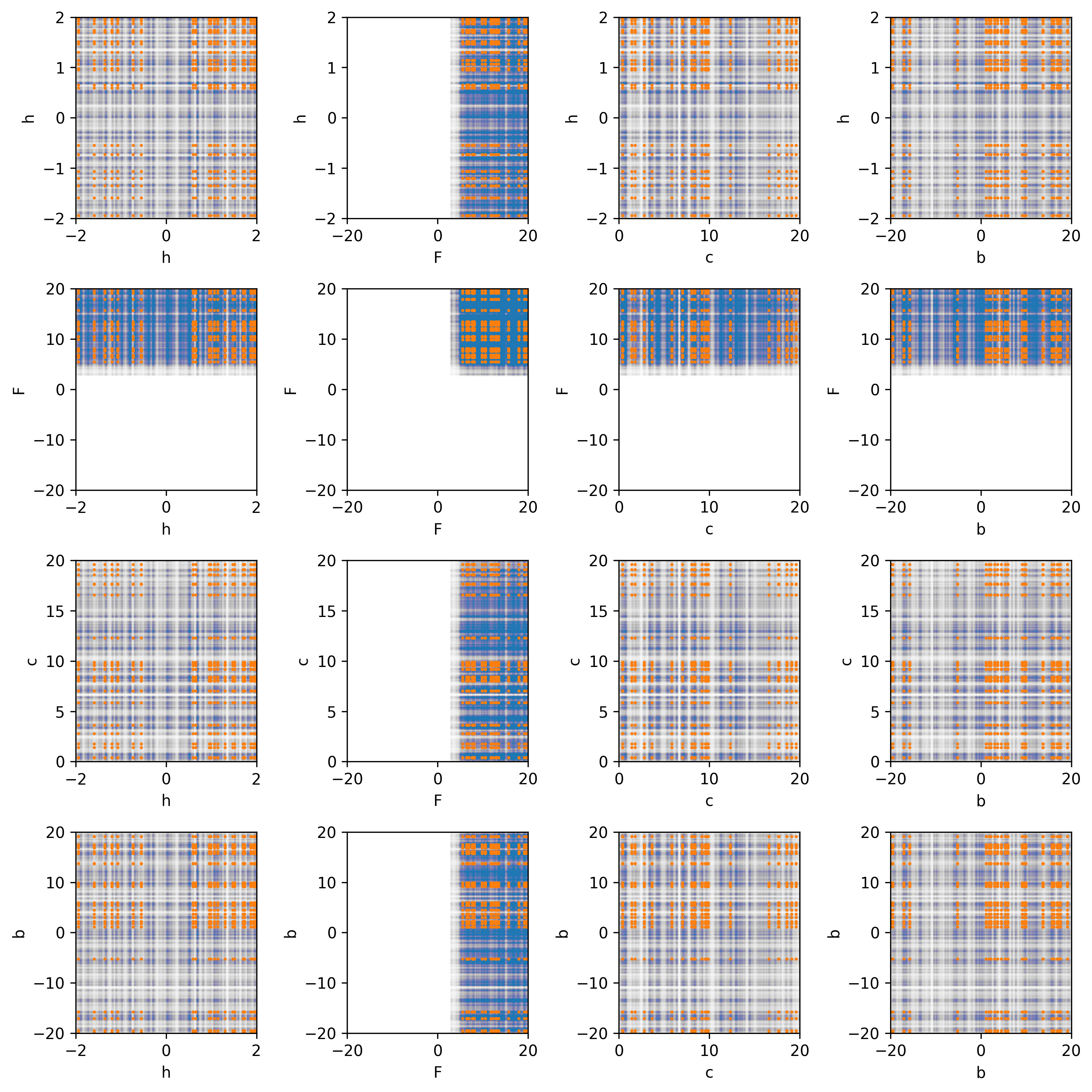}
        \par\vspace{3ex}
        \footnotesize (b) NROY space after the first vs. the second wave.
    \end{minipage}

    \vspace{1.5em} 

    \begin{minipage}[b]{0.48\textwidth}
        \centering
        \includegraphics[width=\textwidth]{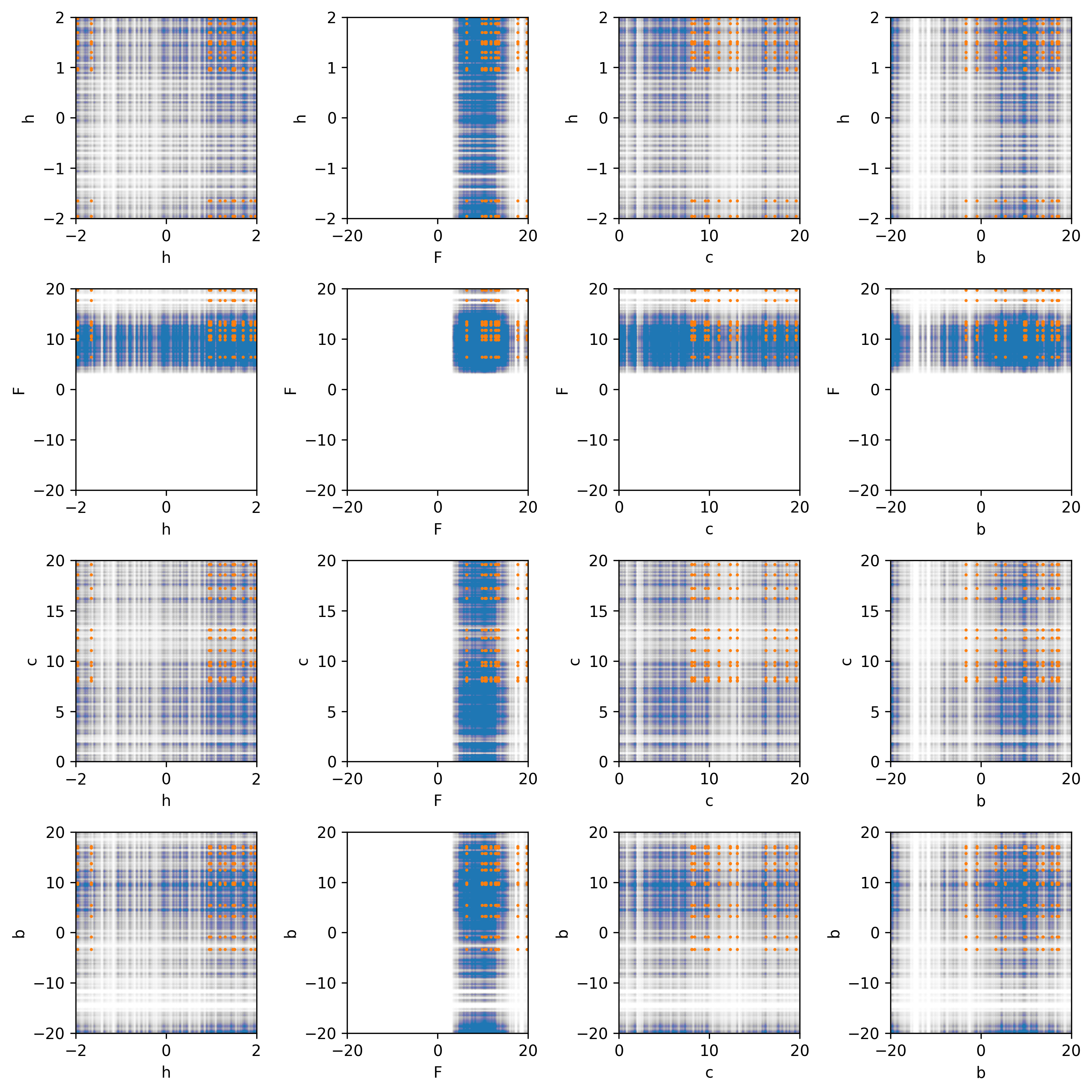}
        \par\vspace{3ex}
        \footnotesize (c) NROY space after the second vs. the third wave.
    \end{minipage}
    \hfill
    \begin{minipage}[b]{0.48\textwidth}
        \centering
        \includegraphics[width=\textwidth]{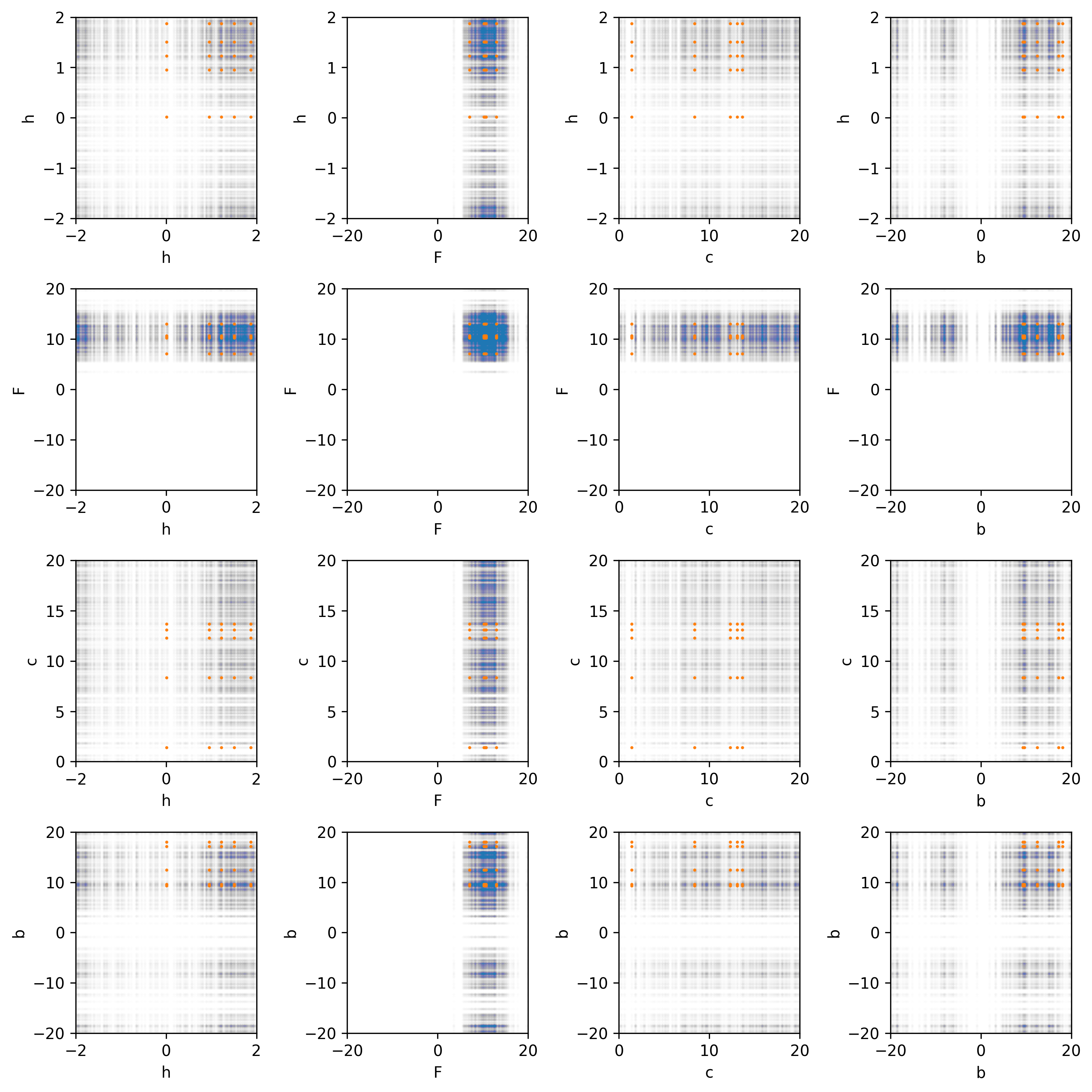}
        \par\vspace{3ex}
        \footnotesize (d) NROY space after the third vs. the fourth wave.
    \end{minipage}

    \vspace{0.5em}

    \caption{Evolution of the NROY space from the initial LHS sampling of the parameter space until convergence after four waves for the best NPQC trial with the following hyperparameter configuration specified in \cref{tab:HPO}: $(N=6, L=2, \nosamplepoints = 1 \times 10^4, \trainkernelonlyonce = 0, \implausibilitythresholdmin \approx 1.096, \implausibilitythresholdmindecayfactor \approx 0.382, \maxnoimplausibilities = 0, \convergencethresholdmetrics = 0.25, \randomnessseed = 42)$. For every combination of L96 model parameters, the NROY space is obtained by projecting to the respective two-dimensional reduced parameter space. The old (previous) NROY space is colored in blue, the updated one in orange. For readability, the number of drawn points is limited to $500$. For the forcing $F$, the HM already finds a very restricted feasible subset of the parameter space in the first wave. The other parameters are more complicated to assess. Even the final NROY space for $h$, $c$, and $b$ spans almost the full associated parameter ranges, with only very few non-colored regions.}
    \label{fig:HMNROYSpace}
\end{figure*}

For the exemplary best NPQC run with the hyperparameters configured as shown in \cref{tab:HPO}, \cref{fig:HMNROYSpace} visualizes the different stages of the NROY space from the initial LHS sampling of the parameter space until convergence is reached after four waves.

\begin{figure*}
    \centering
    \includegraphics[width=\textwidth]{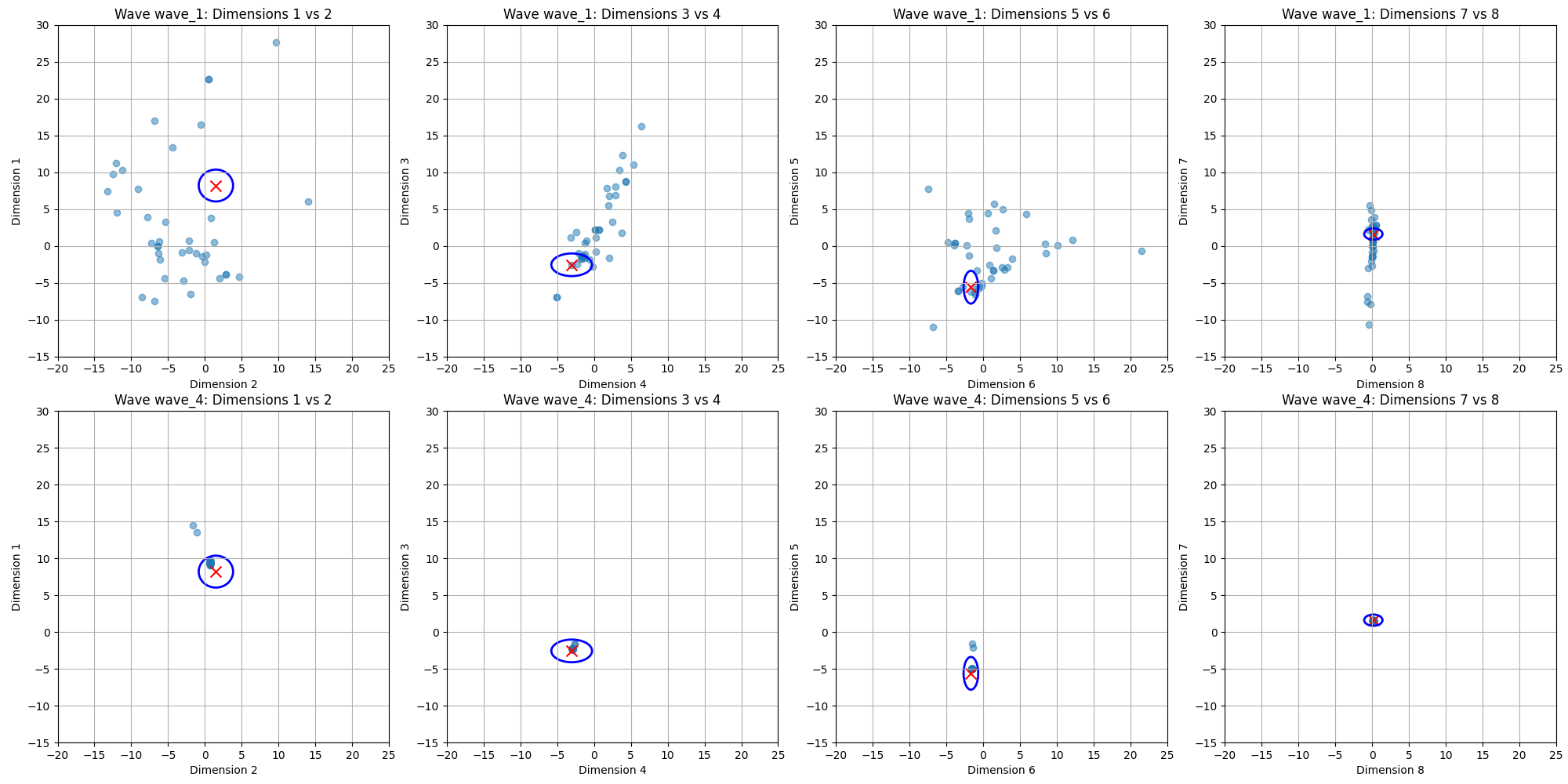}
    \caption{Comparison of the distribution of the PCA-reduced design point metrics between the first wave (upper row) and the last wave (lower row) for the best NPQC trial with the following hyperparameter configuration specified in \cref{tab:HPO}: $(N=6, L=2, \nosamplepoints = 1 \times 10^4, \trainkernelonlyonce = 0, \implausibilitythresholdmin \approx 1.096, \implausibilitythresholdmindecayfactor \approx 0.382, \maxnoimplausibilities = 0, \convergencethresholdmetrics = 0.25, \randomnessseed = 42)$. The multi-dimensional points are plotted by taking out neighboring principal components (dimensions). Red crosses and blue ellipsoids indicate the values corresponding to the parameter truth and the observational uncertainty. By chance, only a few points lie inside initially. The last wave, on the other hand, shows almost no points outside the target regions.}
    \label{fig:HMMetricsConvergence}
\end{figure*}

A detailed comparison of the design point distribution in the first and the last wave, and how close their PCA-reduced metrics are to the observational uncertainty, can be found in \cref{fig:HMMetricsConvergence}.

